\documentclass[12pt,preprint]{aastex}

\newcommand{\etal}{\textit{et~al.}\/}
\newcommand{\ie}{\textit{i.e.}\/}
\newcommand{\eg}{\textit{e.g.}\/}

\shorttitle{Evaluation of the ALMA Prototype Antennas}
\shortauthors{Mangum \etal}

\begin{document}
\title{Evaluation of the ALMA Prototype Antennas\footnote{The performance results presented in this publication were part of a
comprehensive technical evaluation process used to evaluate the ALMA
prototype antennas which concluded in April 2005.}}

\author{Jeffrey G.~Mangum}
\affil{National Radio Astronomy Observatory, 520 Edgemont Road,
  Charlottesville, VA  22903, USA}
\email{jmangum@nrao.edu}

\author{Jacob W.~M.~Baars}
\affil{Max-Planck-Institut f\"ur Radioastronomie, Auf dem H\"ugel 69,
  D-53121 Bonn, Germany and \\
  European Southern Observatory, D-85748 Garching, Germany}
\email{jacobbaars@arcor.de}

\author{Albert Greve}
\affil{Institut de Radio Astronomie Millim\'etrique, 300 rue de la
  Piscine, Domaine Universitaire, 38406 Saint Martin d'H\'eres,
  France}
\email{greve@iram.fr}

\author{Robert Lucas}
\affil{Institut de Radio Astronomie Millim\'etrique, 300 rue de la
  Piscine, Domaine Universitaire, 38406 Saint Martin d'H\'eres,
  France}
\email{lucas@iram.fr}

\author{Ralph C.~Snel}
\affil{SRON Netherlands Institute for Space Research, Sorbonnelaan 2,
3584 CA UTRECHT, The Netherlands}
\email{R.Snel@sron.nl}

\author{Patrick Wallace}
\affil{Rutherford Appleton Laboratory, Chilton, Didcot, Oxon OX11 0QX,
  UK}
\email{ptw@star.rl.ac.uk}

\and

\author{Mark Holdaway}
\affil{National Radio Astronomy Observatory, 949 North Cherry Avenue, 
  Tucson, AZ  85721, USA}
\email{mholdawa@nrao.edu}

\slugcomment{To appear in the September 2006 issue of PASP}

\begin{abstract}

The ALMA North American and European prototype antennas have been
evaluated by a variety of measurement systems to quantify the major
performance specifications. Nearfield holography was used to set the
reflector 
surfaces to 17~$\mu$m RMS. Pointing and fast switching performance was
determined with an optical telescope and by millimeter wavelength
radiometry, yielding $2^{\prime\prime}$ absolute and
$0.6^{\prime\prime}$ offset pointing accuracies. Path length
stability was measured to be $\lesssim 20~\mu$m over 10~minute time
periods using optical measurement devices. Dynamical
performance was studied with a set of accelerometers, providing data
on wind induced tracking errors and structural deformation.
Considering all measurements made during this evaluation, both
prototype antennas meet the major ALMA antenna performance
specifications.

\end{abstract}

\section{Introduction}
\label{intro}

In the early stages of the collaboration between the National Radio
Astronomy Observatory (NRAO) and the European Southern Observatory
(ESO) toward 
the joint design, construction and operation of a large millimeter
array (now called the Atacama Large Millimeter Array, or ``ALMA'',
project), it was agreed that each of the 
partners would acquire from industry a prototype of the planned 12 m
diameter antennas. The specifications of these antennas are beyond the
current state of the art in accurate reflector antenna
technology. Considering that 64 of these antennas would eventually be
needed, this step appeared warranted to mitigate performance risk and
increase competition among prospective manufacturers. It was also agreed
to place the antennas next to each other at a suitable site and
subject them to an extensive evaluation program to be carried out by a
joint team of scientists and engineers drawn from the institutes of
the ALMA collaboration.

The two ALMA prototype antennas are located at the site of the Very
Large Array (VLA) in New Mexico. This site provides a reasonable
compromise between the quality of the atmosphere, necessary for
millimeter wavelength operation, and ease of construction and
operation. The latter is provided by the existing infrastructure of
the VLA site. The atmospheric characteristics however allow only
operation at relatively long millimeter wavelengths during the dry
winter months. Thus, the antennas have been equipped with evaluation
receivers for the wavelength bands near 3 and 1.2 mm. 

The evaluation of the antennas has been carried out by the Antenna
Evaluation Group (AEG), consisting of experienced ``antenna
integrators and commissioners'' from both organizations, Associated
Universities Incorporated (AUI)/National Radio Astronomy Observatory
(NRAO) and the European Southern Observatory (ESO). The charge of the
AEG was to subject both antennas to a series of 
identical tests that would indicate the compliance (or not) of the
antennas with the specifications. The core of the AEG is composed of
the authors of this paper.

\section{Description of the Prototype Antennas}
\label{antdescript}

\begin{figure}
\resizebox{\hsize}{!}{
%\centering
%\includegraphics[scale=0.48,angle=0]{f1}
\includegraphics[angle=0]{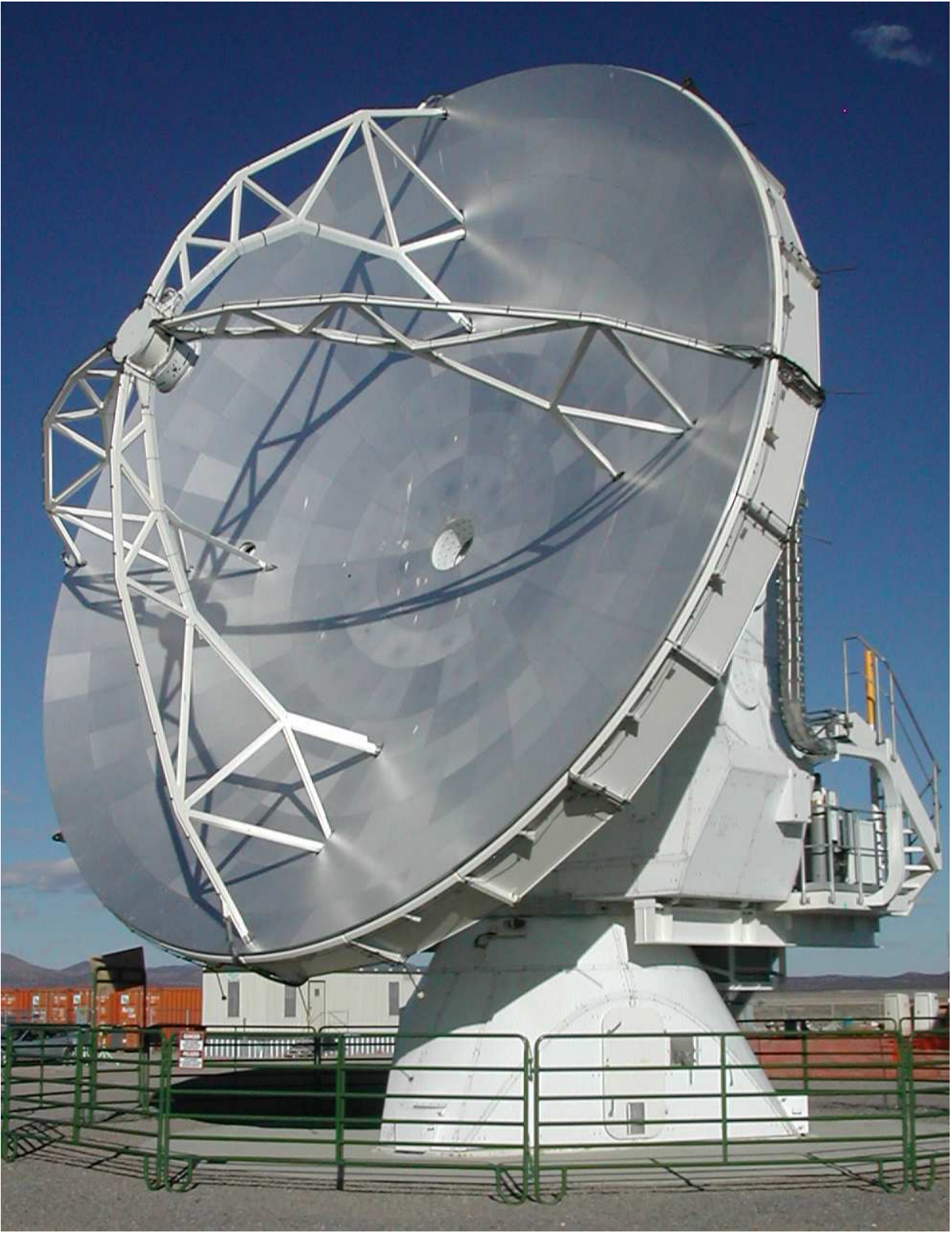}} % FOR PS VERSION ONLY
\caption{The VertexRSI ALMA prototype antenna.}
\label{fig:vertexrsi}
\end{figure}

\begin{figure}
\resizebox{\hsize}{!}{
%\centering
\includegraphics[angle=0]{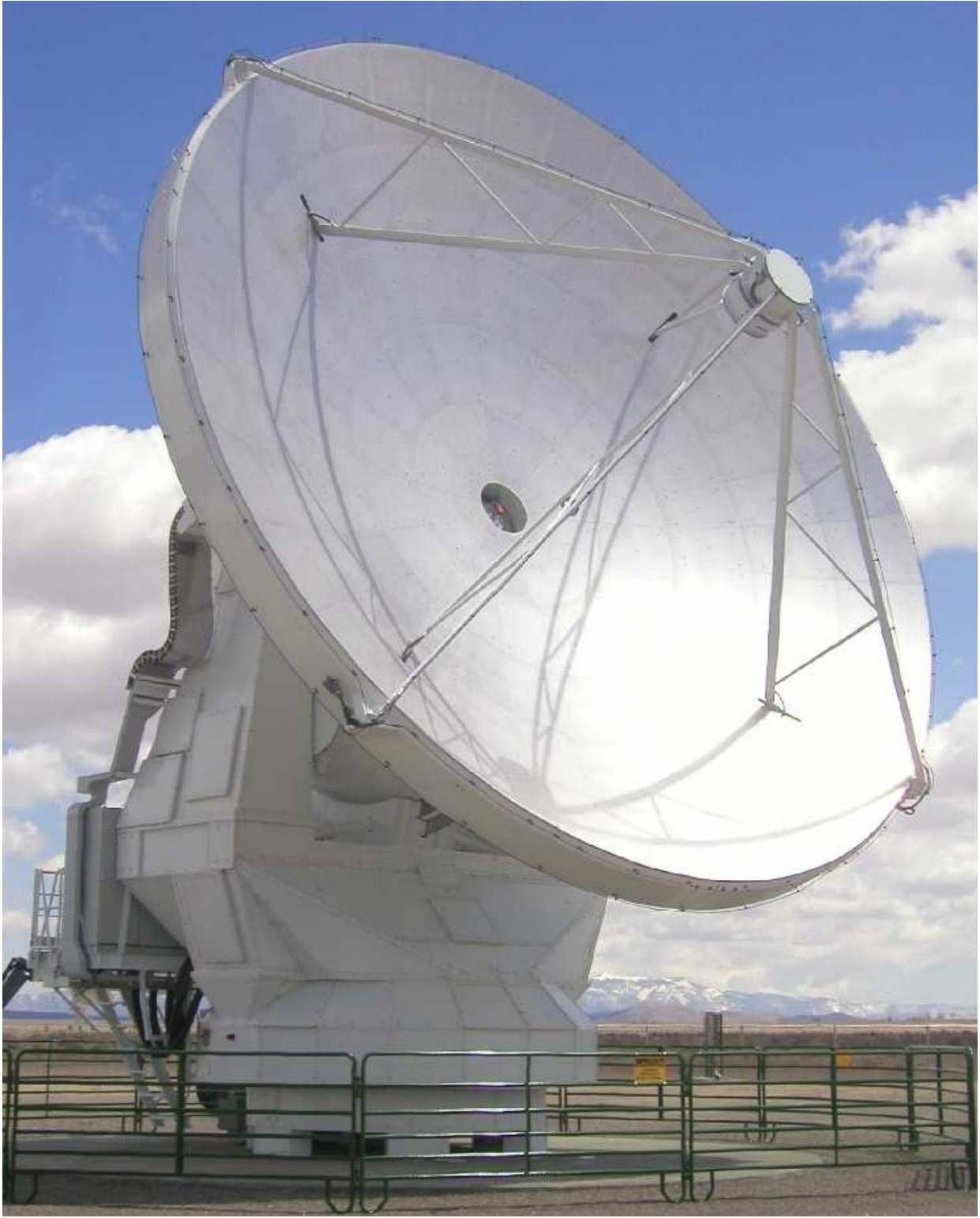}}
\caption{The AEC ALMA prototype antenna.}
\label{fig:aec}
\end{figure}

The ALMA prototype antennas are alt-azimuth mounted Cassegrain
reflector systems of 12~m diameter with a reflector surface and
pointing accuracy 
suitable for observations in the 0.3 mm submillimeter band. VertexRSI
delivered an antenna to AUI/NRAO (see Figure~\ref{fig:vertexrsi}) and
ESO obtained an antenna from the consortium Alcatel/EIE (AEC; see
Figure~\ref{fig:aec}). The main characteristics of the antennas
are summarized in Table~\ref{tab:antennas}.

\begin{table}
\centering
\caption{Main Properties of the ALMA Prototype Antennas}
\begin{tabular}{|c|p{6cm}|p{6cm}|}
\hline
\textbf{Property} & \textbf{VertexRSI} & \textbf{Alcatel/EIE (AEC)} \\
\hline\hline
\textit{Base/Yoke/Cabin} & Insulated Steel & Insulated Steel/Insulated
Steel/ CFRP \\
\hline
\textit{BUS} & Al honeycomb with CFRP plating, 24 sectors, open back,
covered by removable GFRP\tablenotemark{a}
sunshades & Solid CFRP plates, 16 sectors, closed back 
sectors glued and bolted together. \\ 
\hline
\textit{Receiver Cabin} & Cylindrical; Invar/Thermally-Stabilized Steel & CFRP; Direct connection cabin to BUS \\
\hline
\textit{Base} & 3-point support - bolt connection with foundation
     & 6-point base support - flanged attachments with foundation\\
\hline
\textit{Drive} & Gear and Pinion & Direct drives on both axes with linear motors \\
\hline
\textit{Brakes} & Integrated on servo motor & Hydraulic disk brakes \\
\hline
\textit{Encoders} & Absolute (BEI) & Incremental (Heidenhain) \\
\hline
\textit{Panels} & 264 Panels, 8 rings,  machined Al, open back, 8
adjusters (3 lateral/5 axial) per
panel & 120 Panels, 5 rings. Al-honeycomb core with replicated Ni
skins, Rh coated. 5 adjusters per panel \\
\hline
\textit{Apex/Quadripod} & CFRP structure, $+$ configuration & CFRP structure
, $\times$ configuration \\
\hline
\textit{Focus Mechanism} & Hexapod (5 DOF) & Three axes (x,y,z) mechanism \\
\hline
\textit{Total Mass} & $\sim 108$ tonnes & $\sim 80$ tonnes \\
\hline
\textit{Mass Dist (El/Az)} & 50\%/50\% & 35\%/65\% \\
\hline
\end{tabular}
\tablenotetext{a}{Glass Fiber Reinforced Plastic.}
\label{tab:antennas}
\end{table}

The two antennas show interesting differences. To minimise the
structural deformations due to temperature changes, both make
extensive use of carbon-fiber reinforced-plastic (CFRP) for the
back-up structure (BUS) of the reflector. The BUS is a box-structure,
thus avoiding the need to design and fabricate the intricate joints of
a space-frame structure in CFRP. The AEC group has realised the entire
elevation structure (receiver cabin and connection to BUS) in CFRP,
while VertexRSI applies insulated steel for the cabin and Invar for
the connection cone to the BUS.  For the drive systems, VertexRSI uses
a traditional pinion and gear rack system, while AEC uses a direct
linear drive, applied for instance in the Very Large Telescope (VLT)
optical telescopes or the Berkeley Illinois Maryland Association
(BIMA) antennas. 

There are significant differences in the choice of size and
technology of the reflector panels. The VertexRSI antenna has
relatively small panels, made of machined aluminum and chemically
etched to achieve the required effective scattering of solar heat
radiation. They have separate axial (5) and lateral (3) adjustable
supports. The AEC panels are of a novel design by Media
Lario. Electroformed nickel skins, replicated from a machined steel
mold, are bonded to a 20~mm thick aluminum honeycomb core. To improve
the thermal behaviour under sunlight, the surface is coated with
200~nm of rhodium. These larger panels have 5 adjustable axial
supports, which also provide sufficient stiffness in the lateral
direction. Because of the more extensive use of the relatively
lightweight CFRP, the AEC antenna is significantly lighter than the
VertexRSI antenna.

\section{Major Specifications}

The ALMA antennas will be used to a shortest wavelength of about
0.3~mm while located at 5000~m altitude in Chile under rather extreme 
conditions of strong wind and sunshine. Consequently, the requirements
on the accuracy and stability of the reflector surface contour and of
the pointing and tracking behaviour are very high. In fact, no radio
telescope of comparable size has been designed to the specifications
listed below.

It has been the main task of the AEG to establish the compliance of
the ALMA prototype antennas with the major performance specifications
(see Table \ref{tab:results}).

These specifications must be satisfied under all attitude angles of
the antenna and all environmental conditions, in particular wind of 6
m/s (day) and 9 m/s (night), as well as full solar illumination from
changing directions. 

Because of the interferometric mode of operation of ALMA, some
unusual specifications have been introduced, such as path length
variations and the capability of very fast switching between two
relatively close points on the sky. It is beyond the scope of this
paper to elaborate on the reasons for these specifications.

\section{Evaluation Results}
\label{eval}

The major performance specifications, results, and the
primary operating conditions during which these specifications apply,
for the ALMA prototype antennas are listed in Table~\ref{tab:results}. 
In the following we describe how the performance of the
ALMA prototype antennas was derived within the context of each
performance specification.  The technical details that describe the
various measurement systems and techniques used to derive these
results can be found in \citet{Mangum2004}, \citet{Baars2006},
\citet{Greve2006}, \citet{Holdaway2004}, \citet{Snel2006}, 
\citet{Wallace2004}, \citet{Mangum2004b}, and \citet{Mangum2004a}. In
the following we describe the measurements which lead to the summary
results listed in Table~\ref{tab:results}.

\begin{table}[h!]
\centering
\caption{Prototype Antenna Specifications and Performance Overview}
\begin{tabular}{|l|p{4cm}|l|l|}
\hline
Property & Specification & \multicolumn{2}{c|}{Performance} \\
&& VertexRSI & AEC \\
\hline
Surface Accuracy & 25~$\mu$m RMS with 20~$\mu$m RMS goal &
$16\pm5$~$\mu$m & $17\pm5$~$\mu$m \\ 
Absolute Pointing & 2~arcsec over all sky & 1.3--1.8~arcsec & 2.0--2.6
arcsec \\ 
Offset Pointing & 0.6~arcsec over 2 degrees radius & 0.3--1.1~arcsec &
0.3--0.8~arcsec \\ 
Fast Switching & 1.5 degree move in 1.5~seconds, settle to 3~arcsec
peak pointing error & 1.5--1.8~seconds & 1.4--1.8~seconds \\ 
Path Length Stability & 15/20~$\mu$m~(Nonrepeatable/Repeatable) &
$\leq$(15,21,32)~$\mu$m$^a$ & $\leq$(15,18,30)~$\mu$m$^a$ \\ 
\hline
\multicolumn{4}{|l|}{Primary Operating Conditions: $-20~C \leq T_{amb}
  \leq +20~C$; $\Delta T_{amb}\leq$0.6/1.8 C} \\
\multicolumn{4}{|l|}{~~in 10/30~minutes; $V_{wind} \leq$ 6/9 m/s
  (day/night); Full solar loading} \\
\multicolumn{4}{|l|}{$^a$~For $\Delta$t$<$ (3,10,30)~minutes,
  respectively.} \\ 
\hline
\end{tabular}
\label{tab:results}
\end{table}

\section{Reflector Surface Accuracy}

\subsection{Near-Field Holography Measurements}

\begin{figure}
\resizebox{\hsize}{!}{
%\centering
\includegraphics[angle=-90]{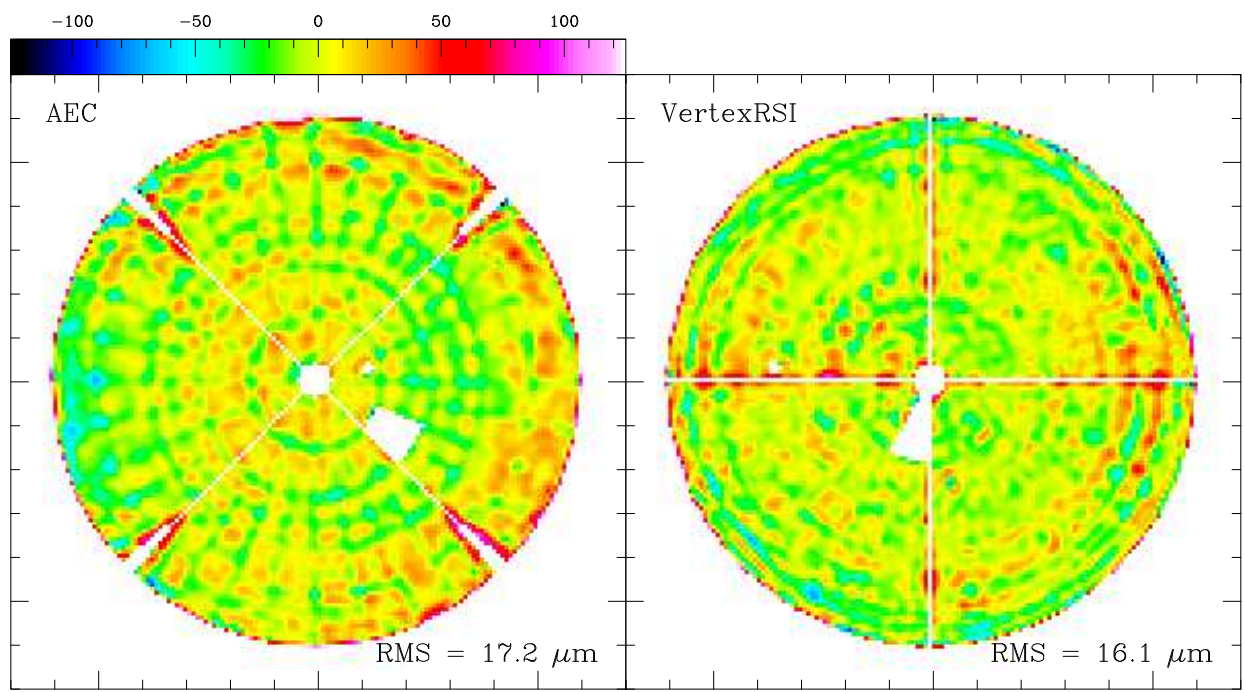}}
\caption{Typical final holographic surface map of the AEC (left) and
    VertexRSI (right) prototype antennas.  Note that the ``$\times$''
    (AEC) and ``+'' (VertexRSI) feed leg structure.  The poor
    measurement results obtained near the feed leg structures of each
    antenna are due to the difficulties encountered with holographic
    measurements near these structures.}
\label{fig:holo}
\end{figure}

The near-field holographic method was used to measure and set the
surface to an accuracy of better than 20~$\mu$m RMS. A transmitter on
a 50 m high tower at a distance of 310 m provided the signal at 79 or
104 GHz. See \citet{Mangum2004} and \citet{Baars2006} for further
details describing this holographic technique.  Figure~\ref{fig:holo}
shows a typical surface error map derived from these holographic
measurements. 

The holographic measurement and setting of both antennas was performed
immediately after the antennas became available for evaluation. During
a period of about one year the antennas were subjected to a number of
hard loads, like fast switching tests, drive system errors resulting
in strong vibrations, and high-speed emergency stops. Also,
the influence of wind and diurnal temperature variations on the
surface stability was a point of concern. It was therefore decided to
close the evaluation program with a second holographic measurement of
the reflector surfaces. This was done in December 2004 to February
2005 during relatively good atmospheric conditions. It was during
these measurements that we discovered that we had not properly taken
care of a correction explained in \citet{Baars2006}. This
correction, which did not affect the final interpretation of the
prototype antenna surface accuracy performance, was subsequently
applied properly.  The final surface maps are shown in
Figures~\ref{fig:holosVertex} and \ref{fig:holosAEC}. Both antennas
were set to a surface accuracy of 16--17~$\mu$m RMS.

\subsubsection{Overview}
\label{holooverview}

\paragraph{VertexRSI Antenna:}
\label{vrsiresults}

The antenna was delivered with a nominal surface error of 80~$\mu$m
RMS, as determined from a photogrammetric measurement. Our first
holography map showed an RMS of approximately 85~$\mu$m. A first
setting of the surface resulted in an RMS of 64~$\mu$m. In four more
steps of holographic measurement followed by adjustment the surface
error decreased to 19~$\mu$m RMS.

\begin{figure}
\centering
\includegraphics[scale=0.65]{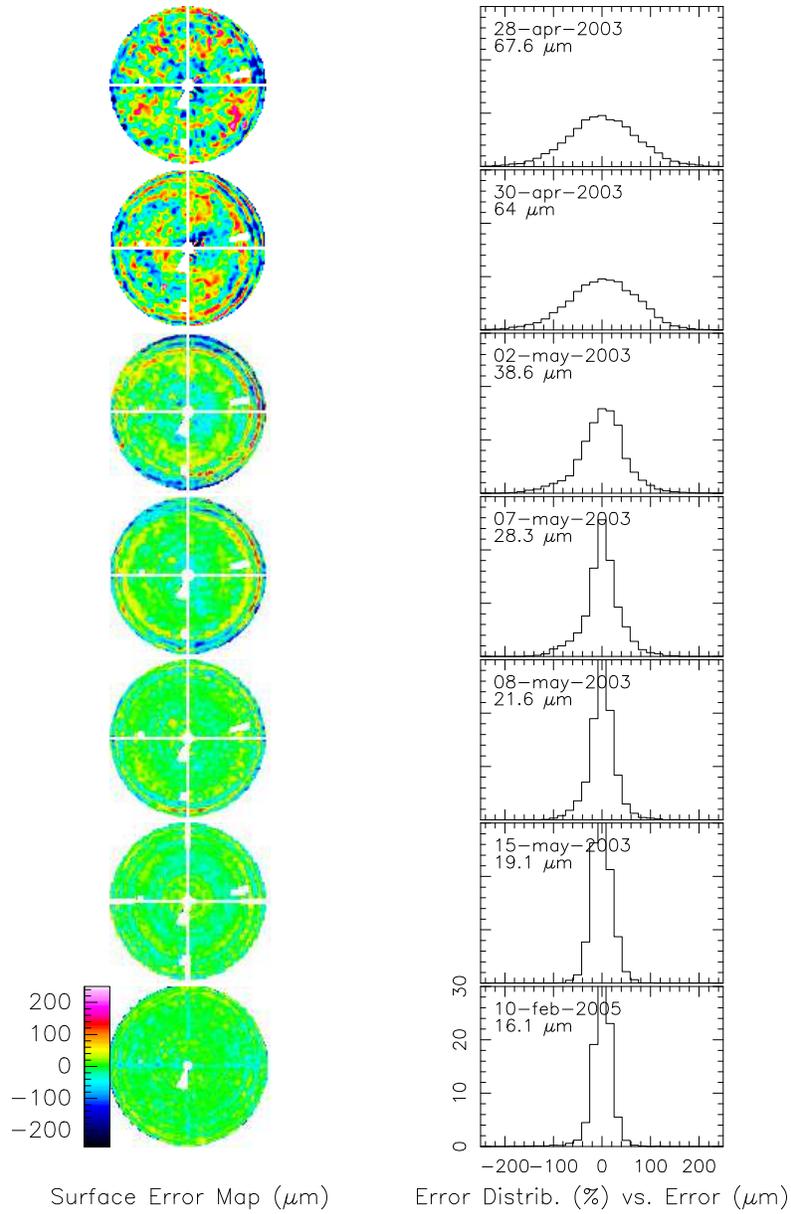}
\caption{Sequence of surface error maps with intermediate panel
  setting for the VertexRSI antenna. The surface contours are shown on
  the left side; the error 
  distribution on the right. The white cross and the small white areas
  represent the quadripod and a few faulty panels and were not
  considered in the calculation of the RMS error.  Progression to the 
  final surface RMS included holography system checkout, so does not
  represent the time required to set the VertexRSI antenna to its
  final surface.}
\label{fig:holosVertex}
\end{figure}

The sequence of surface error maps, along with the RMS and the error
distribution, is shown in Figure~\ref{fig:holosVertex}.
As allowed in the specification, we have applied a weighting over 
the aperture proportional to the illumination pattern of the feed. 
This essentially diminishes the influence of the surface errors in 
the outer areas of the reflector. The white areas in the surface error
maps are the quadripod, optical pointing telescope, and a few bad
panels, that could not be set accurately.  All of these structures
were left out of the calculation of the final overall RMS value.

With increasing accuracy the presence of an artefact in the outer area
of the aperture became apparent. There is a ``wavy'' structure in the
outer section with a ``period'' too large to be inherent in the
panels. Experiments with absorbing
material showed that it was not caused by multiple reflections. The
effect can be described by a DC-offset in the central point of the
measured antenna map, \ie\ some saturation on the point with the
highest intensity. By adjusting this offset in the analysis software,
most of the artefact could be removed.  This has been done with the final
data.  The additional set of follow-up holography measurements in
December 2004 -- February 2005 did not suffer from this signal
saturation, and no artefact was observed in these follow-up maps.
Checks of the holography hardware indeed suggest that the 2003
holography measurements did experience a small amount of signal
saturation.

\begin{figure}
\centering
\includegraphics[scale=0.5]{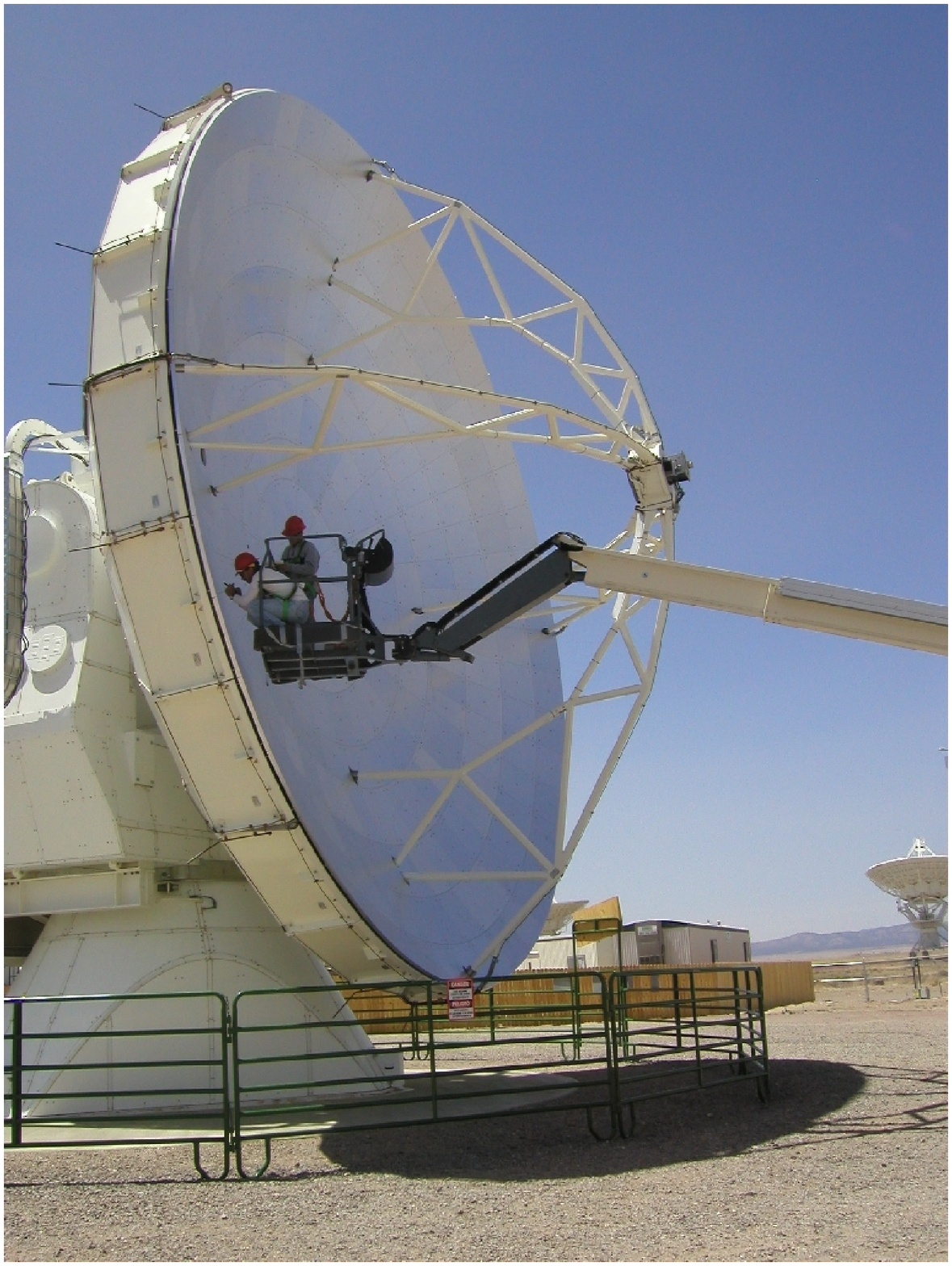} \\
\includegraphics[scale=0.5]{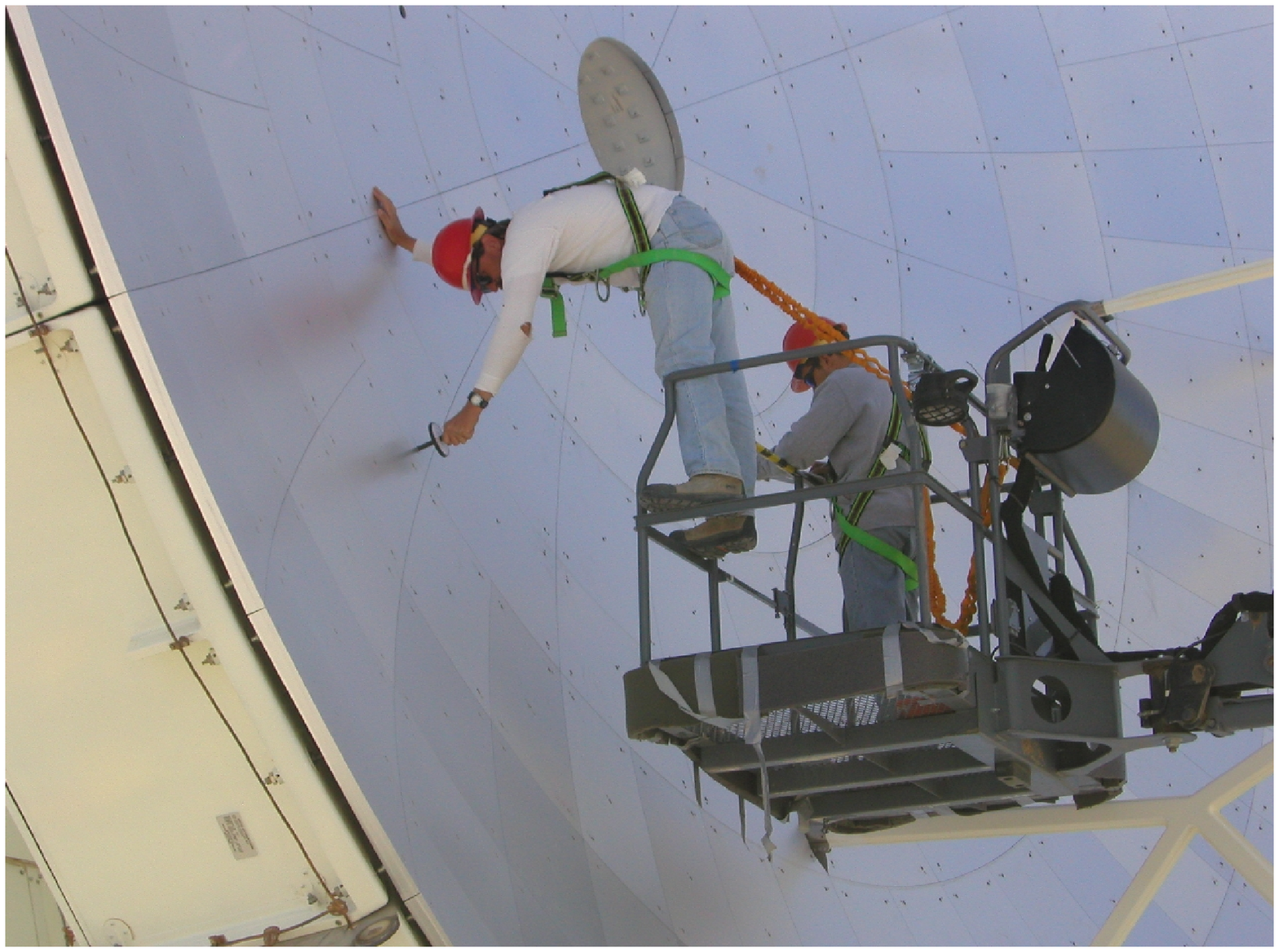} \\ % FOR PS VERSION ONLY
\caption{Panel adjustment of the VertexRSI prototype antenna.}
\label{fig:holosetting}
\end{figure}

The adjustments were done with a simple tool. Two people on a man-lift
approached the surface from the front, where the adjustment screws are
located (see Figure~\ref{fig:holosetting}). The time needed for an
adjustment of the total of 1320 adjusters was 8 hours, as required by
the specification.

The best surface maps were obtained at night. During the spring 2003
period they consistently show an RMS of about 20~$\mu$m. Daytime maps
tend to be somewhat worse; typical values of the RMS lie between 20
and 25~$\mu$m. Part of this is certainly due to the atmosphere, even
over the short path length of 315 m.

\paragraph{AEC Antenna:}
\label{aecresults}

The apex structure of the AEC antenna does not enable us to mount the
holography receiver inside the apex structure cylinder, as in the case
for the VertexRSI antenna. Thus in this case the receiver was bolted to
the flange on 
the \textit{outside} of the apex structure. Consequently, the feedhorn was
brought to the required position by a piece of waveguide of about
500~mm length. This caused significant attenuation in the received
signal from the reflector to the mixer. Considering the available
transmitter power, we concluded that this would not jeopardise our
measurement accuracy significantly.

The AEC antenna surface was set by the contractor with the aid of a
Leica laser-tracker. The RMS of the surface was reported by the
contractor to be 38~$\mu$m. After this measurement an accident caused
the elevation structure to run onto the hard stops at high speed. The
contractor decided to repeat the surface measurement and obtained an
RMS of 50~$\mu$m with some visible ``astigmatism'' in the surface. 

\begin{figure}
\centering
\includegraphics[scale=0.70]{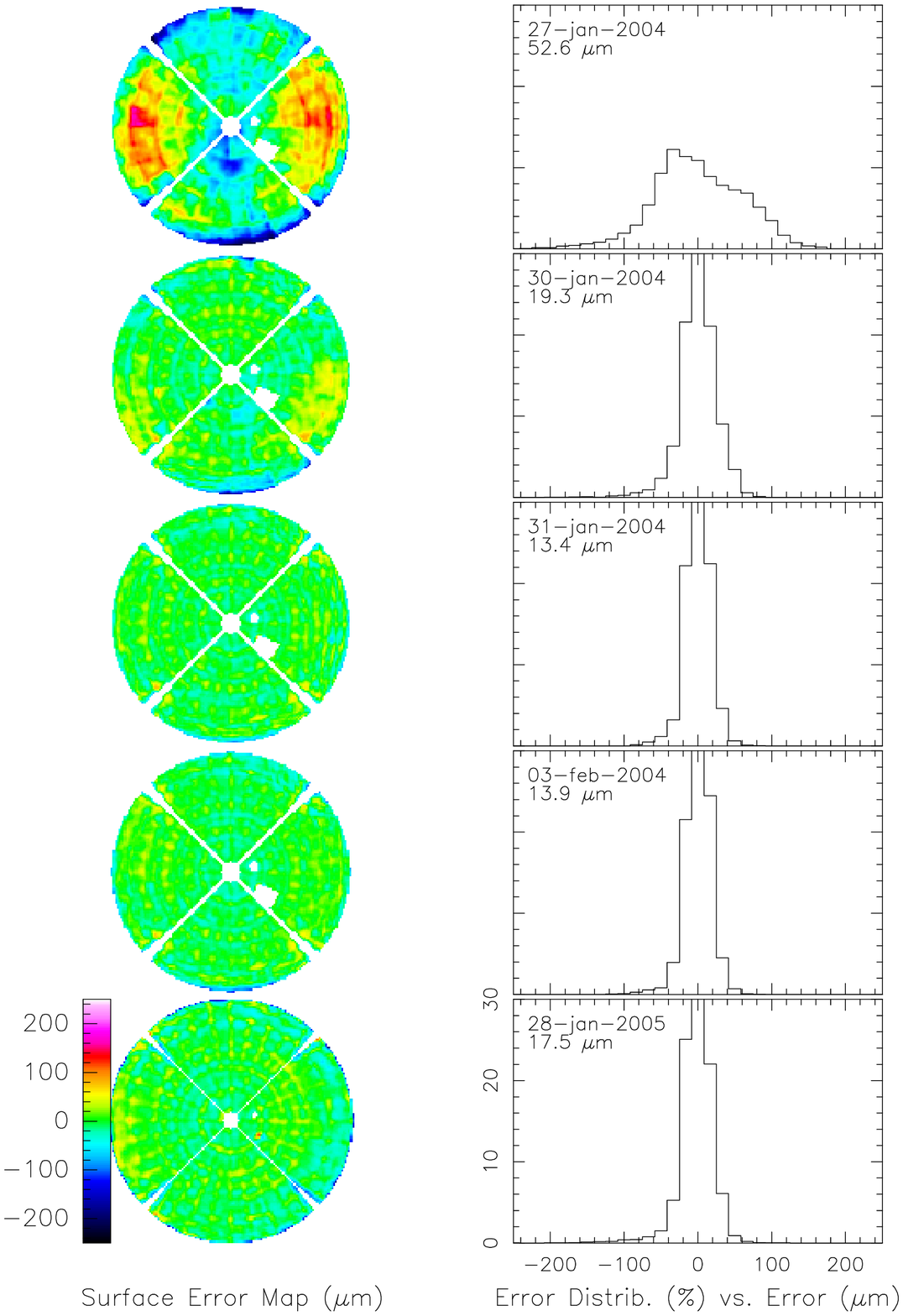}
\caption{Sequence of surface error maps with intermediate panel
  setting for the AEC antenna. The surface contours are shown on the
  left side; the error 
  distribution on the right. The white cross and the small white area
  represent the quadripod and a faulty panel and were not considered
  in the calculation of the RMS error.} 
\label{fig:holosAEC}
\end{figure}

Our first holography map indicated an RMS of 55~$\mu$m with a clearly
visible astigmatism. We could identify the high and low regions with
those on the final AEC measurement. With two complete
adjustments we surpassed the goal of 20~$\mu$m. A third partial adjustment
improved the surface RMS to about 14~$\mu$m . There is no indication of
the ``artefact'' seen in the VertexRSI antenna. There is one panel
with a large 
deviation over part of the area. This is believed to have been caused
during the measurement and setting procedure by the contractor. We
have not included this panel in the computation of the final RMS
value. The results of the consecutive adjustments are summarised in
Figure~\ref{fig:holosAEC}.  The last panel in this figure shows the
final map after the repeated measurement and setting in January 2005.

The adjustments were done with a tool provided by the contractor. It
was similar to the one used by us on the VertexRSI antenna, but it was
calibrated in ``turns'' rather than in micrometres. 
Again two people on a manlift approached the surface
from the front, where the adjustment screws are located. The time
needed for an adjustment of the total of 600 adjusters was 7 hours,
well within the specification of 8 hours.

\clearpage
\subsubsection{Temporal Surface Stability}

\paragraph{VertexRSI Antenna:}
\label{vrsisurfstab}

\begin{figure}
\centering
\includegraphics[scale=0.65,angle=-90]{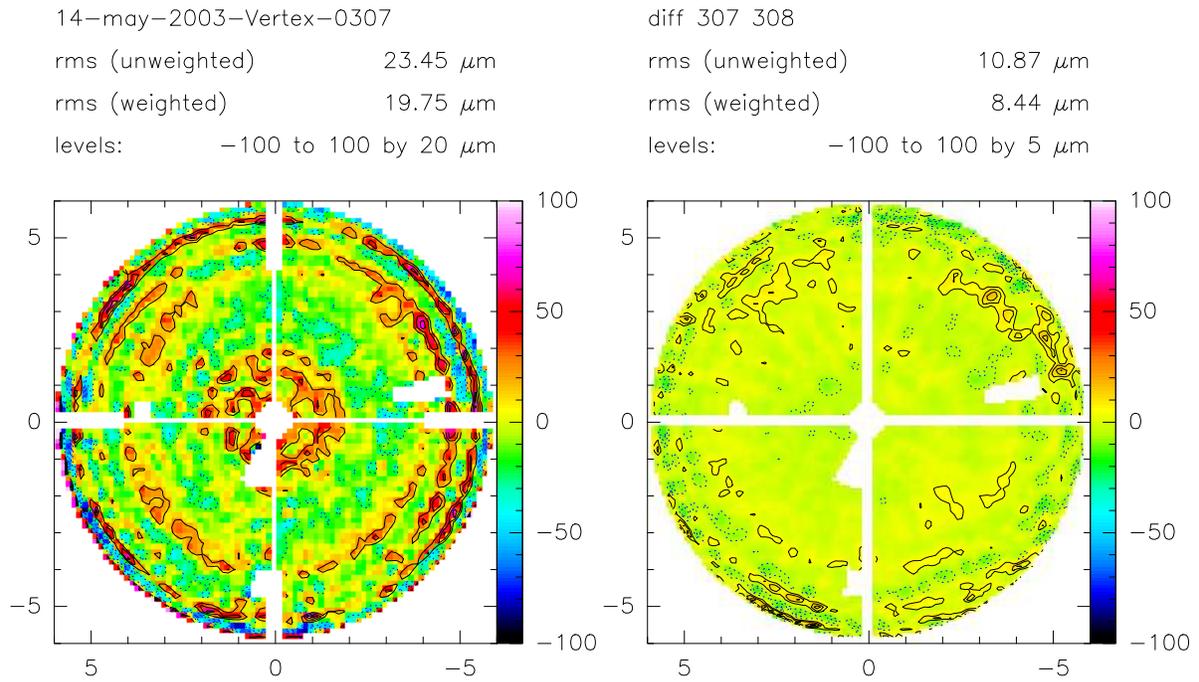}
\caption{Example of the repeatability of the VertexRSI holography
  measurements. The map on the right is the difference between the one
  at left and a map made one hour afterwards. The RMS of the
  difference maps is about 8~$\mu$m, which is commensurate with the
  expected value due to noise and atmospheric fluctuations.}
\label{fig:VertexDiff}
\end{figure}

To estimate the accuracy and repeatability
of the measurements, we produced difference maps between successive
measurements throughout the measurement period.  The RMS difference
between consecutive maps 
is typically less than 10~$\mu$m, and usually $\sim 8$~$\mu$m. An example of a
difference map is shown in Figure~\ref{fig:VertexDiff}.  The map of
measurement number 307 is shown on the left, while the right hand side
shows the difference between map 307 and 308, made about one hour
later.

We have made many maps after the final setting of the surface while
changing the orientation of the antenna with respect to the Sun. 
Maps taken over a few consecutive days were obtained at widely
different temperatures, while also the wind conditions varied over
time. The measured RMS error during a 30 hour period in early May 2003
varied between 20 and 23~$\mu$m. During this period the wind was calm
($<5$~m/s) and temperature changes of 15~C were encountered. A later 5
day series in mid June gave RMS errors from 22-26~$\mu$m with
temperature variation up to 20~C, wind speeds up to 10 m/s and
periods of full sunshine. Most of this increase is believed to be due
to the deteriorating atmospheric conditions at the VLA site during
summer, when the humidity was significantly higher than normal. The
much better results of 17~$\mu$m obtained during the cold and dry
winter period also point to a significant atmospheric component in the
spring and summer results. 

However, some of the changes will be caused by temperature and
wind. To increase the RMS from 20 to 22~$\mu$m, the ``additional''
component has a magnitude of 9~$\mu$m RMS. Such a contribution can be
expected from the calculated values of 4~$\mu$m each for wind and
temperature for the panels, and 5~$\mu$m for wind and 7~$\mu$m for
temperature for the BUS. These numbers are all within the
specification. Actually, the measured differences are close to those
expected from the estimated accuracy of the holography measurement and
the measured RMS differences in consecutive maps of about 8~$\mu$m.

\paragraph{AEC Antenna:}
\label{aecsurfstab}

\begin{figure}
\centering
\includegraphics[scale=0.65,angle=-90,bb=49 1 437 684]{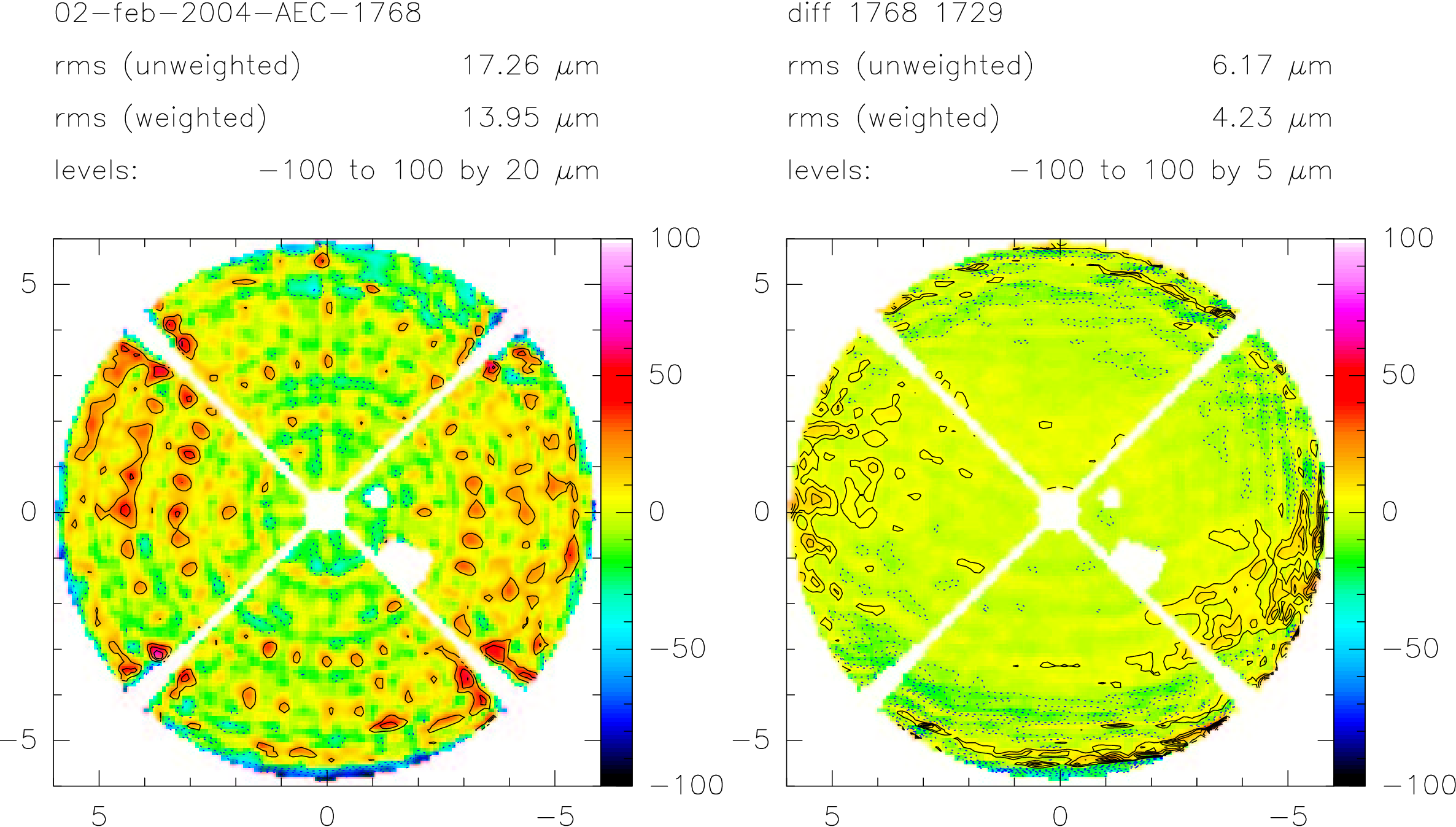}
\caption{Example of the repeatability of the AEC holography
  measurements. The map on the right is the difference between the one
  at left and a map made one hour afterwards. The RMS of the
  difference maps is about 5~$\mu$m.} 
\label{fig:AECDiff}
\end{figure}

In Figure~\ref{fig:AECDiff} we show one of the final results and a
difference map of this and the following measurement, made one hour
later. The difference map indicates a repeatability of $\sim 5 \mu$m
RMS. There is no indication of the wavy structure in the outer region of
the aperture, as was the case for the VertexRSI antenna. We ascribe
this to the lower signal level due to the long piece of waveguide
between feed and mixer.

We made a series of 16 maps over a period of more than two
days in early February 2004. Temperatures ranged from $+2$ to $-10$~C,
while the wind was mostly calm with some periods of speeds up to 10
m/s. During one day there was full sunshine. The measured RMS error is 
very constant with a peak to peak variation of less than 2~$\mu$m on an
average of 14~$\mu$m. The differences are fully consistent with the
allowed errors under environmental changes and also of the same order
as the measurement accuracy. We believe that the significantly better
overall result is mainly due to the much drier and more stable
atmosphere during these measurements as compared with the summer data
from the VertexRSI antenna.

%\clearpage
\subsubsection{Surface Stability with Changes in Ambient Temperature}

\paragraph{VertexRSI Antenna:}
\label{vrsisurftemp}

\begin{figure}[h!]
\centering
\includegraphics[scale=0.80]{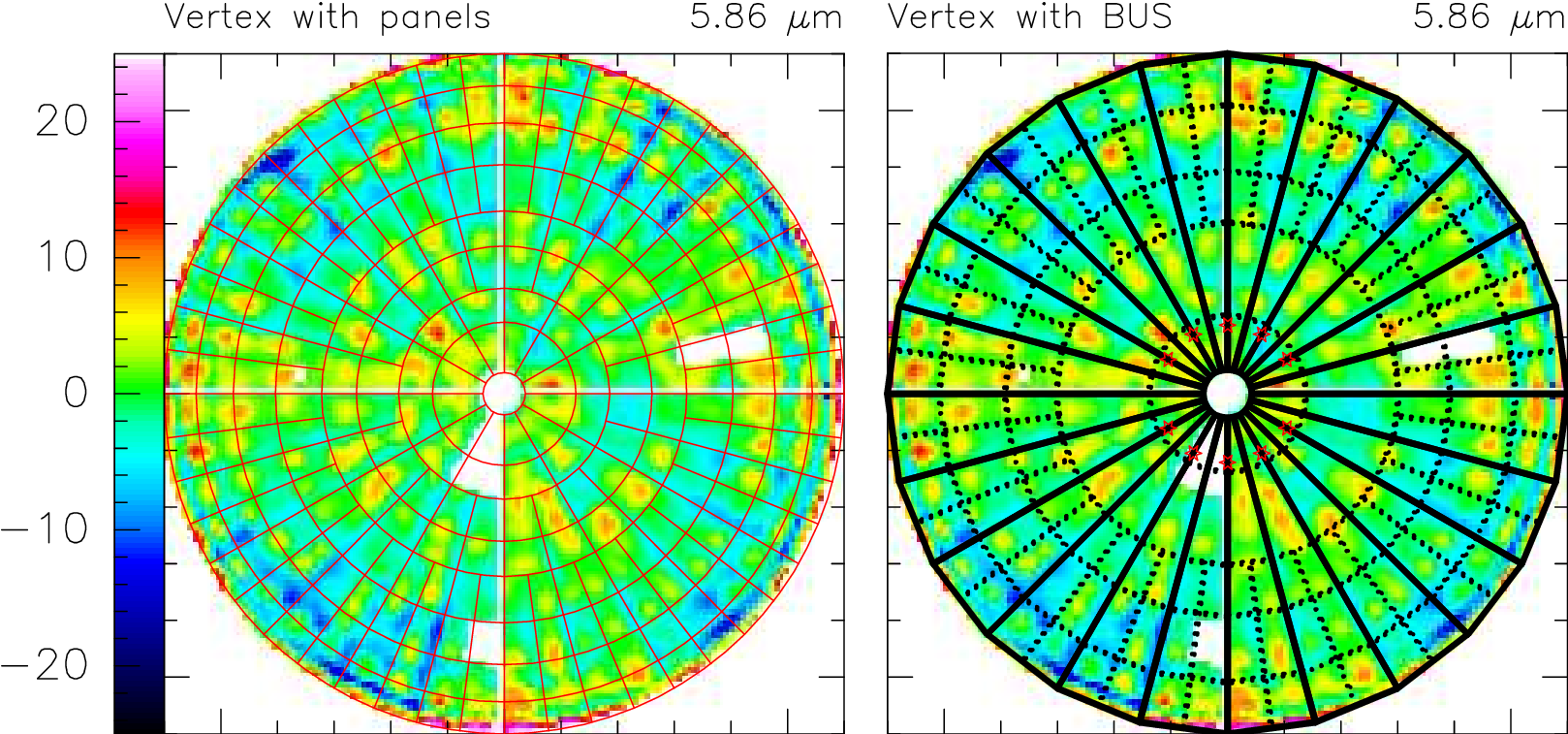}
\caption{VertexRSI temperature span plots with panels (left)
  and BUS (right) boundaries overlain.  To put these
  measurements on the same scale the data have been uniformly scaled
  to a temperature difference of 10~C.  The RMS
  surface deformation over this 10~C temperature difference is
  indicated in the upper right.}
\label{fig:vrsitfullspan}
\end{figure}

Further analysis of the VertexRSI time series data presented in
\S\ref{vrsisurfstab} has been made to extract the dependence of the
antenna surface to changes in ambient temperature.  The
analysis entailed the following:

\begin{itemize}
\item For each series the average of all maps in the series was
  subtracted from all the maps in the series. This 
suppresses dependencies on any long term effect.
\item In each series the maps were divided into several temperature
  ranges; in each temperature range the average map was computed.
\item Finally the difference between the coldest and warmest range
  averages were computed for each series.  From these data we derived
  the surface deformation caused by a temperature change of 10~C, as
  shown in Figure~\ref{fig:vrsitfullspan}.
\end{itemize}

From this analysis we note that:

\begin{itemize}
\item Many of the samples were for temperatures outside the primary
  operating range of $-20$ to $+20$~C.
\item The results for the spring 2003 and winter 2005 periods are very
  consistent with each other.
\item The magnitude of the RMS deformations is $\sim 0.6$-$0.7\mu$m/K.
  This value seems high.  It is not clear to which degree it should be
  split into panels and BUS and between absolute and gradient terms in
  the surface error budget. Assuming all effects to be linear with
  temperature, and antennas set at 0~C, we would expect RMS temperature
  contributions of $\sim 13\mu$m for the VertexRSI antenna at either
  end of the operational temperature range ($+20$~C or $-20$~C). 
\item The thermal deformations for the VertexRSI antenna appear to be
  a mixture of BUS (the BUS sector edge is sticking out at higher
  temperature) and panels, not all panels deforming equally. 
\item From the BUS overlay drawing (Figure~\ref{fig:vrsitfullspan}) it
  appears that there is a print-through of the connections of the BUS
  with the Invar cone (the turnbuckles). The small red islands on (or
  just outside) the second stippled ring are at the diameter of the
  outer support of the BUS on the cone. We are not certain whether
  they are at or in between the connections to the cone. In both cases
  such a print-through might reasonably be expected.
\item The BUS is also supported on the cone towards the center outside
  the inner stippled circle by about one fifth of the distance between
  the two stippled circles. Also there we see clear red islands of
  print-through.  These effects are less pronounced in the ``lower''
  section of the dish; actually they seem most pronounced in the left
  and right quadrants. We don't understand why this would be the
  case.
\item There are clear, narrow ``valleys'' along the radials where the
  sectors of the BUS are joined, especially in the outer half of the
  radius. They are not equally strong all around, but visible in most
  cases. The inner part is less clear, presumably because of the
  already existing effects of the support points. 
\item From the map with the panel outline, and ignoring the islands
  connected to the BUS, as argued above, there is some but not very
  much evidence for individual panel deformations. It is most likely
  the case that individual panel deviations are caused more by forces
  originating in the stiffer BUS than from the panel itself.
\end{itemize}

\paragraph{AEC Antenna:}
\label{aecsurftemp}

\begin{figure}[h!]
\centering
\includegraphics[scale=0.50]{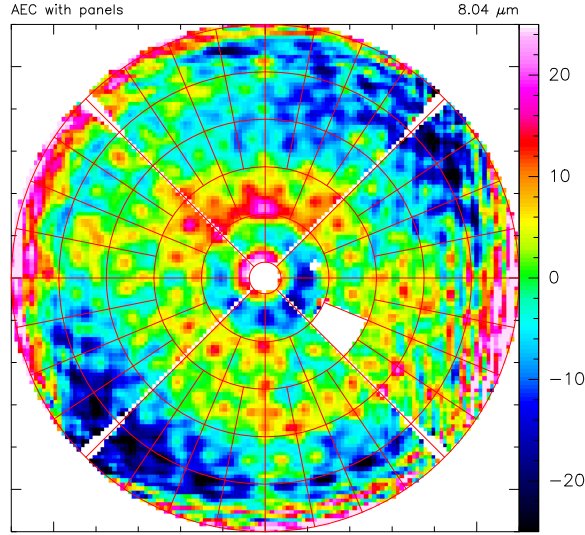}
\caption{AEC temperature span plot with panel boundaries overlain.
  The data have been uniformly scaled to a temperature difference of
  10~C.  The RMS surface deformation over this 10~C temperature
  difference is indicated in the upper right.}
\label{fig:aectfullspan}
\end{figure}

Further analysis of the AEC time series data presented in
\S\ref{aecsurfstab} has been made to extract the dependence of the
antenna surface to changes in ambient temperature.  The
analysis entailed the same sequence of calculation as those presented
for the VertexRSI antenna (see \S\ref{vrsisurftemp}).  
Figure~\ref{fig:aectfullspan} shows the temperature span assuming a
uniform 10~C temperature difference.

From this analysis we note that:

\begin{itemize}
\item The sampled temperature range is completely inside the primary
  operating condition ambient temperature range of $-20$ to $+20$~C.
\item The magnitude of the RMS deformations is $\sim 0.8\mu$m/K.
  This value seems high.  It is not clear to which degree it should be
  split into panels and BUS and between absolute and gradient terms in
  the surface error budget. Assuming all effects linear with
  temperature, and antennas set at 0~C, we would expect RMS temperature
  contributions of $\sim 16\mu$m for the AEC antenna at either
  end of the operational temperature range ($+20$~C or $-20$~C).
\item The large-scale deformation at 45 degrees position angle (which
  looks like astigmatism), along with the additional deformation of
  the inner ring, are hard to explain. These deformations are perhaps
  due to the temperature gradients in the cabin and the BUS, coupled
  to the quadripod connection points.  The vertical stripes on the
  edges of the AEC maps are an artefact of the measurement not related
  to temperature.
\item In principle there should be little print-through structure in
  the difference maps, due to the fact that the BUS design is more
  homogeneous, being all CFRP, and it should have a continuous
  transfer of forces to the cabin, which is also the same material.
\item We don't see any individual panel deformation.
\end{itemize}

\subsection{Near-Field Holography Measurement Conclusions}
\label{holoconclusions}

\begin{enumerate}
\item The holography system has functioned according to specification
  and has enabled us to measure the surface of the antenna reflector
  with a repeatability of better than 10~$\mu$m.
\item As shown in Figures~\ref{fig:holosVertex} and
  \ref{fig:holosAEC}, we have set both antenna surfaces to an
  accuracy of 16-17~$\mu$m RMS. This will provide an aperture
  efficiency of about 65 percent of that of a perfect reflector at the
  highest observing frequency of 950 GHz.
\item The small differences in the surface maps obtained over several
  days of measurement are consistent with the measurement
  repeatability and at best marginally significant. If taken at face
  value, they indicate that the deformations of the reflector under
  varying wind and temperature influence are fully consistent with,
  and probably well within, the specification.  This excellent
  behaviour over time is more important than the actual achieved
  surface setting. We stopped iteration of the settings after having
  achieved the goal of less than 20~$\mu$m.
\item The measured variations with ambient temperature changes appear
  somewhat large.  They correspond to a thermal contribution to the
  surface error of 10--16~$\mu$m at the boundaries of the
  operational range ($-$20 and $+$20~C), assuming a setting at 0~C. This
  is slightly more than the contribution in the ``ALMA error budget''
  and also more than the prototype antenna contractors have budgeted. On the
  other hand, taking BUS, panel and adjuster error contributions all
  together leads to approximate agreement with these measurements.
\item Within the primary operating range for ambient temperature of
  $-20$ to $+20$ C the surface error budget of the production antennas
  allows for a BUS contribution of 8~$\mu$m due to absolute
  temperature change and 7~$\mu$m for gradients. The panels are
  allowed 4~$\mu$m for absolute temperature and temperature gradient
  each. If we assume that the surface is set at 0~C to a measured value
  of 15~$\mu$m, which is the value we achieved, and we add
  the 15~$\mu$m maximum of the temperature deformation, the resulting
  error is 22~$\mu$m. This is more than the goal, but well within the
  specification.  High mountain sites, including Chajnantor, generally
  experience wind. This will dampen temperature gradients in the
  antenna structure and hence decrease the resulting structural
  deformation.
\end{enumerate}

\subsection{Accelerometer Measurements of BUS Deflection}
\label{accelsurf}

Accelerometers are used to measure accelerations in the inertial 
coordinate system of the antenna, allowing determination of rigid body
motion of the elevation structure, and a few low-order distortions of the 
BUS. In addition, motion of the subreflector structure with respect 
to the BUS can be measured.  For the BUS deformation measurements
10 accelerometers were placed on the BUS in the following
configuration (Figure~\ref{fig:accelconfig}):

\begin{figure}[!h]
\resizebox{\hsize}{!}{
%\centering
%\includegraphics[scale=0.30]{f11}
\includegraphics[scale=0.30]{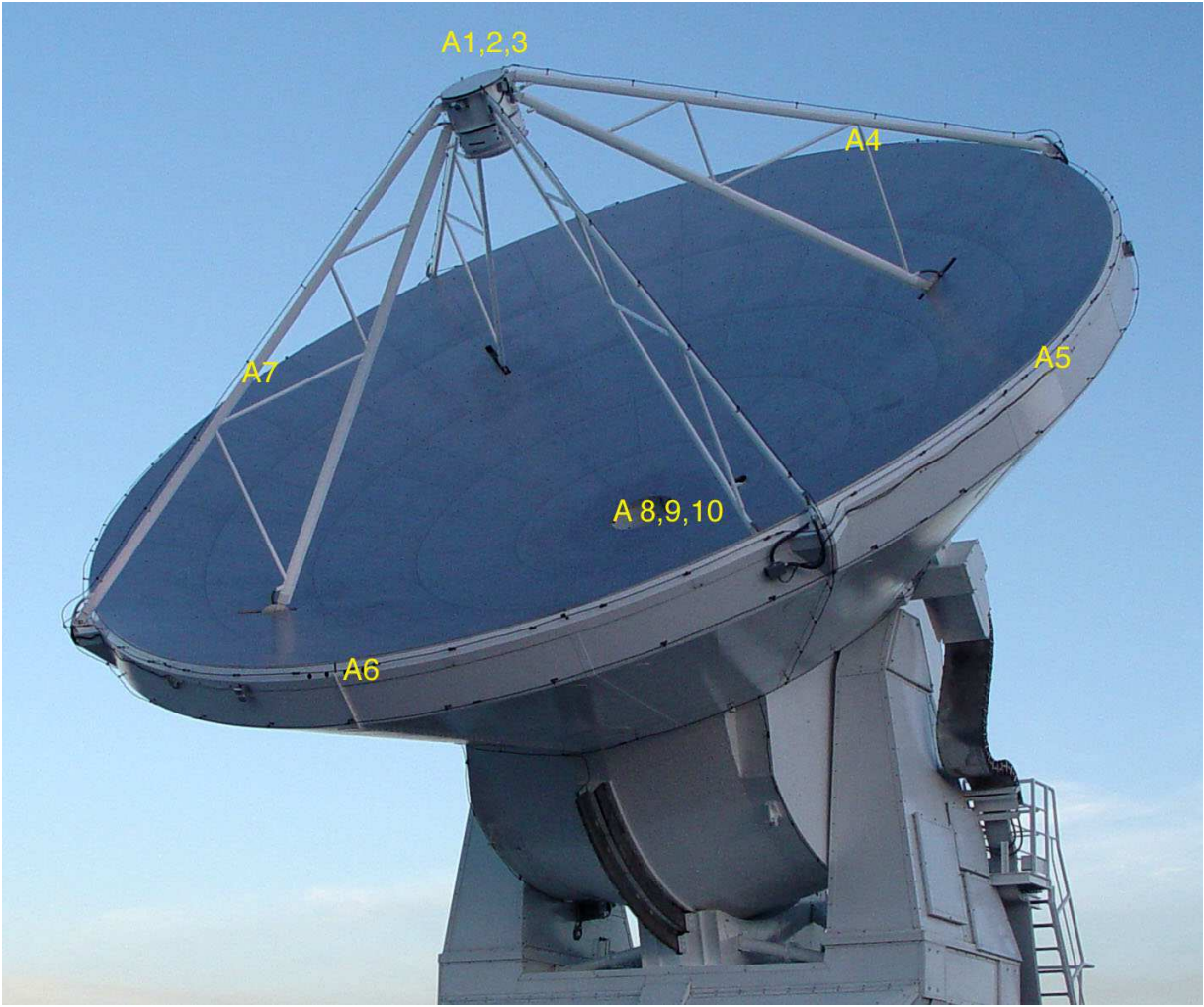}} % PS VERSION ONLY
\caption{Placement of the accelerometers on the antenna for BUS
  deflection (surface accuracy) measurement.}
\label{fig:accelconfig}
\end{figure}

\begin{itemize}
\item 3 accelerometers as a 3-axis sensor on the subreflector
  structure (numbered 1, 2, and 3 in Figure~\ref{fig:accelconfig})
\item 4 accelerometers along the rim of the BUS in boresight direction
  (numbered 4, 5, 6, and 7 in Figure~\ref{fig:accelconfig}) 
\item 3 accelerometers as a 3-axis sensor on the receiver flange Invar
  ring (numbered 8, 9, and 10 in Figure~\ref{fig:accelconfig})
\end{itemize}

The nature of the accelerometers used here limits accurate
displacement measurement to timescales of at most 10 seconds or
frequencies of at least 0.1 Hz.  Since this is well below the lowest
eigenfrequencies of the antennas, this is sufficient to determine
dynamic antenna behaviour.  See \cite{Snel2006} for further
information about the ALMA accelerometer measurement system. 

Over the 0.1 to 30 Hz sensitivity range of the accelerometers, the
changes in reflector surface shape through measurements of the focal
length (Zernike polynomial Z3 ``power'', ``defocus'' term, n=2, m=0)
and ``+'' astigmatism (Zernike polynomial with n=2, m=2) can be
measured to better than a few micrometers.

\subsection{BUS Deflection Measurement Conclusions}
\label{busdeflect}

For 9 m/s wind over time scales of 15~minutes, the VertexRSI antenna
surface is stable to 5.3~$\mu$m astigmatism RMS at the rim of the BUS, and
2.2~$\mu$m defocus. The AEC antenna surface is stable to 6~$\mu$m
astigmatism RMS at the rim of the BUS, and 5~$\mu$m defocus. For both
antennas, surface stability is dominated by the stiffness of the BUS
for wind excitation at low frequencies. The measured values are in
reasonable agreement with the error budget numbers from the
contractors. Note that these values do not include gravitational and
thermal effects.  During sidereal tracking the astigmatism and defocus
deformations remain below 1~$\mu$m for both antennas.

A frequently used operational mode of ALMA is ``fast switching'', in
which the antennas are pointed alternately at the observed source and
a nearby calibration source 1-2 degrees away typically at 10~second
time intervals. The antennas are designed to achieve such a switch
within 1.5~seconds. It is of interest to measure the dynamical
behaviour of the antenna structure during such switching
cycles. During the strongly accelerated and decelerated movement we
measure negligible BUS rim deformations of the order of one millimeter, which
decrease to a few micrometers after the new position has been
reached. 

During ``On-The-Fly'' (OTF) observing, the antenna scans a region of
sky back and forth at 0.5 deg/sec speed. During this movement we
measure astigmatism at the reflector edge and a defocus movement of a
few micrometers each. Interferometric mosaicing at 0.05 deg/sec
causes barely measurable astigmatism and defocus stability of about
1~$\mu$m.

Within the sensitivity of these measurements it can be stated that both
antennas exhibit dynamical deformations of the reflector surface that
are small and in agreement with the surface error budget.  Also, as 
these measurements demonstrate, very small dynamical deformations can
reliably be measured with this accelerometer setup. We consider this
method, pioneered by \cite{Ukita2002} and also used on an optical
telescope by \cite{Smith2004}, of great potential for the study of
the dynamical behaviour of large and highly accurate telescopes.

\section{Absolute and Offset Pointing}
\label{pointing}

In the following we describe the analysis of the pointing performance of the
two ALMA prototype antennas. The analysis is (mostly) based on optical
pointing observations carried out between October 2003
and May 2004 and radio pointing measurements carried out during March
2004 (VertexRSI) and May 2004 (AEC). The optical and radio
pointing data analysis was carried out using the proprietary
TPOINT telescope pointing-analysis software.
The AEC results are more limited than those from the VertexRSI antenna
and cover a shorter time span, because of late handover of that
antenna.

This pointing analysis studies three aspects of the antennas:

\begin{enumerate}
\item All-sky blind pointing accuracy.  In the optical case:
\begin{itemize}
\item Find the model that best describes the all-sky
      pointing for the antenna under study, in its
      ``optical pointing test configuration''.
\item Find the best coefficient values for use
      in the 7-term ALMA ``standard internal model'' (\S\ref{internalmod}).
\item Estimate the best all-sky optical pointing
      performance that could be expected operationally
      (\S\ref{optpoint}).
\item Assess the stability of the optical pointing model
  (\S\ref{optmodstability}).
\end{itemize}
In the radio case:
\begin{itemize}
\item Find the model that best describes the all-sky pointing
      for the antenna in question when operating in its radio
      configuration (in particular at 95~GHz).
\item Estimate the best all-sky pointing performance that could be
      expected operationally.
\end{itemize}
\item Assess the smoothness of tracking (optical, and for one antenna only).
\item Determine the ability to offset accurately across short
  distances (optical only).
\end{enumerate}

\begin{figure}
\centering
\includegraphics[scale=0.8,angle=-90]{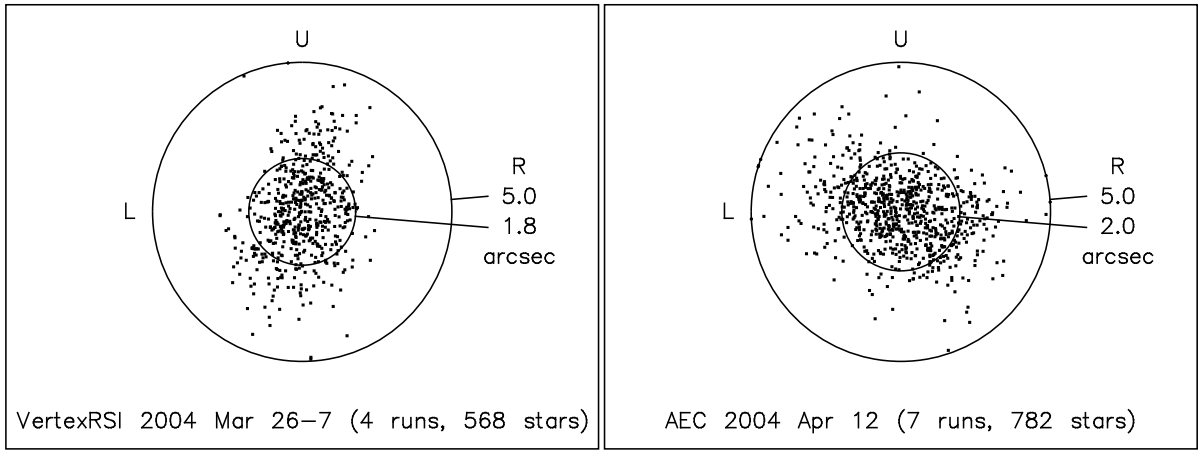}
\caption{Left: Sample VertexRSI (left) and AEC (right)
    OPT pointing residual plots.  Shown are the measurement residuals
    following application of the respective pointing models, each
    comprising 13 terms. The outer ring marks a radial residual of 5~arcsec,
    while the inner ring indicates the RMS pointing residual for each
    antenna.}
\label{fig:optpoint}
\end{figure}

\begin{figure}
\centering
\includegraphics[scale=0.40,angle=0]{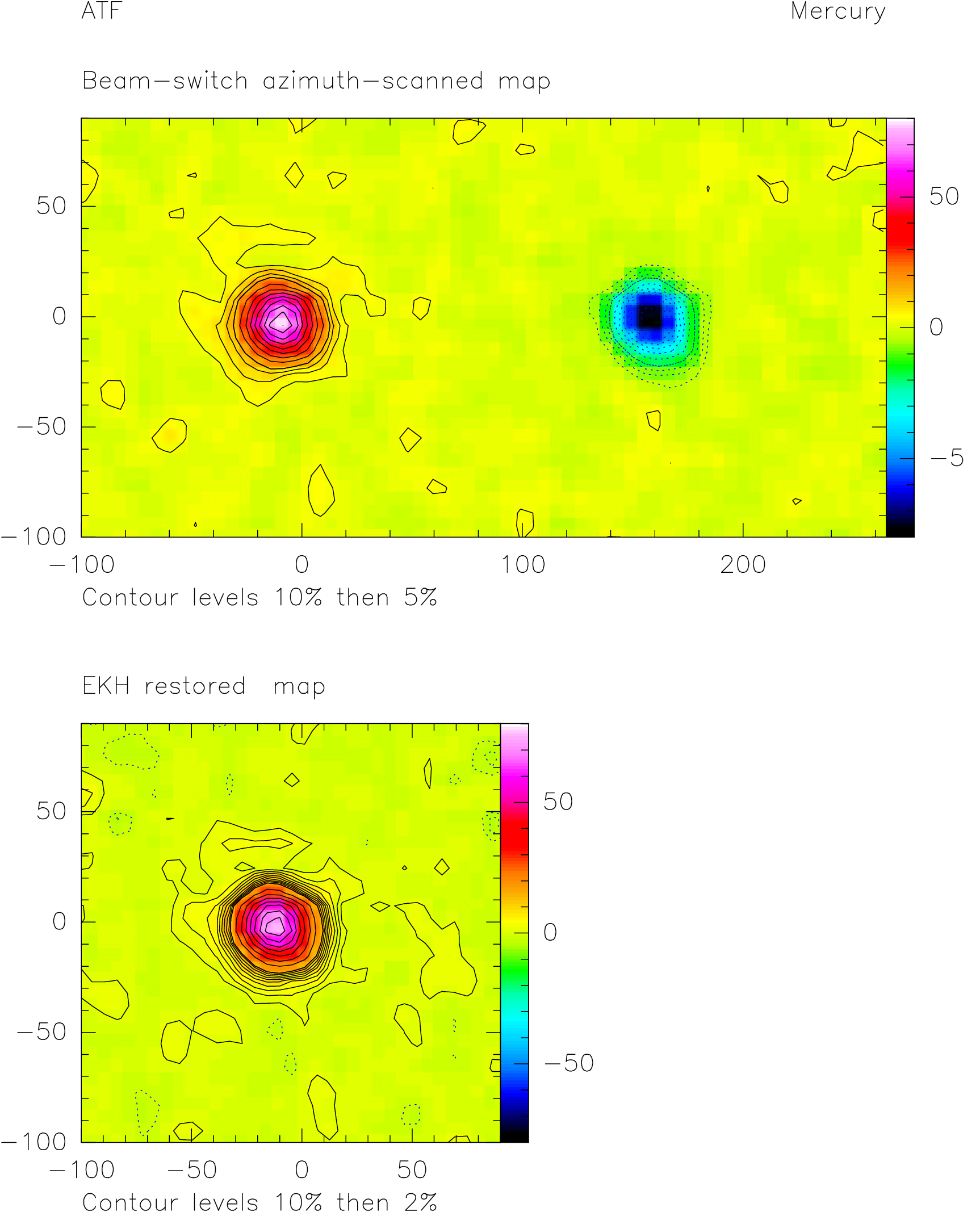}
\includegraphics[scale=0.40,angle=0]{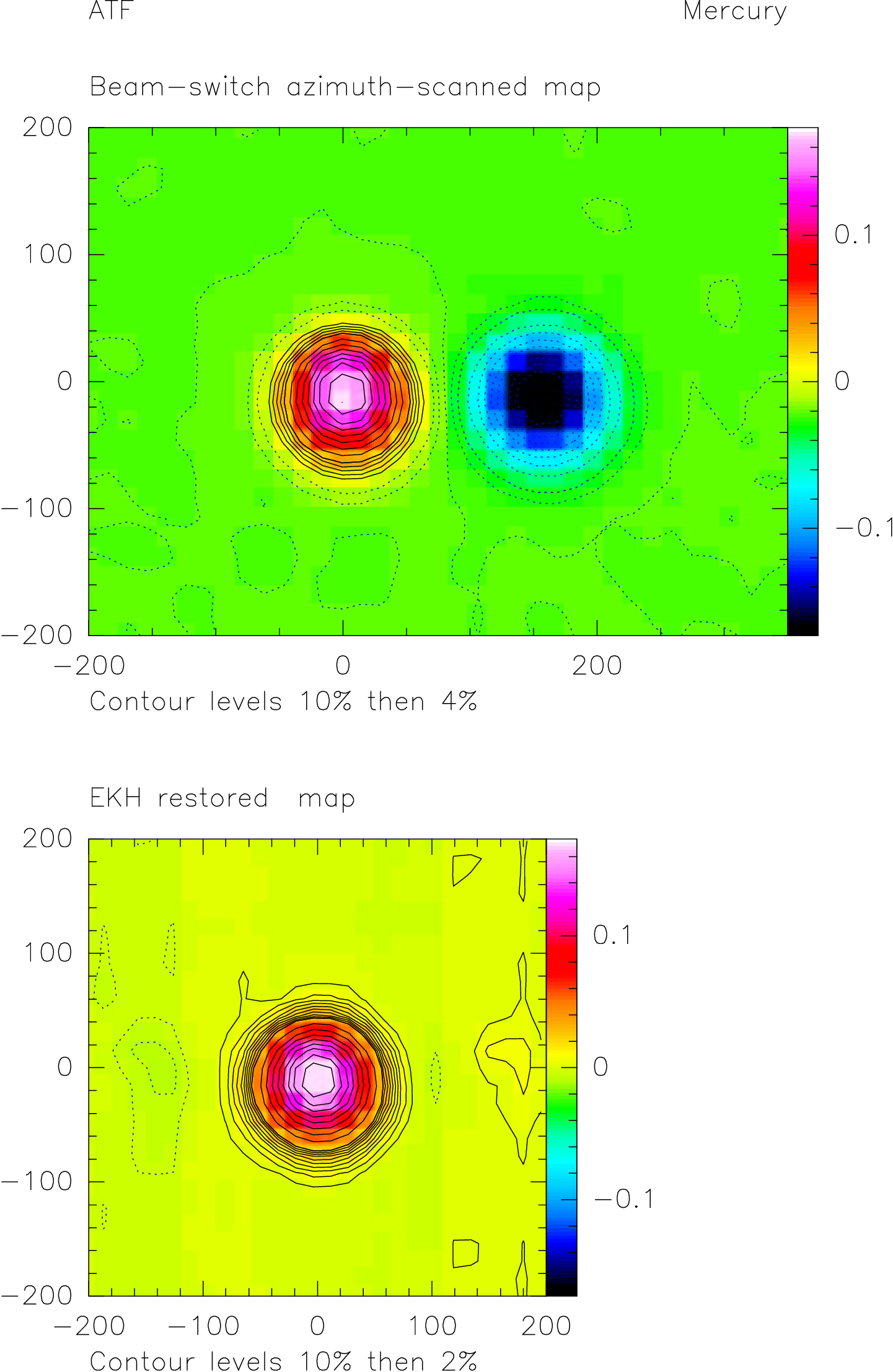}
\caption{Beam-switched radiometric measurements of the planet
  Mercury.  Left: VertexRSI measurement made at 265 GHz.  Mercury was
  at an angular distance of 16 degrees from the Sun, the planet
  angular diameter was 5.1~arcsec, and the contour levels are 5,10,20,
  ..., 90~\%.  Top: raw beam-switched map (beam separation 160
  arcsec); bottom: map restored using the EKH (\cite{Emerson1979})
  algorithm; contour 
  levels are 2, 4, ..., 16, 18, 20, 30, ..., 90\%. The low-level
  structure located $\sim 35$~arcsec seen in both restored and
  unrestored maps is due to a slight defocus in the Y-coordinate.
  Right: AEC measurement made at 95 GHz.  Mercury was at an angular
  distance of 25 degrees from the Sun, the planet angular 
  diameter was $7.2^{\prime\prime}$, and  contour levels are $4, 8,
  12, 16, 20, 30, ... 90\%$. Top: raw beam-switched map (beam
  separation $160''$); bottom: map restored with the EKH algorithm;
  contour levels are $2, 4, 8,..., 16, 18, 20, 30, ... 90\%$.}
\label{fig:mercury}
\end{figure}

In all, the pointing analysis of the ALMA prototype antennas relied on
measurements made with three distinct systems:

\begin{itemize}
\item An optical pointing telescope (OPT)
  mounted within the BUS of each antenna.  The OPT is a NRAO-designed
  refracting telescope composed of a 4 inch lens mounted within a
  rolled-Invar tube (\cite{Mangum2004a}).  The detector used was a
  commercial CCD camera coupled to a video frame grabber.
  Figure~\ref{fig:optpoint} shows 
  two typical absolute (all-sky) pointing residual plots obtained with
  the OPT system.
\item Radiometric measurements at 1 and 3 mm wavelength.  Figure
  \ref{fig:mercury} shows a sample single dish image obtained
  with this receiver system.
\item An accelerometer system (\S\ref{accelpoint}).
\end{itemize}

Both prototype antennas were designed to meet the pointing
specifications with the aid of various metrological correction
systems.  These metrology systems were in general unsuccessful at
improving the pointing performance of the prototype antennas, as
described in \S\ref{metrology}.

\subsection{Pointing Models}
\label{pointingmodels}

\subsubsection{TPOINT Modeling}

If a telescope or antenna is pointed at a star, there is in
general a difference between the true direction of the incoming
light and the demands to the mount servos, due to mechanical
misalignments, flexures and other imperfections.  By observing
stars all over the sky and logging the actual and commanded
positions (or data from which these angles can be deduced), 
a mathematical model can be developed that estimates the
servo inputs required to point the antenna accurately in a given
direction.  The TPOINT pointing analysis software reads
a file containing the logged observations and provides graphics and
modeling tools that allow a mathematical description of the
pointing errors to be created.  The resulting pointing model has
then to be built into the telescope or antenna control system, so
that the demands to the servos are appropriately adjusted and
accurate dead-reckoning acquisition of celestial targets brought
about.

For any two-axis ``gimbal'' mount, the basic \textit{a~priori}
model consists of seven terms:
two encoder index errors, two non-perpendicularities, two
azimuth-axis tilt components, and Hooke's Law flexure.  All of
these terms have a clear mechanical interpretation and, in
general, all will be present to some degree.
This basic model usually performs well,
typically leaving only a small
remaining systematic residue to be mopped up using
the repertoire of other TPOINT terms.  Most of the
latter set out to represent plausible mechanical effects---for
example harmonic terms to deal with
run-outs and mechanical deformation errors---though it is also possible to
introduce unashamedly empirical terms such as polynomials
should there be no alternative.

\begin{table}[h!]
\centering
\begin{minipage}{14cm}
\caption{The Seven Basic Pointing Model Terms}
\begin{tabular}{|l|c|c|p{5cm}|}
\hline
Term & \multicolumn{2}{c|}{Correction Formula} & Nominal Cause \\
& $\Delta Az$ & $\Delta El$ & \\
\hline\hline
IA & $-\textrm{IA}$ & $\ldots$ & \footnotesize{Az encoder zero point offset} \\
\hline
IE & $\ldots$ & $+\textrm{IE}$ & \footnotesize{El encoder zero point offset} \\
\hline
HECE & $\ldots$ & $+\textrm{HECE} \cos{E}$ &
 \footnotesize{Hooke's Law vertical flexure} \\
\hline
CA & $-\textrm{CA}\sec{E}$ & $\ldots$ &
 \footnotesize{Non-perpendicularity between the boresight and El axis} \\
\hline
NPAE & $-\textrm{NPAE} \tan{E}$ & $\ldots$ &
 \footnotesize{Non-perpendicularity between the Az and El axes} \\
\hline
AN & $-\textrm{AN}\tan{E}\sin{A}$ & $-\textrm{AN}\cos{A}$ &
 \footnotesize{Az axis offset/misalignment north-south (north = positive)} \\
\hline
AW & $-\textrm{AW}\tan{E}\cos{A}$ & $+\textrm{AW}\sin{A}$ &
 \footnotesize{Az axis offset/misalignment east-west (west = positive)} \\
\hline
\end{tabular}
\label{tab:fterms}
\end{minipage}
\end{table}

The functional form of the basic 7-term model is given in
Table~\ref{tab:fterms}.  Note that:

\begin{itemize}

\item The sign of the correction terms is such that the expressions should
  be subtracted from the true direction in order to generate mount
  demands.

\item The coefficient values in service will either be those from
  the most recent set of observations or averages of several recent runs.

\item The expressions are approximate, adequate when the coefficients
  are small and/or the target is not too close to the zenith.
  TPOINT in fact fits rigorous vector formulations that do not have
  these limitations.
\end{itemize}

\subsubsection{The Standard ALMA Internal Model}
\label{internalmod}

\begin{figure}
\centering
\includegraphics[scale=0.55,angle=-90]{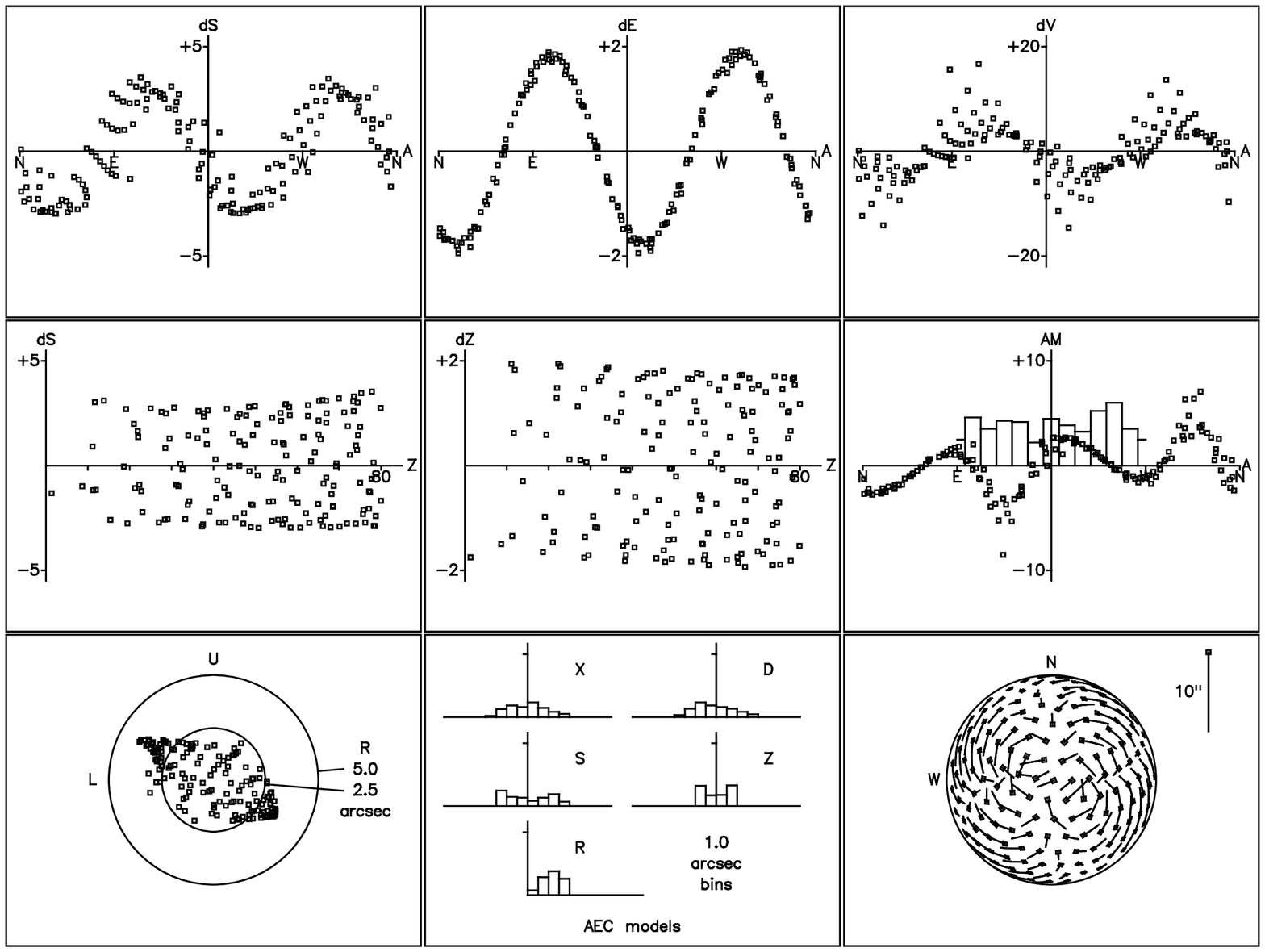}
\includegraphics[scale=0.55,angle=-90]{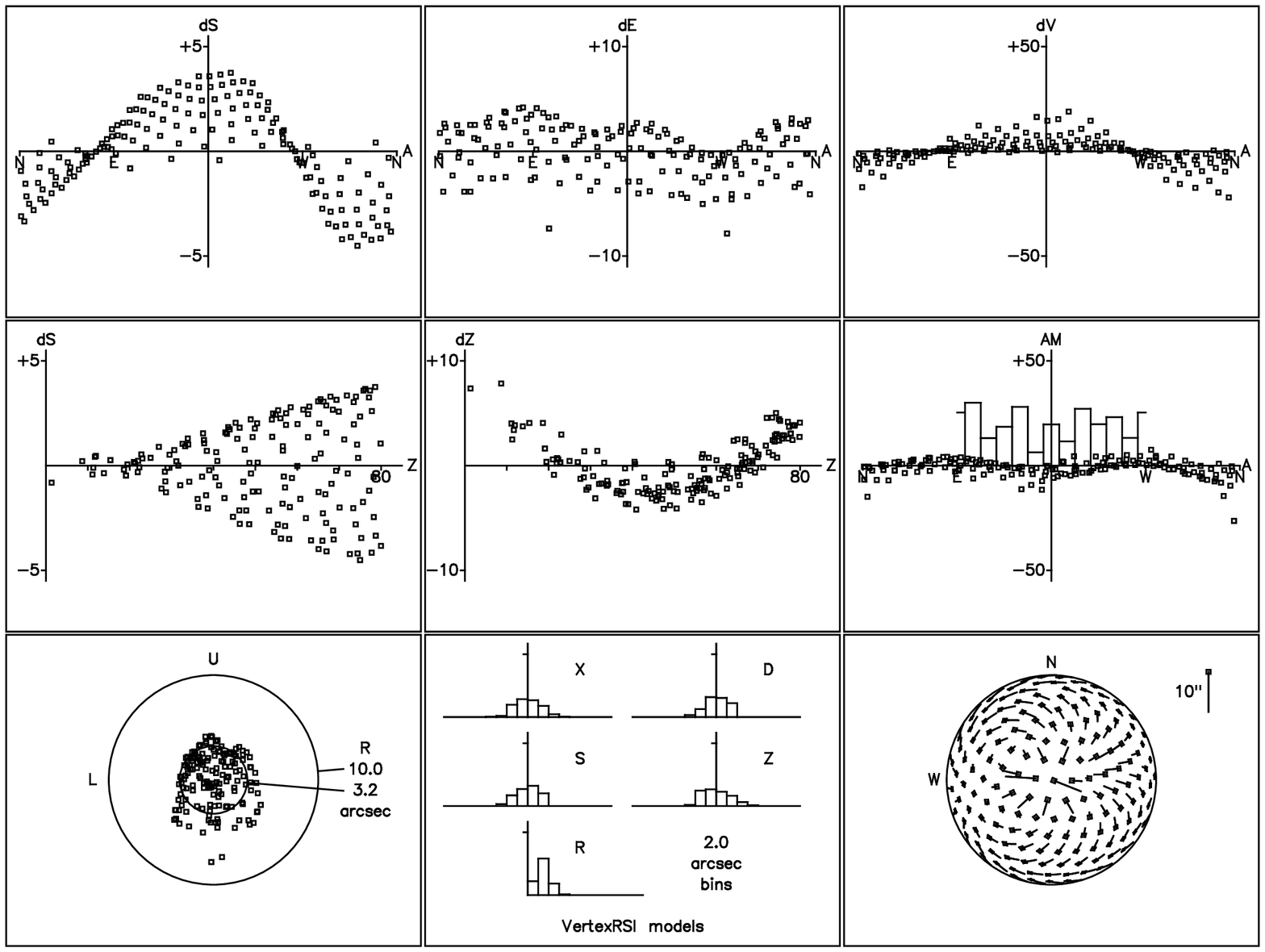}
\caption{The predicted pointing model residuals using the AEC (top)
  and VertexRSI (bottom) 7-parameter pointing models.  The panels
  show, left-to-right: (top row): $\Delta A \cos(E)$, $\Delta E$, and
  inferred Az/El nonperpendicularity versus A; (middle row): $\Delta A \cos(E)$
  and $\Delta Z$ (zenith angle) versus Z, and azimuth axis tilt versus A;
  (bottom row): pointing residual versus (A,E), pointing residual
  histogram versus all coordinates, and pointing residuals on an
  all-sky pointing error vector map.  The limitations of these simple
  models are apparent.}
\label{fig:7mod}
\end{figure}

The \textit{a~priori} 7-term model (Table~\ref{tab:fterms})
is also that specified by ALMA for internal implementation in the
low-level antenna controllers, principally as an aid to
commissioning. Once the final TPOINT model is available, the
optimum coefficients to be used with this basic 7-term
model can be estimated by subtracting the full model from a set of
points distributed evenly over the sky, then fitting the 7-term model.
The final RMS from such an experiment represents the maximum
achievable performance of the
ALMA internal model for the antenna in question, while the residual plots
illustrate the antenna's
pointing idiosyncrasies.  This experiment was carried
out for both antennas, using OPT data, for illustrative purposes.  For
the AEC antenna, the best-fit sky RMS was 2.46~arcsec.  For the
VertexRSI antenna, the best-fit sky RMS was 3.25~arcsec.  The
limitations of this simple 7-parameter model are illustrated in
Figure~\ref{fig:7mod}.

\subsection{The ALMA Optical Pointing Telescopes}
\label{opt}

Optical pointing systems have become standard equipment on
millimeter and submillimeter telescopes.  There are benefits and
limitations associated with the reliance on optical pointing systems
on radio telescopes:

\begin{enumerate}
\item Benefits:
  \begin{enumerate}
  \item There are many more bright stars that can be used as optical
  pointing sources than radio pointing sources.
  \item With a CCD camera, optical data acquisition can be done on
  time scales as short as a few tenths of a second, as opposed to the
  tens of seconds time scales necessary for radio pointing measurements.
  \item Positions of optical stars can be derived to sub-arcsecond
  accuracy, which is a much higher precision than that obtainable
  through single antenna radio measurements.
  \end{enumerate}
\item Limitations:
  \begin{enumerate}
  \item Differences between the position and behavior of the optical
  and radio signal paths can make the derivation of the radio pointing
  coefficients from optical pointing measurements difficult.
  \item Optical pointing is not possible during overcast weather
conditions.
  \item Optical pointing cannot be used in the daytime, unless the
  system has been designed with near-infrared sensitivity.
  \end{enumerate}
\end{enumerate}

\noindent{Optical} pointing systems find application in such areas as:

\begin{itemize}
\item Pointing and tracking diagnosis
\item Pointing coefficient derivation and monitoring
\item Auto-guiding
\end{itemize}

Table \ref{tab:design} lists the ALMA prototype optical telescope
system specifications (\cite{Mangum2004a}).

\begin{table}[h!]
\small
\centering
\caption{Prototype ALMA Optical Telescope Specifications}
\medskip
\begin{tabular}{|l|l|}
\hline
Telescope & Meade $4^{\prime\prime}$ model 102ED Apochromatic
Refractor \\
          & f/9 (920mm focal length) \\
          & NRAO-design rolled Invar tube and mount \\
Additional Lenses & X2 Barlow \\
Filters & IR continuum ($\lambda >$ 700nm) and clear focus correction \\
Camera     & Cohu 4920-3000 camera \\
           & Extended IR sensitivity to 900nm \\
           & 56 dB dynamic range \\
           & $768\times 494$ pixels\\
           & $8.4\times 9.8$~$\mu$m pixel size \\
           & $6.4\times4.8$~mm image area \\
Frame Grabber & Imaging Technologies PC-Vision \\
              & $1024\times1024$ array capability \\
              & $640\times480$ sampling (RS-170) \\
              & 2MB VRAM \\
              & 4ms frame transfer \\
              & 48 dB dynamic range \\
Control PC & COTS \\
Control Software & NRAO design \\
\hline\hline
Diffraction Limit & $1.^{\prime\prime}2$ (for $\lambda$ = 500 nm)\\
Plate Scale & 0.1121~arcsec/$\mu$m \\
Effective Plate Scale 
& $1.^{\prime\prime}12\times 1.^{\prime\prime}12$ per detector pixel \\
Field of view & $12^{\prime}\times 9^{\prime}$ \\
Sensitivity & $\geq 7^{th}$ magnitude in 5~second exposure \\
Mount location & Backup structure (requires access hole in panel) \\
Mount & Three-point Invar; accessible \\
\hline
\end{tabular}
\label{tab:design}
\end{table}

\subsubsection{Optical Telescope Measurement System Limitations}

\begin{enumerate}

\item The optical seeing at the VLA site is generally poor.  To
  quantify this seeing contribution to the OPT measurements a
  sequential multiple star measurement, where each star measurement
  was repeated three times, was used to derive seeing and/or 
  short timescale antenna positioning errors.  These measurements
  suggest a seeing/vibration contribution of $\sim$1.5~arcsec, in line
  with the results from tracking tests.  With the OPT's 100~mm
  aperture, the seeing manifests itself as large and rapid movements
  of the image, making optical measurement of the all-important
  tracking and offsetting performance almost impossible.

\item The OPT design, which was always going to be a major challenge, is
itself marginal for this application, given the impressive
performance of the antennas. The comparatively short focal length means
that the image is under-sampled, and in some of the tracking tests
quantization effects are evident, making performance limitations much
below 1~arcsecond difficult to detect. It is far from obvious how to
deal with this; a refracting OPT of (say) twice the aperture would
have been prohibitively expensive, and also very large, whereas a
reflector (a commercial Schmidt-Cassegrain for example) would probably
have had insufficient optical stability.

\end{enumerate}

\subsection{Optical Pointing Telescope Measurements}
\label{optpoint}

\subsubsection{Data Acquisition}
\label{dataacq}

The pointing measurement data logging procedure evolved during the
antenna evaluation process, ultimately leading to a very
self-consistent and robust scheme for position calculation and
logging.  We refer to this final position calculation and logging
scheme as ``PMDR'' (Pointing Model Done Right).  

\begin{figure}
\centering
\includegraphics[scale=0.75]{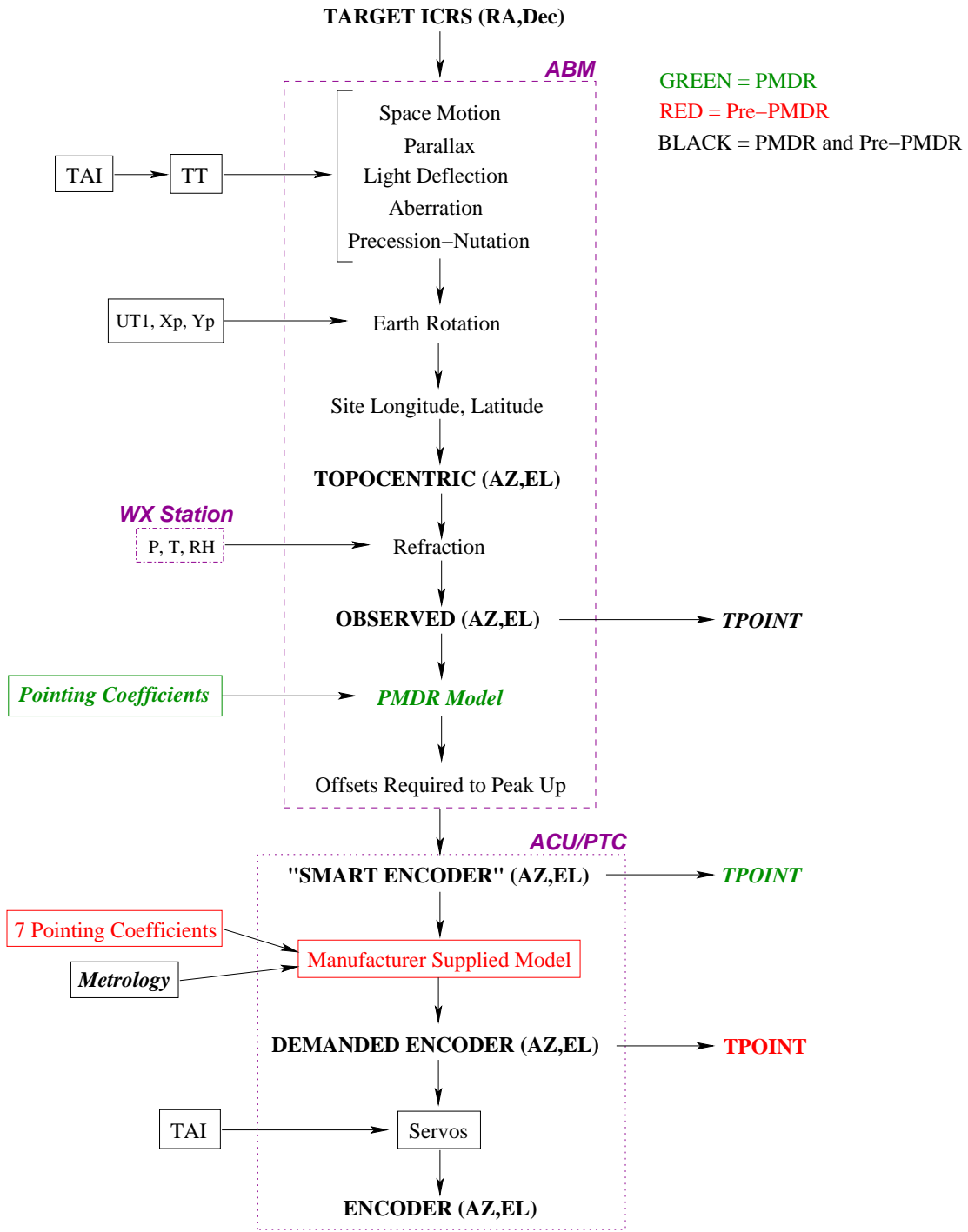}
\caption{Positioning calculation flow for the pre-PMDR and PMDR
         systems.  The roles of the respective computers are
         shown boxed.}
\label{fig:tpoint-flow}
\end{figure}

The two schemes (pre-PMDR and PMDR) are illustrated in
Figure~\ref{fig:tpoint-flow}.  In the pre-PMDR mode the inputs to the
TPOINT pointing analysis are (i)~the ``observed'' azimuth and
elevation (\ie\ the true line of sight) and (ii)~the encoder
readings from the mount axes.  The TPOINT analysis produces updated
pointing coefficients for the ALMA internal 7-term model, calculated
to minimize the offsets required to peak up on a new source and,
equivalently, the drifts during subsequent tracking.  Note, however,
that coefficients obtained in this way presume zero metrology input;
they are not suitable for use with metrology enabled, and the fitting
statistics (\eg\ RMS) do not reflect any benefits coming from
the metrology system.  Moreover, if use with the manufacturer's
supplied control system is envisaged, the TPOINT model is limited to
the seven standard terms (from Table~\ref{tab:fterms}), and additional
forms of correction are not available.  This final point proved to be a major
handicap, hampering the pointing performance characterization of the
ALMA prototype antennas and limiting the ultimate performance (see
Figure~\ref{fig:7mod}).

These shortcomings were addressed by the PMDR scheme (see the green items
in Figure~\ref{fig:tpoint-flow}).  In this scheme, the quantities
logged for input to TPOINT are (i)~the observed (Az,El) as before but
this time (ii)~the commanded inputs to the manufacturer's model and
metrology systems, the latter constituting a ``smart encoders''
interface.  The link between the two is a separate TPOINT model that
can contain any required terms, not just the seven basic terms.  In
fact it is advantageous to disengage the manufacturer's model, so that
the TPOINT model provides the whole correction apart from the
metrology inputs: the end result is the same and the TPOINT model then
represents absolute physical estimates -- nonperpendicularies,
misalignments, flexures, \textit{etc.}  All measurements quoted in the
following analysis used the PMDR positioning and logging scheme.

\subsubsection{All-Sky Optical Pointing Data}

The respective data samples for each prototype antenna are listed
below.  For both samples, we note that:

\begin{itemize}
\item The short time span available for evaluating pointing
  performance unfortunately precluded any studies of the long term
  stability of the pointing terms.  
\item A handful of these pointing tests were carried out with
  contractor-delivered metrology systems.  We describe the results of
  these few metrology tests in \S\ref{metrology}.
\item To obtain the total sample of measurements for both antennas,
  individual runs were for the most part accepted or rejected on the
  basis of notes made at the time they were carried out, rather than
  whether they gave better or worse pointing residuals.  However,
  certain runs were clearly aberrant.  These obviously untypical runs
  were rejected.
\end{itemize}

\paragraph{AEC:}

A selection of 63 pointing test runs were carried out during an
intensive period of testing between 2004/03/06, when the antenna first
became available, and 2004/05/12.  There were a total of 7411
observations in this sample.  Following some modifications by the
antenna contractor in December 2004, an additional 
series of measurements were made during the period 2005/01/07 and
2005/03/02.  As these additional measurements are not directly
comparable to those made during the 2004/03/06 through 2004/05/12
period considered here, we defer their analysis to
\S\ref{aec-additional}.

\paragraph{VertexRSI:}

A selection of 34 pointing test runs that utilized the PMDR
positioning system (see \S\ref{dataacq}) were carried out between
2004/03/10 and 2004/05/03, comprising 4059 observations.

\subsubsection{The OPT Core Terms}

In the final optical pointing derived models we have distinguished the
set of floating basic pointing terms---IA, IE, CA, AN, AW---from a fixed
``core'' set of terms that are not only considered to have a specific
mechanical cause but can also be assumed to be fixed in size.
Moreover, for most of these core terms there is no reason to suppose that 
they will be affected by the transition from optical to radio pointing.

\begin{table}
\centering
%\begin{minipage}{14cm}
\caption{OPT Core Pointing Model Terms}
\begin{tabular}{|l|r|l|p{5cm}|}
\hline
Term & Ampl. (VertexRSI/AEC) & Correction Formula & Nominal
Cause \\ \hline\hline
HESE  & NA/$-27.56$ & $\Delta E = +\mathrm{HESE} \sin{E}$
 & \footnotesize{El encoder run-out, sine component} \\
\hline
HECE  & $+30.09$/$-16.85$ & $\Delta E = +\mathrm{HECE} \cos{E}$ &
 \footnotesize{El encoder run-out, cosine component and/or
 vertical flexure} \\
\hline
HASA  & $-1.20$/NA  & $\Delta A = +\mathrm{HASA} \sin{A}$
 & \footnotesize{Az encoder run-out, sine component} \\
\hline
HACA  & $-4.42$/NA  & $\Delta A = -\mathrm{HACA} \cos{A}$
 & \footnotesize{Az encoder run-out, cosine component} \\
\hline
HASA2 & $+0.73$/$-1.70$  & $\Delta A = +\mathrm{HASA2} \sin{2A}$ &
 \footnotesize{Az encoder tilt, sine component} \\
\hline
HACA2 & NA/$+2.94$  & $\Delta A = -\mathrm{HACA2} \cos{2A}$ &
 \footnotesize{Az encoder tilt, cosine component} \\
\hline
HESA2 & NA/$-0.99$  & $\Delta E = +\mathrm{HESA2} \sin{2A}$ &
 \footnotesize{El nod twice per Az rev, sine component} \\
\hline
HECA2 & NA/$+1.53$  & $\Delta E = +\mathrm{HECA2} \cos{2A}$ &
 \footnotesize{El nod twice per Az rev, cosine component} \\
\hline
HASA3 & $+0.38$/NA  & $\Delta A = +\mathrm{HASA3} \sin{3A}$
 & \footnotesize{Az encoder error 3$\times$ per rev, sine component} \\
\hline
HESA3 & $+0.94$/NA  & $\Delta E = +\mathrm{HESA3} \sin{3A}$
 & \footnotesize{El nod 3$\times$ per Az rev, sine component} \\
\hline
HECA3 & $-0.34$/NA  & $\Delta E = +\mathrm{HESA3} \cos{3A}$
 & \footnotesize{El nod 3$\times$ per Az rev, cosine component} \\
\hline
HVSA2 & NA/$-2.25$  & $\Delta A = +\mathrm{HVSA2} \sin{2A} \tan{E}$ &
 \footnotesize{Az/El nonperp variation twice per Az rev, sine component} \\
\hline
HVCA2 & NA/$-2.08$  & $\Delta A = -\mathrm{HVCA2} \cos{2A} \tan{E}$ &
 \footnotesize{Az/El nonperp variation twice per Az rev, cosine component} \\
\hline
NPAE  & $-15.01$/$+28.62$ & $\Delta A = +\mathrm{NPAE} \tan{E}$ &
 \footnotesize{Az/El non-perpendicularity} \\
\hline
\end{tabular}
\label{tab:core}
%\end{minipage}
\end{table}

The core terms for both prototype antennas are listed in
Table~\ref{tab:core}.  For each term we list the nominal mechanical
deformation that produces each term.  Note that:

\begin{itemize}

\item The sign of the correction terms is such that the expressions
  should be subtracted from the true direction in order to generate
  mount demands.

\item The smallest terms are of order 1 and 0.5~arcsec in size for the
  AEC and VertexRSI models, respectively. Their
  significance can be gauged by eliminating the smallest of them,
  namely HESA2 (AEC) and HECA3 (VertexRSI), fitting the full model to
  take up the slack, and plotting the residuals.  The result of this
  experiment shows a distinct $\sin(2A)$ (AEC) and $\cos(3A)$
  (VertexRSI) systematic residual in $\Delta El$ versus Az for the
  respective antennas. Figure~\ref{fig:a_hesa2} shows the graphical
  results for the AEC antenna.

\begin{figure}
\centering
\includegraphics[scale=0.55,angle=-90]{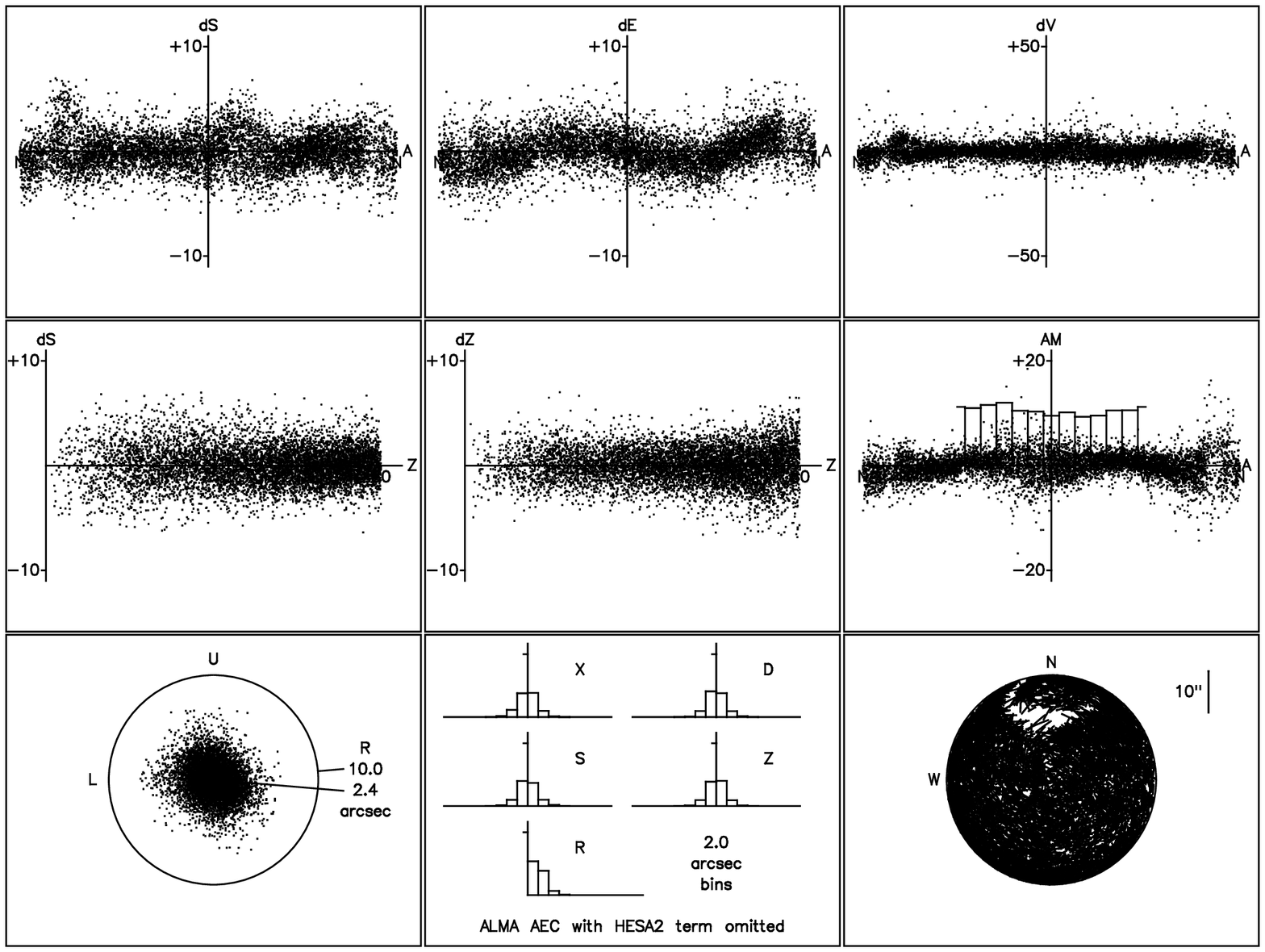}
\caption{TPOINT plot showing the significance of the HESA2
  term for the AEC antenna pointing model.  Note the clearly
  systematic residual in $\Delta El$ versus Az in the top-middle
  panel, suggesting the need for a $\sin(2A)$-dependent term.}
\label{fig:a_hesa2}
\end{figure}

\end{itemize}

\subsubsection{Stability of the OPT Pointing Model}
\label{optmodstability}

In principle, all of the terms of an antenna's pointing model should
maintain constant amplitude for long periods unless something on the 
antenna has been adjusted. This supposition was studied by taking the
individual models for the series of pointing tests on each antenna,
and plotting the coefficient values for the five terms thought most
likely to be affected by mechanical adjustments and drifts.  The
stability of these coefficient values were then studied by plotting
them versus sequence number (in effect versus time). 

For the AEC antenna, the stability of the five basic pointing model
terms over the period 2004/03/06 through 2004/05/12 was ($\Delta$IA,
$\Delta$IE, $\Delta$CA, $\Delta$AN, $\Delta$AW) = ($\pm3$, $\pm5$,
$\pm3$, $\pm1$, $\pm1$)~arcsec, with very little drift of the average
values for each term.  This is very good basic pointing term stability
for an antenna.  The VertexRSI antenna pointing term stability over
a much longer measurement period (2003/10/16 to 2004/03/01) was
($\Delta$IA, $\Delta$IE, $\Delta$CA, $\Delta$AN, $\Delta$AW) =
($\pm3$, $\pm7$, $\pm5$, $\pm5$, $\pm3$)~arcsec, with very little
drift of the average values for most terms (the AN term showed a
moderate $\sim5$ arcsec drift of the average during the measurement period).
The basic pointing model term stability for the VertexRSI antenna was
notably worse than that of the AEC antenna.  Part of this poorer
performance is likely attributed to the failure of the tiltmeter
metrology system (see \S\ref{metrology}), which was designed to correct for
variations in the azimuth axis tilt (AN and AW).

\subsubsection{Additional OPT Measurements of the AEC Antenna}
\label{aec-additional}

Following the all-sky OPT measurements described above the antenna
contractor made some modifications designed to improve the all-sky
pointing performance.  In particular, the attachment points of the
antenna base to its foundation were tightened, and a tiltmeter-based
metrology correction was enabled in an attempt to correct the
suspected instability in the azimuth bearing.  An additional series of
measurements were made during the period 2005/01/07 and 2005/03/02
following these modifications.  From these measurements we note that:

\begin{itemize}

\item Two terms -- HVSA2 and HVCA2 -- were no longer significant and
  need not exist in the pointing model solution.  The HVSA2 and HVCA2
  terms represent changes in the Az/El nonperpendicularity as the
  mount rotates.  The insignificance of these terms suggests that the
  Az axis had become less wobbly.

\item Figure~\ref{fig:aecopt-rms} shows the cumulative AEC optical
  pointing model results for this period.  For each total RMS value
  shown we also show the contributions to this total RMS due to
  cross-elevation and elevation pointing residual.

\begin{figure}
\centering
\includegraphics[scale=0.5,angle=-90]{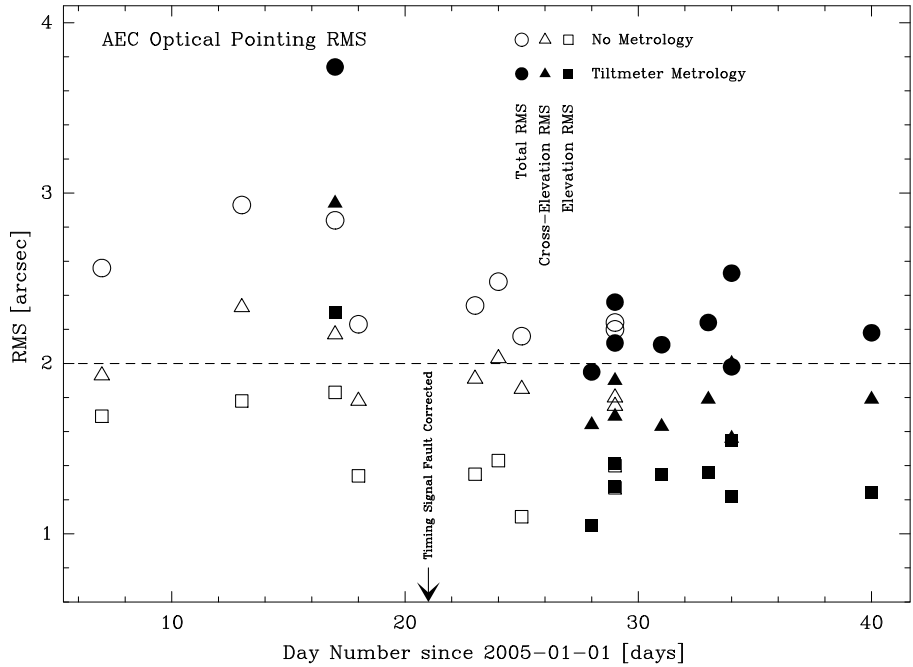}
\caption{AEC optical pointing results for the period 2005/01/07 --
  2005/03/02.}
\label{fig:aecopt-rms}
\end{figure}

\item A large observed shift in the IA and AW pointing terms, relative
  to the previously-derived pointing models reported in
  \cite{Wallace2004}, was ultimately diagnosed as a timing error
  in the monitor and control system of approximately 13~seconds.  This
  timing error does not affect the resultant pointing RMS values
  derived.  On 2005/01/20 this timing error was corrected.

\item A timing signal cable problem was fixed on 2005/01/23.

\end{itemize}

\subsubsection{Pointing Model Recalibration Frequency}

The ALMA pointing specification permits monthly recalibration of the
pointing model. To investigate this,
two representative all-sky OPT runs taken a month apart were chosen.
The first was used to simulate the periodic recalibration of the five
``non-core'' 
terms of the AEC model (IA IE CA AN AW).  This simulated operational
model was then applied to the second data set and the resulting
pointing performance assessed with various terms refitted as might be
done in a quick daily recalibration.

Operationally, the frequent observation of standard stars means there
is an opportunity to adjust at least the boresight offsets more
frequently, and this was simulated by allowing the terms IE and 
CA to float.  They changed by (AEC/VertexRSI) 2.9/14.6 and 1.7/1.5~arcsec
respectively and gave a sky RMS values of 2.44/3.05~arcsec, still
somewhat outside the specification.

If the tilt terms AN and AW were allowed to vary in addition, the
changes were (AEC/VertexRSI) 1.3/1.8~arcsec and 0.0/2.4~arcsec
respectively, and the sky RMS values were 2.22/1.85~arcsec, somewhat
better, and well within the specification for the VertexRSI antenna.
This is an indication of the improvement that correctly working metrology
(base tiltmeters) would offer.

With the azimuth zero point term IA added, completing
the standard five floating terms, the sky RMS reduced only slightly,
to 2.16~arcsec for the AEC antenna, and was unchanged for the
VertexRSI antenna.  However, a significant improvement in the fit to
the AEC measurements, to 1.95~arcsec~RMS, occurred when the Az/El
non-perpendicularity term NPAE was included in the fit.  The
changes in the IA and NPAE terms for the AEC antenna were 11.2 and
12.6~arcsec respectively, and may be evidence of the suspected azimuth
bearing instability in this antenna.

The conclusions are:
\begin{itemize}
\item Some form of rapid recalibration (say daily) will be
needed in addition to any relatively infrequent full calibration.
\item Frequent recalibration of the boresight offsets, IE and CA,
is essential.
\item Recalibration of the tilt terms, AN and AW, is also essential
  for the VertexRSI antenna performance, unless correctly functioning
  base tiltmeters can render this superfluous.
\item Even with frequent recalibration of the boresight offsets,
the AEC antenna performance would not quite reach the 2~arcsec
specification.
\end{itemize}

\subsubsection{Optical Tracking Tests on the VertexRSI Antenna}

A pointing run consisting of observations of a single star for about
45~minutes was used to test antenna tracking performance. Because only
a small region of sky was observed (a line 
about $10^\circ$ long) the data set is not suitable for fitting a
pointing model. Although doing so produces, it turns out, a
sensible model, there is a danger of fitting out some of the
tracking error that the experiment is designed to expose.
Accordingly, an all-sky pointing run acquired during the following day
was used instead. Applying this model to the tracking
test and allowing only a zero-point correction (\ie\ the
TPOINT terms IE and CA) produced a sky RMS of
1.46~arcsec, and the $\Delta A \cos{E}$ and $\Delta E$ tracking
residuals shown in Figure~\ref{fig:v_tracking} (the x-axis is
sequence number and hence time):

\begin{figure}[h!]
\centering
\includegraphics[scale=0.45,angle=-90]{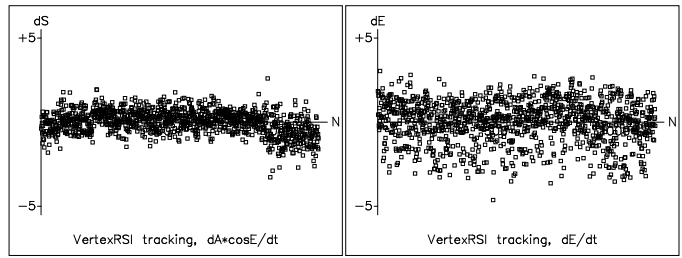}
\caption{Tracking errors for the VertexRSI antenna.  The x-axis is
  sequence number, and hence time.}
\label{fig:v_tracking}
\end{figure}

Note that:

\begin{itemize}
\item The scatter up/down is larger than left-right, and is not
  normally distributed; the larger deviations tend to be downward on
  the plot.
\item Beneath the noise, both axes show slow drifts. However, these
  are very small, barely 1~arcsec peak-to-peak.  They might be the
  underlying errors in the pointing model, and how it varies from
  day to day and (if temperature plays a role) from hour to hour,
  or they could be anomalous refraction, or changing wind pressure.
  They happen too slowly to affect the ability to offset accuracy
  over short timescales, but do perhaps suggest that small adjustments
  to the pointing model on timescales of a few tens of minutes
  may be a wise precaution.
\item The noise is of such high frequency that it is not what would
  usually be thought of as tracking error.  This high frequency noise
  is more a ``jitter'', and is likely caused by the seeing-induced
  movements of the optical image, rather than the smoothness of the
  antenna motion.
\item There is a fairly abrupt change in the left-right tracking near
  the end of the run, and hints of other relatively high frequency
  artefacts. If such events can happen during offsetting maneuvers the
  0.6~arcsec in $2^\circ$ science requirement might be in
  jeopardy. However, if the typical tracking were an illustration of the
  offsetting performance then the 0.6~arcsec figure would be easily
  achievable. In other words, in a typical $2^\circ$ portion of the
  tracking plot the drift is within that figure. The largest slope on
  the elevation plot, for example, corresponds to about 2~arcsec in
  the total of $11^\circ$ of track, or about 0.4~arcsec per $2^\circ$. The
  slopes on the $\Delta A \cos{E}$ track are, if we overlook the
  high-frequency events, somewhat steeper.
\end{itemize}

\subsubsection{All-Sky Optical Pointing Summary}

\begin{enumerate}

\item The rapid slewing of the ALMA antennas leads to data sets
of ample size compared with the five or six variable terms of the
model, and so the RMS figure is a good guide to the population
standard deviation.

\item All-sky optical pointing measurements made over ambient
  temperatures from $-10$ to $+20$~C and during which the wind speed
  ranged up to 17 m/s showed no dependence on either of these site
  conditions.

\item For the AEC antenna, the all-sky RMS pointing accuracy from the
  2004/03/06 through 2004/05/12 period (Figure~\ref{fig:a_rms2}) is
  typically 2.4~arcsec, with the better results in the first half of
  the test period, which was when the more stable IE and CA values
  were seen.

\begin{figure}
\centering
\includegraphics[scale=0.5,angle=-90]{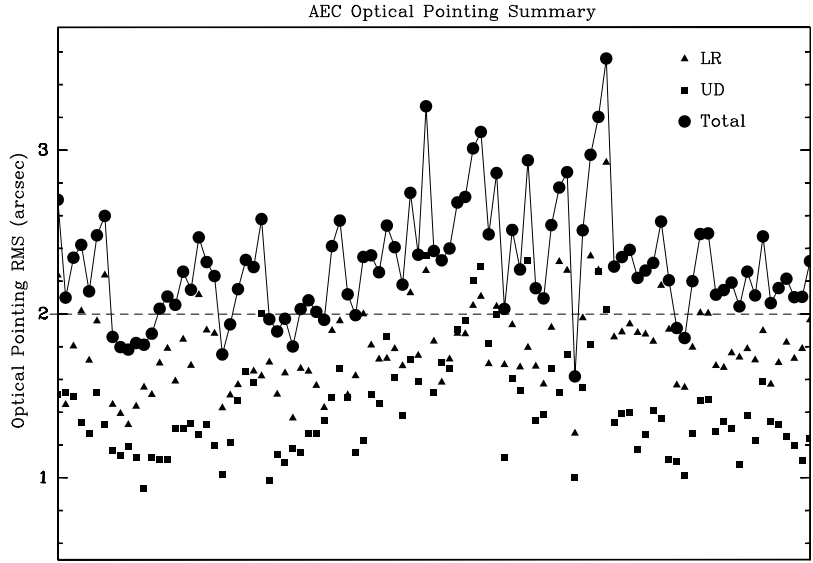}
\caption{AEC antenna OPT all-sky pointing RMS.  The horizontal,
   vertical, and total RMS are shown.}
\label{fig:a_rms2}
\end{figure}

\begin{figure}
\centering
\includegraphics[scale=0.5,angle=-90]{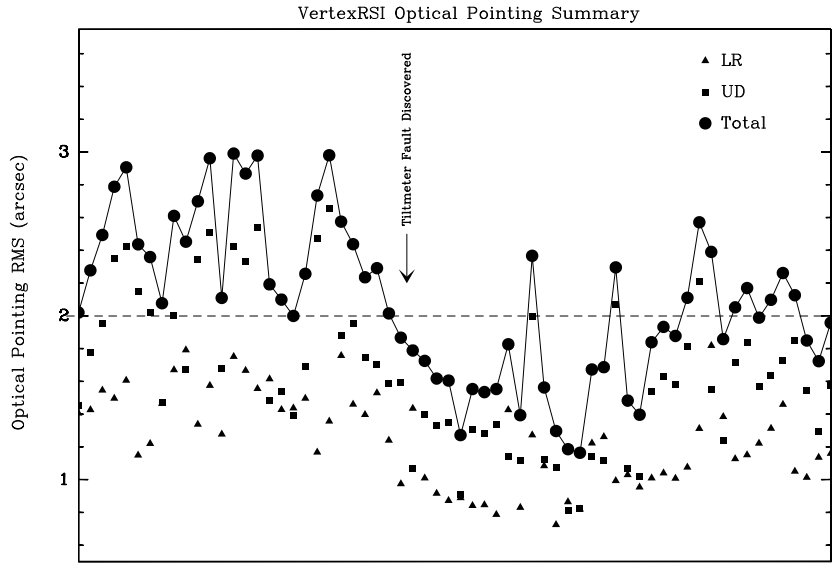}
\caption{VertexRSI antenna OPT all-sky pointing RMS.  The
    horizontal, vertical, and total RMS are shown.}
\label{fig:v_rms2}
\end{figure}

\item In Figure~\ref{fig:a_rms2} the AEC sky RMS is resolved into two
components, left-right and up-down respectively.  It is evident
that the left-right component (i.e.~$\Delta A \cos{E}$) is worse
than the up-down (i.e.~$\Delta E$) component.
This worse azimuth behavior is fairly constant over the test
period, whereas the unexplained decline in elevation performance in
the second half accounts for most of the deterioration in overall
RMS.

\item Figure~\ref{fig:aecopt-rms} shows the AEC all-sky RMS pointing
  accuracy for the period 2005/01/07 and 2005/03/02.  The mechanical
  modifications made by the antenna contractor have produced a
  moderate improvement to the all-sky pointing performance.

\item For the VertexRSI antenna, Figure~\ref{fig:v_rms2} shows that if
  we neglect the earlier, pre-PMDR period, the sky RMS is not much
  affected by the passage of time and by changing conditions. A
  typical result is 2.0~arcsec RMS, and most of the PMDR runs
  delivered better than this, in some cases much better.
  Figure~\ref{fig:v_rms2} also shows the sky RMS resolved into two
  components, left-right and up-down respectively.  It is evident that
  the up-down component (i.e.~$\Delta E$) is worse than the
  left-right component (i.e.~$\Delta A \cos{E}$).  The performance
  variations in the two axes seem to be somewhat correlated, perhaps
  evidence that variations in optical seeing were involved, while the
  occasional poor sky RMS happened when there is a sudden worsening in
  elevation performance.

\item A sequential multiple star measurement, where each star
  measurement was repeated three times, was used to derive seeing and/or
  short timescale antenna positioning errors.  These measurements
  suggest a seeing/vibration contribution of $\sim$1.5~arcsec, in line
  with the results from radiometric tracking tests.  Extending the
  interpretation to account for the seeing contribution suggests that
  the true antenna pointing residual could be as low as 1.7~arcsec for
  the AEC antenna and 0.8~arcsec for the VertexRSI antenna.

\end{enumerate}

\subsection{Radiometric Pointing Measurements}
\label{radioanalysis}

Details of the radiometric test procedures can be found in
\cite{Lucas2004}.  The radiometric data were logged in a different way from
the OPT data and required some pre-processing, but the end results
were similar in the two cases.

\subsubsection{Radiometric Pointing Models and Results}
\label{radmod}

Essentially the same procedure was used for both the OPT and the
radio data. Special TPOINT scripts 
were used to mass-reduce the final selection of files,
with automatic removal of outliers and with no manual
intervention. The residuals from the different runs were standardized
and superimposed, first to look for hitherto undetected terms and then
to determine the coefficient values for a fixed ``core'' group of
pointing coefficients.  

Radiometric pointing measurements were obtained during the period
2004/05/16-25 with the AEC antenna and 2004/03/20-22 with the
VertexRSI antenna.  Some pre-processing of the data was necessary before
conventional TPOINT analysis could begin, to fold the
operational model back into the observations.  Once treated in this
way, the four (AEC) and three (VertexRSI) measurement runs obtained,
which contained (22, 30, 46, 57) and (44, 48, 37) pointing 
measurements, respectively, proved to be sufficiently consistent
simply to concatenate them.  Fitting each one separately was less promising
given the relatively small number of observations in each data set.

For both antennas the first model tried was the OPT model with only
IE, CA and HECE allowed to vary. The first and second terms are simply
the radio boresight offset relative to that of the OPT. The third term
is the Hooke's Law vertical flexure, which will be different for the
radio and OPT cases. With this approach, the residuals were clearly
systematic.

To deal with the elevation errors as a function of elevation, both
HESE (elevation errors proportional to 
$\sin{E}$) and TX (elevation errors proportional to $\tan{Z}$) were
tried, and the former worked best. This suggested some asymmetry in
the BUS, subreflector supports or nutator.  

There were also signs 
that the azimuth zero point correction (IA) and/or the Az/El
non-perpendicularity (NPAE) had changed. The explanation (see
\S\ref{optrad}) seems to be that the 
asymmetric location of the OPT, mainly in the horizontal direction,
makes it vulnerable to elevation-dependent flexures in the BUS, a
sideways deflection of the OPT occurring as the BUS ``opens out'' with
increasing elevation.  It was therefore decided to allow both of these
terms to vary in addition to CA. For the AEC antenna these tests
showed that a substantial 15~arcsec change in IA was needed, and an
additional core term (HESE),
but with
no significant change in NPAE (under 1~arcsec in fact).  For the
VertexRSI antenna, only an 8~arcsec change in IA was needed, but large
changes in HESE (46~arcsec) and NPAE (27~arcsec) were required.

With IA, IE, HESE, HECE and CA allowed to vary for both antennas (and
for good measure AN and AW---these may change slightly depending on
static wind pressure, and as they are rather uncorrelated with other
terms are safe to include) satisfactorily non-systematic residuals 
were obtained.  The final model obtained for both prototype antennas
using this procedure is listed in Table~\ref{tab:radiometric}.

\begin{table}[h!]
\centering
\caption{Radiometric Pointing Model}
\begin{tabular}{|lrr|lrr|}
\hline
\multicolumn{3}{|c}{AEC} & \multicolumn{3}{c|}{VertexRSI} \\
\hline
Coeff &     Value  &  Sigma &  
Coeff &     Value  &  Sigma \\ \hline\hline
IA    &  $-59.41$   & 1.542 &
IA    &  $-8.43$  &  6.495 \\
IE    &  $+0.04$   & 6.614 &
IE    &  $-0.18$  &  8.817 \\
CA    &  $+0.09$   & 1.167 &
CA    &  $+11.76$  &  8.711 \\
AN    &  $+6.91$   & 0.511 &
AN    &  $-0.13$  &  0.609 \\
AW    &  $-1.78$   & 0.421 &
AW    &  $-0.03$  &  0.509 \\
NPAE  &  $+28.62$   & &
NPAE  &  $+4.10$  &  6.588 \\
HESE  &  $+17.84$   & 5.230 &
HESE  &  $+19.24$  &  6.724 \\
HECE  &  $+40.63$   & 4.882 &
HECE  &  $+25.01$  &  6.554 \\
HASA2 &  $-1.70$   & &
HASA  &  $-1.20$  & \\
HACA2 &  $+2.94$   & &
HACA  &  $-4.42$  & \\
HESA2 &  $-0.99$   & &
HASA2 &  $+0.73$  & \\
HECA2 &  $+1.53$   & &
HACA3 &  $+0.38$  & \\
HVSA2 &  $-2.25$   & &
HESA3 &  $+0.94$  & \\
HVCA2 &  $-2.08$   & &
HECA3 &  $-0.34$  & \\ \hline
\multicolumn{3}{|r|}{Sky RMS =   4.35} & \multicolumn{3}{r|}{Sky RMS =   4.77} \\
\multicolumn{3}{|r|}{Popn SD =   4.45} & \multicolumn{3}{r|}{Popn SD =
  4.94} \\ 
\hline
\end{tabular}
\label{tab:radiometric}
\end{table}

For both antennas a plot of the $\Delta A \cos{E}$ and $\Delta E$ 
errors against observation number shows evidence of drifts during each
of the pointing runs.
For the VertexRSI antenna, the average drift for each of the three
runs was just under 0.2~arcsec per observation.  If the data from the
three VertexRSI radiometric pointing runs are simply concatenated,
without individual boresight adjustments, 
drifts in both $\Delta A \cos{E}$ and $\Delta E$ can be seen.

To derive the ``best radiometric pointing model'' from these
measurements \textit{without any drift corrections}, we used an
analysis procedure similar to that used for the optical pointing
measurement reductions.  The HESE, HECE and NPAE coefficients were
iteratively refined, and the individual pointing runs were normalized
by freeing them from their preferred IE, CA, AN and AW values.  The
normalized data was then combined and then fitted for the five (AEC)
and four (VertexRSI) core terms
plus IA, HESE and HECE (and NPAE for the VertexRSI antenna). The
resulting ``best radiometric pointing model'' for the AEC and
VertexRSI antennas is shown in Table~\ref{tab:bestrad}.

\begin{table}[h!]
\centering
\caption{Radiometric/Optical Pointing Model Comparison}
\begin{tabular}{|lrr|lrr|}
\hline
\multicolumn{3}{|c}{AEC} & \multicolumn{3}{c|}{VertexRSI} \\
\hline
  Term &   Radiometric &  Optical &
  Term &   Radiometric &  Optical \\
\hline
  IA    &  $-59.41$    &  $-74.2$ &
  IA    &  $-250$     &  $-258.0$ \\
  IE    &  $+147$     &  $-695.3$ &
  IE    &  $+288$     &  $+433.0$ \\
  CA    &  $+273$     &  $-829.5$ &
  CA    &  $+319$     &  $+173.0$ \\
  AN    &  $+6.91$    &  $+7.7$ &
  AN    &  $-29$      &  $-31.0$ \\
  AW    &  $-1.78$    &  $-0.8$ &
  AW    &  $+15$      &  $+14.0$ \\
  NPAE  &  $+28.62$   &  $+28.62$ &
  NPAE  &  $+12.14$   &  $-15.01$ \\
  HESE  &  $+17.84$   & &
  HESE  &  $+17.95$   &   $-27.56$ \\
  HECE  &  $+40.63$   &   $-16.85$ &
  HECE  &  $+23.16$   &   $+30.09$ \\
  HASA2 &  $-1.70$    &   $-1.70$ &
  HASA  &  $-1.20$    & \\
  HACA2 &  $+2.94$    &   $+2.94$ &
  HACA  &  $-4.42$    & \\
  HESA2 &  $-0.99$    &   $-0.99$ &
  HASA2 &  $+0.73$    & \\
  HECA2 &  $+1.53$    &   $+1.53$ &
  HACA3 &  $+0.38$    & \\
  HVSA2 &  $-2.25$    &   $-2.25$ &
  HESA3 &  $+0.94$    & \\
  HVCA2 &  $-2.08$    &   $-2.08$ &
  HECA3 &  $-0.34$    & \\
\hline
\end{tabular}
\label{tab:bestrad}
\end{table}

\begin{figure}
\centering
\includegraphics[scale=0.60,angle=-90]{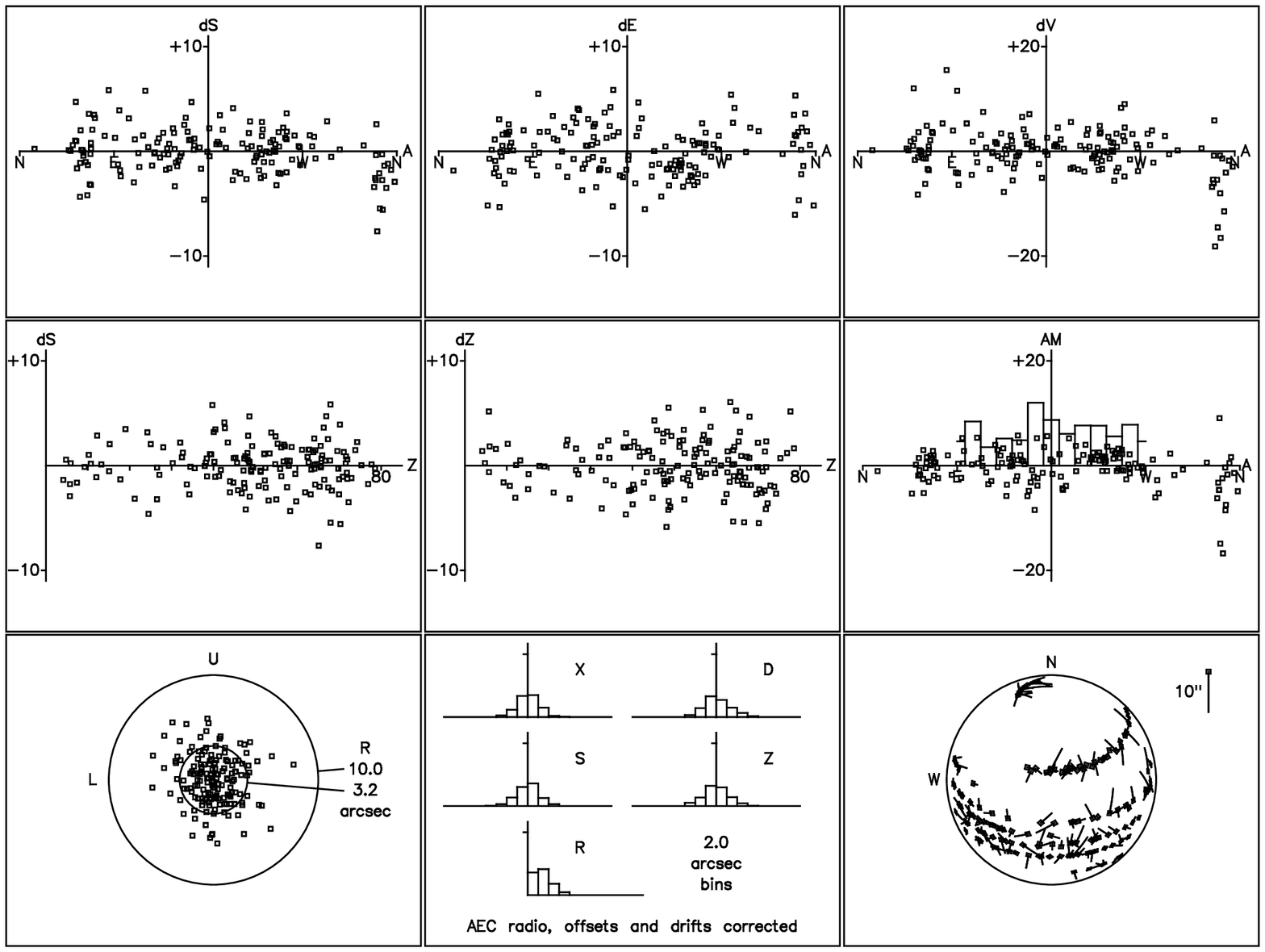} \\
\includegraphics[scale=0.60,angle=-90]{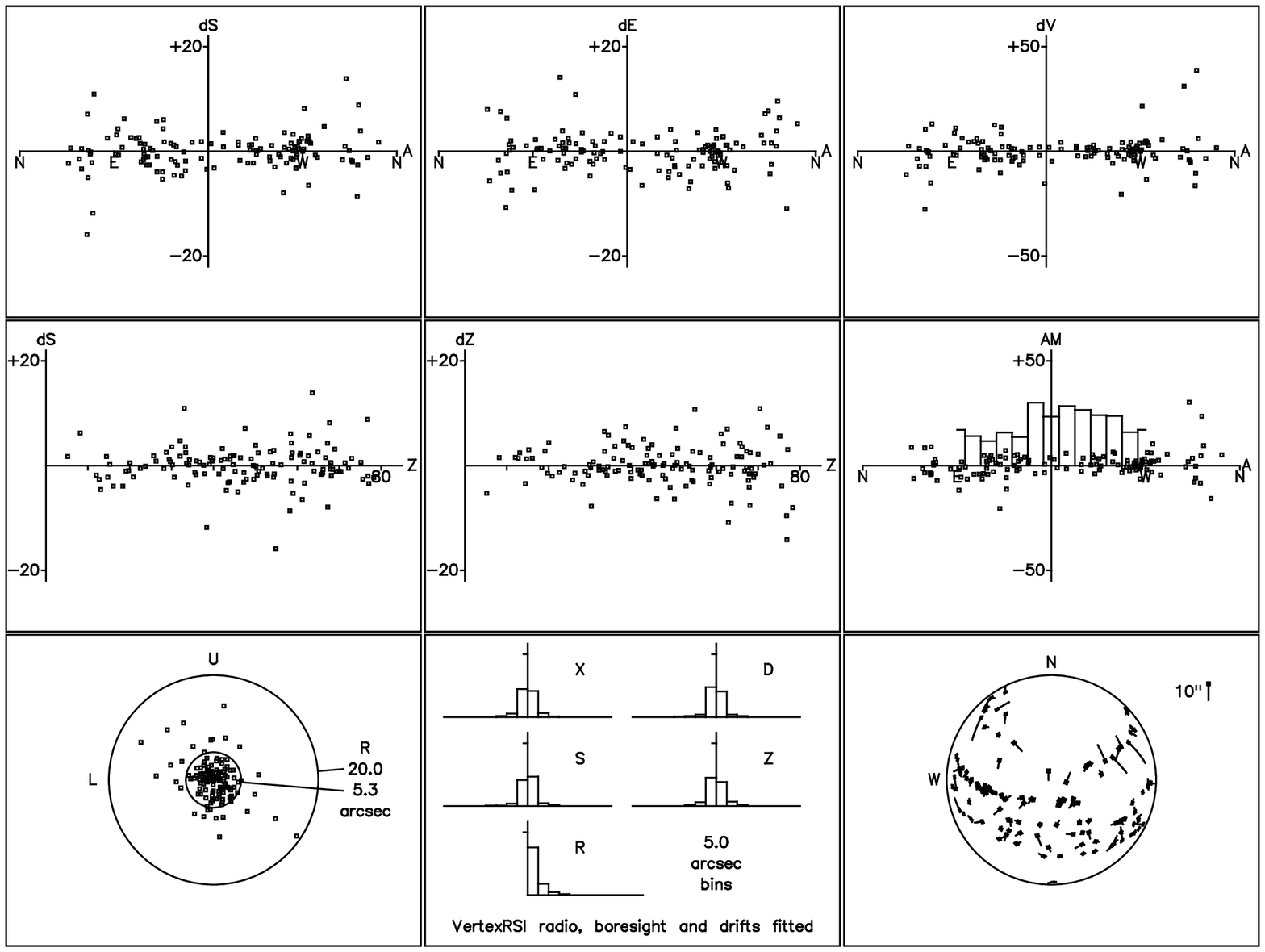}
\caption{Superimposed best radiometric pointing model residuals for
  the AEC (top) and VertexRSI (bottom) antennas.  See
  Figure~\ref{fig:7mod} for a description of the quantities plotted in
  each panel above.  The pointing model terms in this fit are listed
  in Table~\ref{tab:bestrad}, which yielded an overall 3.2 and
  5.3~arcsec RMS for the AEC and VertexRSI antennas, respectively.
  Note that these results are dominated by measurement error.}
\label{fig:radio4}
\end{figure}

For each antenna its model was applied individually to each run by
fitting only IE (plus drift), CA (plus drift), AN and AW and the
results combined (Figure~\ref{fig:radio4}).  The overall 3.2~arcsec
and 5.3~arcsec RMS figures for the AEC and VertexRSI antennas,
respectively, should be treated with some caution: superimposing
individually fitted runs is a particularly optimistic way to present
pointing results.

An analysis of the same three days of data was done (\cite{Lucas2004})
using the Plateau de Bure pointing model software. The results are in
full agreement. No systematic effect was detected: the results are
dominated by measurement errors.

\subsubsection{Optical/Radiometric Pointing Correspondence}
\label{optrad}

The location of the OPT within the backup structure of the antenna
requires that one consider how the measurement axes of both systems
are related.
In the pointing model for the antenna concerned, for an ideal OPT
fixed to the radio telescope surface and assuming mechanical symmetry
about the axis of the parabola, one 
would expect only the boresight offset (IE and CA) and vertical
flexure (HECE) terms in the pointing model to change.  However,
on both antennas, preliminary analysis of the radiometric
pointing showed signs 
of the azimuth zero point correction (IA) and/or the Az/El
non-perpendicularity (NPAE) having changed.  The likely
explanation of such effects is that the asymmetric location of the
OPT, mainly in the horizontal direction, makes it vulnerable to
elevation-dependent flexures in the BUS, a sideways deflection of the
OPT occurring as the BUS ``opens out'' with increasing elevation. A
deflection proportional to $\sin{E}$ would produce a term functionally
identical to NPAE, while the $\cos{E}$ phase would produce an apparent
change in IA.

For a perfect parabola at elevation $45^\circ$, the aperture of the
antenna will change into being more elongated in azimuth at the zenith
and more elongated in elevation at the horizon.  Since the ALMA OPTs
are located azimuthally off-axis, the boresight of the OPTs are
expected to be pointed to more positive azimuths at the
zenith, and to more negative azimuths at the horizon.  One
would expect, then, the following additional harmonic elevation and
azimuth terms to the OPT pointing correction with respect to the
radiometric pointing model:

\begin{eqnarray}
\Delta E &=& a \sin{E}~~\textrm{(same as HESE)} \nonumber\\
\Delta E &=& b \cos{E}~~\textrm{(same as HECE)} \nonumber\\
\Delta A &=& c \left(\frac{\sin{E}}{\cos{E}}\right) \nonumber\\
         &=& c \tan{E}~~\textrm{(same as NPAE; equivalently a CA
  contribution proportional to $\sin{E}$)} \nonumber\\
\Delta A &=& d \left(\frac{\cos{E}}{\cos{E}}\right) \nonumber\\
         &=& d~~\textrm{(same as IA; equivalently a CA
  contribution proportional to $\cos{E}$)} \nonumber
\end{eqnarray}

\noindent{This} indicates that one should expect fixed differences
between the radiometric and optical determinations for IA, NPAE, HESE,
HECE. 

Taking the VertexRSI antenna as an example,
a mechanical verification of this theory can be derived from the
measured axial focus change due to BUS deflection (see
\cite{Greve2006}).  The VertexRSI BUS axial focus is measured to
deflect by $+1.7$~mm over elevation range $0^\circ$ to $90^\circ$.  Using an
80\% radiometric taper leads to a deflection of $+1.35$~mm.  With the
following constants:

\begin{eqnarray}
x &\equiv& \textrm{Distance from antenna axis to OPT = 3494 mm}\nonumber \\
f &\equiv& \textrm{Antenna nominal focus = 4875 mm}\nonumber
\end{eqnarray}

\noindent{we} find that, for a perfect parabola:

\begin{eqnarray}
x^2 &=& 2pz \nonumber \\
\tan{\alpha} &=& \frac{x}{2f} \nonumber \\
\frac{\delta\alpha}{\delta f} &=& 13.43~\textrm{arcsec} \nonumber \\
\delta\alpha(\delta \textrm{f = 1.35~mm}) &=& \textrm{18~arcsec} \nonumber
\end{eqnarray}

\noindent{This} corresponds well to the combined measured shift in
NPAE ($+24\arcsec$) and IA ($-7\arcsec$) between our optical and
radiometric pointing models for the VertexRSI antenna.

\subsubsection{Radiometric Seeing}
\label{radseeing}

We have investigated the influence of refractive pointing errors (or
``radiometric seeing'') on our radiometric tracking measurements.
Measurements of the differences between radiometrically-derived
positions and their corresponding encoder positions while tracking a
source for 15-20~minutes strongly indicate the existence of a
refractive pointing error term (see \cite{Holdaway2006}).
These refractive fluctuations result in an $\sim0.6$~arcsec
radiometric positioning jitter.  The magnitude of the inferred
atmospheric pointing structure function at 1~s (the atmospheric
crossing time of a 12~m antenna) is predicted to within 10\% by
theoretical arguments about refractive pointing jitter and Atmospheric
Phase Interferometer data.

\subsection{Optical Offset Pointing Tests}

To investigate the accuracy of the offsetting performance, OPT
observations were made where the antennas were switched repeatedly
between a number of stars over distances of about $2^\circ$.  Four
(AEC) and seven (VertexRSI) such runs were conducted, consisting of 
2 to 5 and 3 to 7 stars, respectively.  The 
analysis of these observing runs proceeded as follows:

\begin{enumerate}
\item A fit and removal of the standard 13-parameter pointing model.
  In this model all parameters except the boresight offsets IE and
  CA were held fixed.
\item Outlier removal.
\item Concatenate the resulting files and perform a final fit to a
  model consisting only of the boresight position and drift terms IE,
  CA, A1S and A1E.  This exhaustive removal of position and drift
  effects is appropriate here because we are investigating neither
  tracking nor all-sky pointing.  The individual stars in the 
  field are all equally affected by these measures, with any
  systematic effects associated with the transition from one star
  to another left as a residual.
\item For each star in this concatenated data set fit a model
  consisting of IE and CA alone.  This last step reveals any
  systematic offsetting errors.
\end{enumerate}

\begin{figure}
\centering
\includegraphics[scale=0.6,angle=-90]{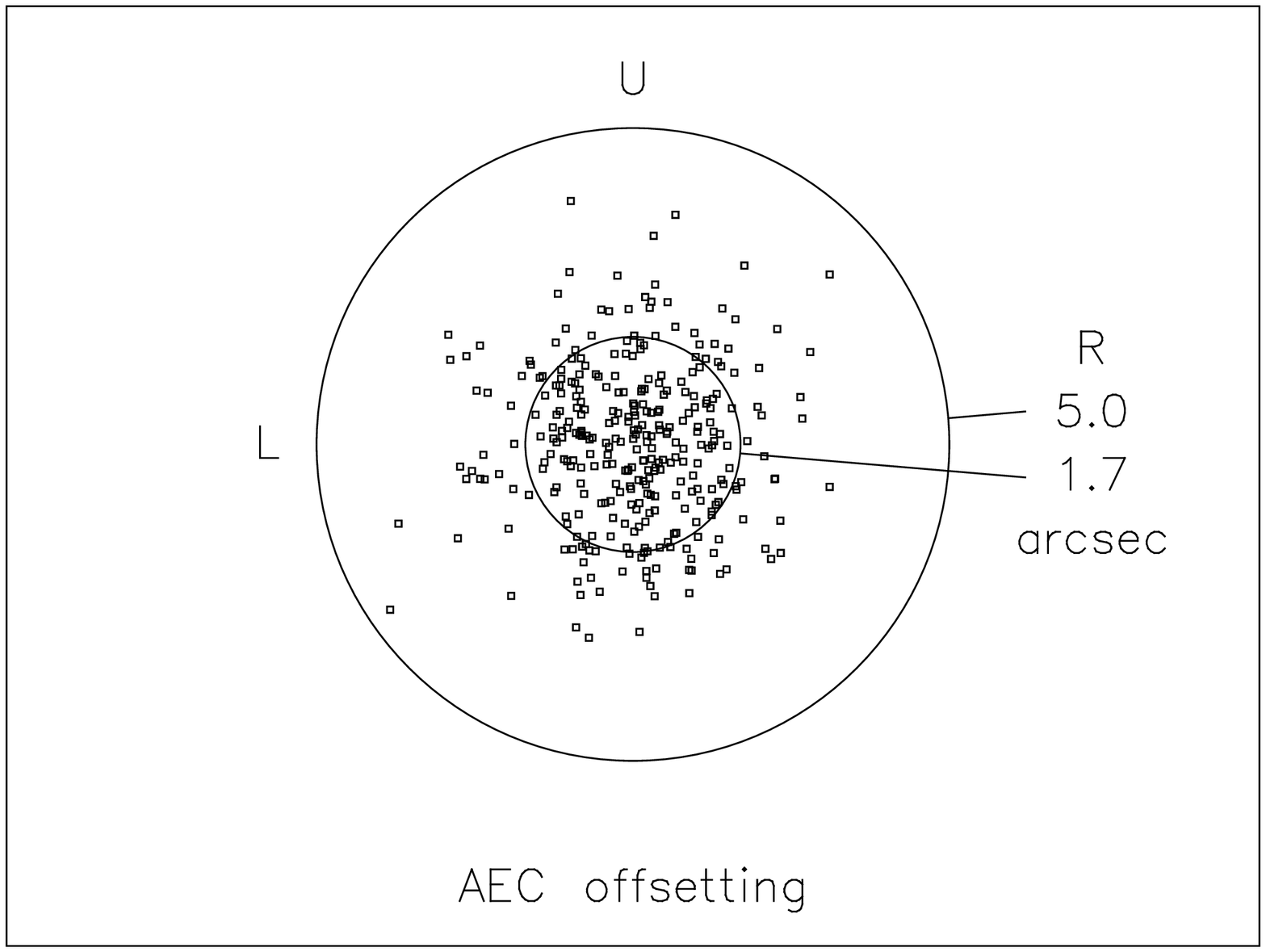}
\includegraphics[scale=0.6,angle=-90]{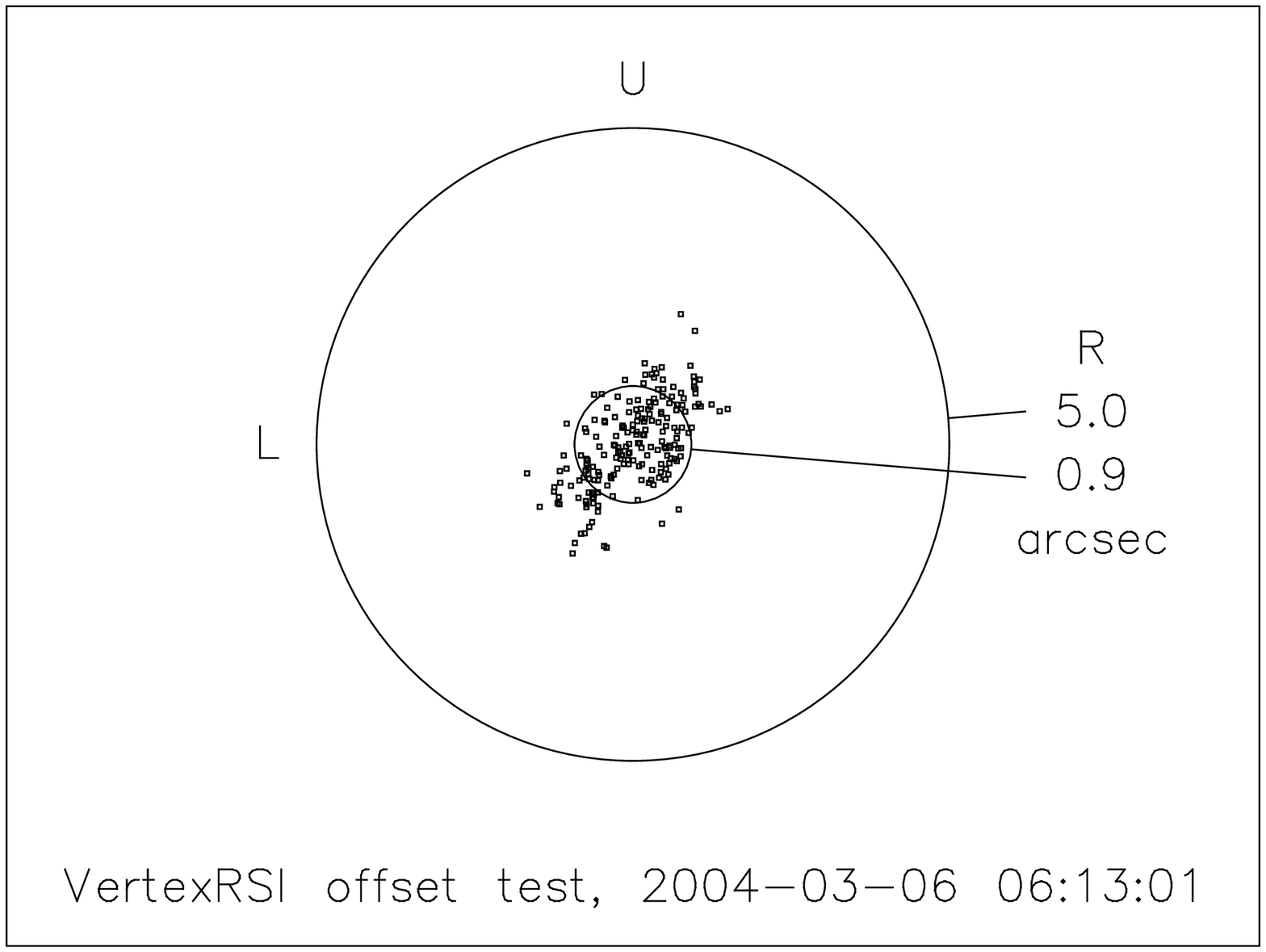}
\caption{Sample AEC (top) and VertexRSI (bottom) offset pointing
   residuals.} 
\label{fig:offs}
\end{figure}

The displacements of each star from the mean were then examined.
Figure~\ref{fig:offs} shows sample 
offset pointing residuals for each antenna resulting from these
analyses.  The majority of the offset optical pointing runs for both
antennas resulted in a derived offset pointing performance within the
0.6~arcsec specification.  A couple of these optical pointing runs
resulted in offset pointing residuals in the 0.9 to 1.1~arcsec range.  
These results appear to show that both antennas just miss the
0.6~arcsec offsetting specification. However, it must be borne in mind
that not only were the experiments delicate and vulnerable to many
sources of error, but also that the results were achieved without the
metrology systems being active.

\subsection{Metrology Performance}
\label{metrology}

The AEC prototype metrology system originally comprised three
components: 

\begin{itemize}
\item Temperature sensor system to measure the pointing offsets
  due to thermal deformation of the structure.
\item Tiltmeter system to measure pointing offsets due to tilts in
  specific structural elements.
\item Laser interferometer system to measure structural changes in the
  fork arms of the antenna.
\end{itemize}

Only the temperature sensor and tiltmeter systems were installed
on the prototype antenna.  Attempts by AEC to qualify the temperature
sensor and tiltmeter metrology systems were unsuccessful.  The
limited number of measurements that the AEG made with these metrology
systems active lead to a poorer pointing performance for the antenna,
which is a clear indication that these metrology systems were not
functioning properly.

The VertexRSI antenna was designed to meet the pointing specifications
with the aid of two separate metrology systems:

\begin{itemize}
\item A displacement sensor system mounted on the CFRP reference
  structure near each elevation encoder.  This system was designed to
  measure the tilt component of the fork arm structure and BUS which
  cannot be measured by the encoders themselves.
\item A system composed of two tiltmeters in the pedestal of the
  antenna.  These tiltmeters are designed to measure the tilt of the
  azimuth axis.
\end{itemize}

Well into the evaluation the tiltmeter metrology system readout was
found to drift by many tens of arcseconds over timescales of up to one
minute.  It was not clear what was causing these drifts.  This could
be a problem with the software that applies these tiltmeter
measurements to the antenna positioning, or to a mechanical problem
with the mounting of the tiltmeters themselves. Once this problem was
discovered, all pointing measurements were made with this metrology
system disabled. 

A number of measurements which included the corrections applied by the
displacement sensor metrology system were made.  Unfortunately, these
measurements were not extensive enough to evaluate the effectiveness
of the displacement metrology system.  At the very least, this system
did not make the pointing performance of the VertexRSI antenna worse.

We should note that even though none of the delivered metrology
systems appeared to improve the pointing performance of either
antenna, both antennas met the ALMA pointing specifications.
Apparently the antennas on their own outperform their designed
pointing performance specifications.

\subsection{Evidence of Tripod Print-Through on the VertexRSI Antenna}
\label{v_tripod}

\begin{figure}
\centering
\includegraphics[scale=0.45,angle=-90]{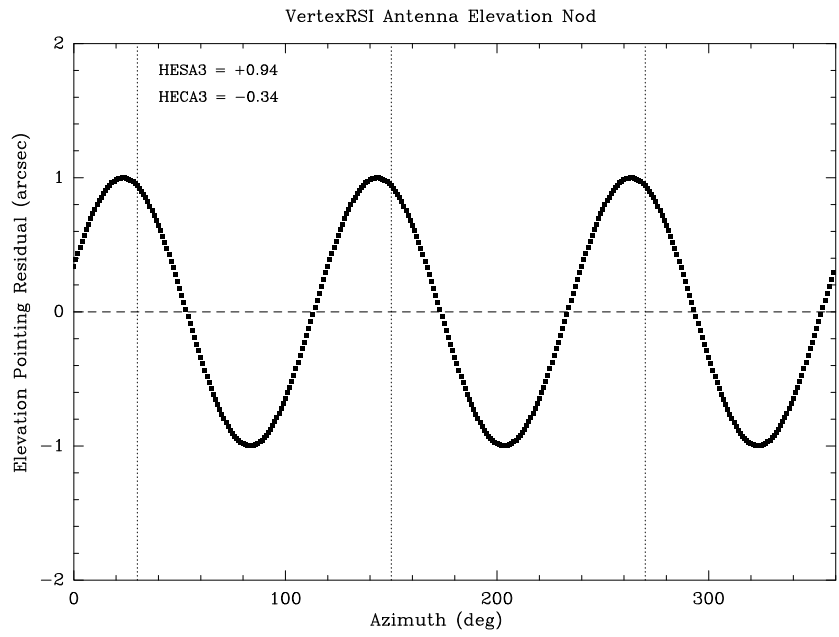}
\caption{The HESA3 and HECA3 pointing model terms, showing
  possible tripod print-through in the azimuth motion of the antenna.
  The azimuth positions of the three mount points of the VertexRSI
  antenna are shown as dotted lines.}
\label{fig:v_nod}
\end{figure}

To assess the viability of the low-level ($\lesssim 1$~arcsec) terms
in our pointing models we have checked the physical association of two
of these minor terms with a plausible mechanical deformation on the
VertexRSI antenna.
The portion of the OPT pointing model involving \textit{HESA3} =
$+0.94$ and \textit{HECA3} = $-0.34$ shows that the antenna nods in
elevation as it rotates 
in azimuth, executing three cycles of nodding per turn of azimuth.
This may be evidence of print-through from the tripod support into
the azimuth bearing. The effect can be illustrated by removing all
but those two fixed terms from the model, and then applying the
model to a grid of stars all over the sky. The resulting plot of
elevation corrections versus azimuth is shown in
Figure~\ref{fig:v_nod}.

The azimuths of the three pedestal legs would be aligned to the maxima
of the nodding if there were no HECA3 term. The misalignment to the
tripod leg positions is $\arctan{(\frac{0.34}{0.94})}/3$, or about
$7^\circ$.  The sign of the 
correction is such that when the antenna is pointing over a pedestal leg
the star is about 1~arcsec higher in the sky than indicated by the
elevation encoder reading.

To further confirm the mechanical interpretation of this $\sin{3A}$
pointing term, measurements with the tiltmeter installed above the Az
bearing were made at a slow speed (0.1~deg/s) and $\pm260^\circ$ rotation in
Az.  These measurements show a single-angular sinusoidal response
[$\sin{A}$] due to the inclination of the Az axis, and a three-angular
sinusoidal response [$\sin{3 A}$] due to a repeatable wobble in the
inclination of the antenna. The three-angular wobble is attributed
to a wobble of the Az bearing. The three-angular wobble amounts to
$\pm1.5$~arcsec;
the extrema of the wobble are
correlated with the location of the three-corner support of the
pedestal. After elimination of the components $\sin{A}$ and
$\sin{3A}$, the residual deviation amounts to less than
$\pm0.5$~arcsec (peak-to-peak).

The location of the extrema of the residuals from the $\sin{A} +
\sin{3A}$ fit is further evidence that the three-angular wobble is a
print-through of the pedestal mount.  The best-fit values to the
measurement given by the following: 

$$a0 + a1 \sin{(A + b1)} + a3 \sin{(3A + b3)}$$

\begin{figure}
\centering
\includegraphics{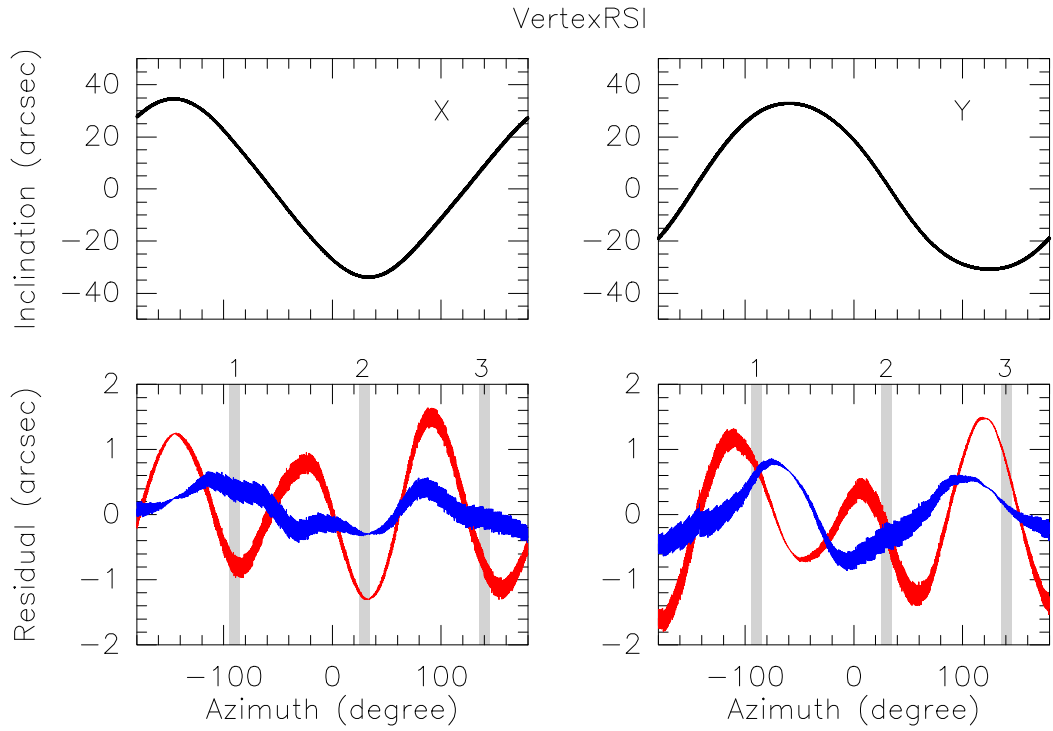}
\caption{VertexRSI antenna measurement of --\,180$^{\rm o}$ to 
+\,180$^{\rm o}$ AZ rotation. The left panel is the x--direction of the 
tiltmeter, the right panel is the y--direction. The red curves are the 
residuals of the best--fit a$_{0}$ + a$_{1}$\,sin(A + b$_{1}$), the blue 
curves are the residuals of the best--fit a$_{0}$ + a$_{1}$\,sin(A + b$_{1}$) 
+ a$_{3}$\,sin(3\,A + b$_{3}$). The gray lines show the location of the 
pedestal corners (1,2,3).}
\label{fig:azwobble}
\end{figure}

\noindent{are} listed in Table \ref{tab:azwobble} and shown in
Figure~\ref{fig:azwobble}.

\begin{table}
\centering 
\caption{VertexRSI Azimuth Rotation Tilt Measurements}
\begin{tabular}{|llll|}
\hline
Parameter & 2004 Feb 25  &  2004 Feb 25  &  Repeat 2004 Apr 10 \\
\hline \hline
Response & Linear  &   Linear + 3Az  & Linear + 3Az \\
a1 (arcsec)   & $-$32.9  & $-$32.8   &  $-$31.2     \\
b1 (arcsec)   & 56.659   & 56.811    &  59.912   \\
a3 (arcsec)   & $\ldots$ & $-$1.10   &  $-$1.13    \\
b3 (arcsec)   & $\ldots$ & $-$10.840 &  $-$4.638   \\
RMS (residual) (arcsec) & 0.85 & 0.30 &   0.25    \\
\hline \hline
\end{tabular}
\label{tab:azwobble}
\end{table}

We conclude that the telescope/azimuth bearing has a 3-Az wobble
which seems to be very stable, and which is recovered and corrected for
in the pointing model. The location of the 3-Az term suggests a
print-through of the pedestal mount.

\subsection{Accelerometer Measurements of Pointing Performance}
\label{accelpoint}

The accelerometer system described in \S\ref{accelsurf} and
\cite{Snel2006} was used to derive three components to the antenna
pointing performance:

\begin{itemize}
\item Cross-elevation pointing (left and right accelerometers on the
  rim of the BUS)
\item Elevation pointing (top and bottom accelerometers on the rim of
  the BUS)
\item Subreflector structure translations with respect to the BUS
  (four accelerometers on the rim of the BUS and one on the receiver
  flange).
\end{itemize}

In the following we characterize antenna pointing performance during
several operational modes:

\begin{itemize}
\item Pointing toward a fixed (Az,El) without tracking (called
  ``Stationary Pointing'').  In this mode the drives are powered and the
  brakes are released. Wind effects are optimally investigated in this
  mode, as any effects of antenna shake due to position updates are
  minimal in this mode. 
\item Sidereal tracking.  During sidereal tracking the antenna drives
  are constantly updating the azimuth and elevation positions,
  including the azimuth and elevation speed, in order to achieve a
  smooth tracking motion. 
  Sidereal tracking was checked for 30
  azimuth/elevation combinations spread evenly over the sky. Calm wind
  conditions were chosen, in order to clearly separate effects due to
  the drive system from those due to wind.
\item Interferometric mosaicing and On The Fly (OTF) mapping at 0.05
  and 0.5 deg/s scan speeds, respectively.
\end{itemize}

During the accelerometer measurements simultaneous wind speed and
direction measurements were used to eliminate the effects of the local
wind spectrum.  Wind-related antenna performance was calculated for
the same wind spectrum used for the design of the antennas. Our
measurements provide us with the structural stiffness for wind
excitation at frequencies between 0.1 and 3 Hz. We extrapolate from
these to lower frequencies, using the low frequency component of the
specified wind power spectrum to predict the pointing jitter due to
wind for time scales up to 15~minutes (0.001 Hz). Because we could not
measure the tracking jitter at these low frequencies, a conservative
value equal to the observed tracking jitter at 0.1 Hz is used at 0.001
Hz as well.

\begin{figure}
\centering
\includegraphics[scale=0.4]{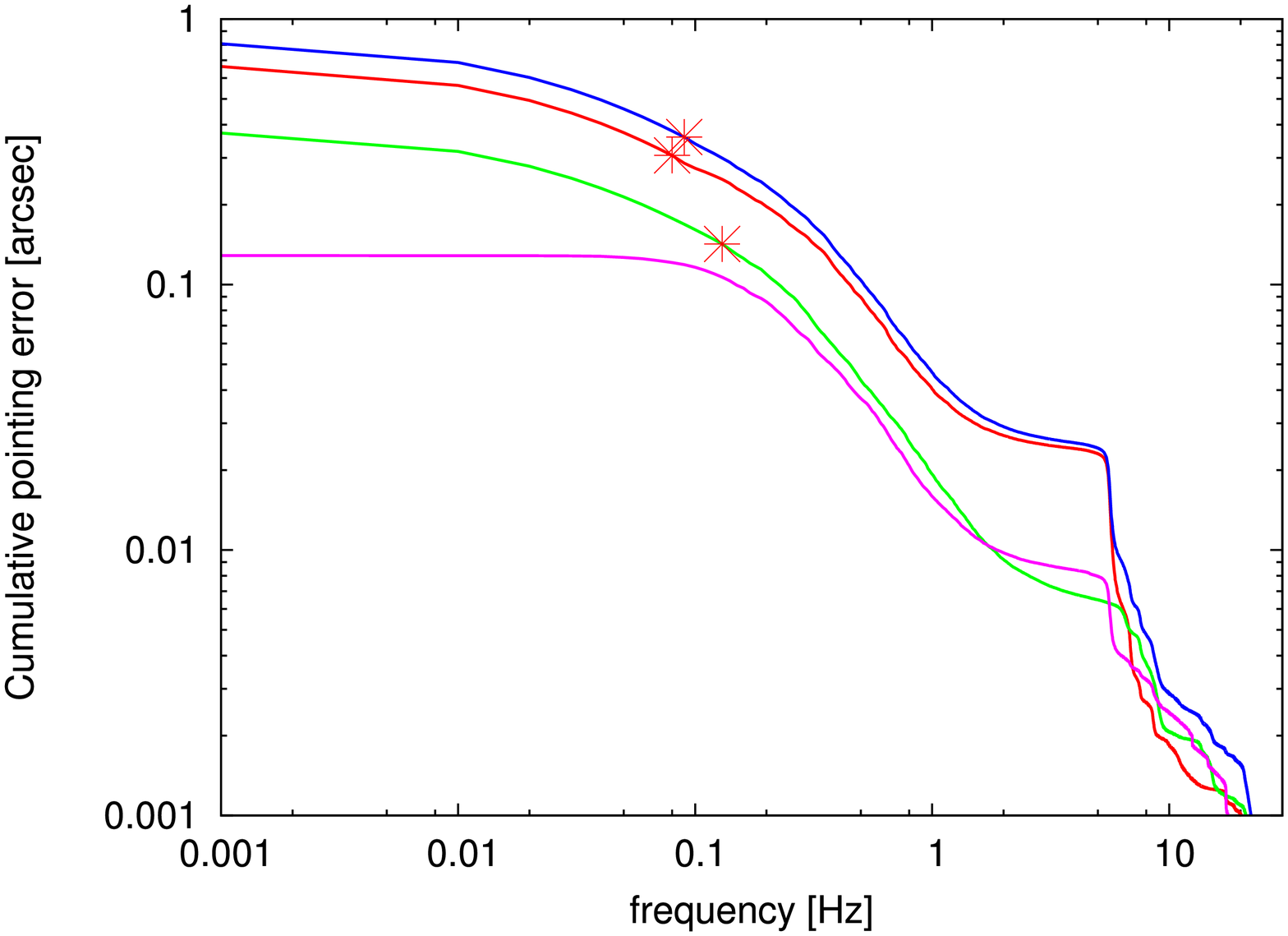}\\[60pt]
\includegraphics[scale=0.4]{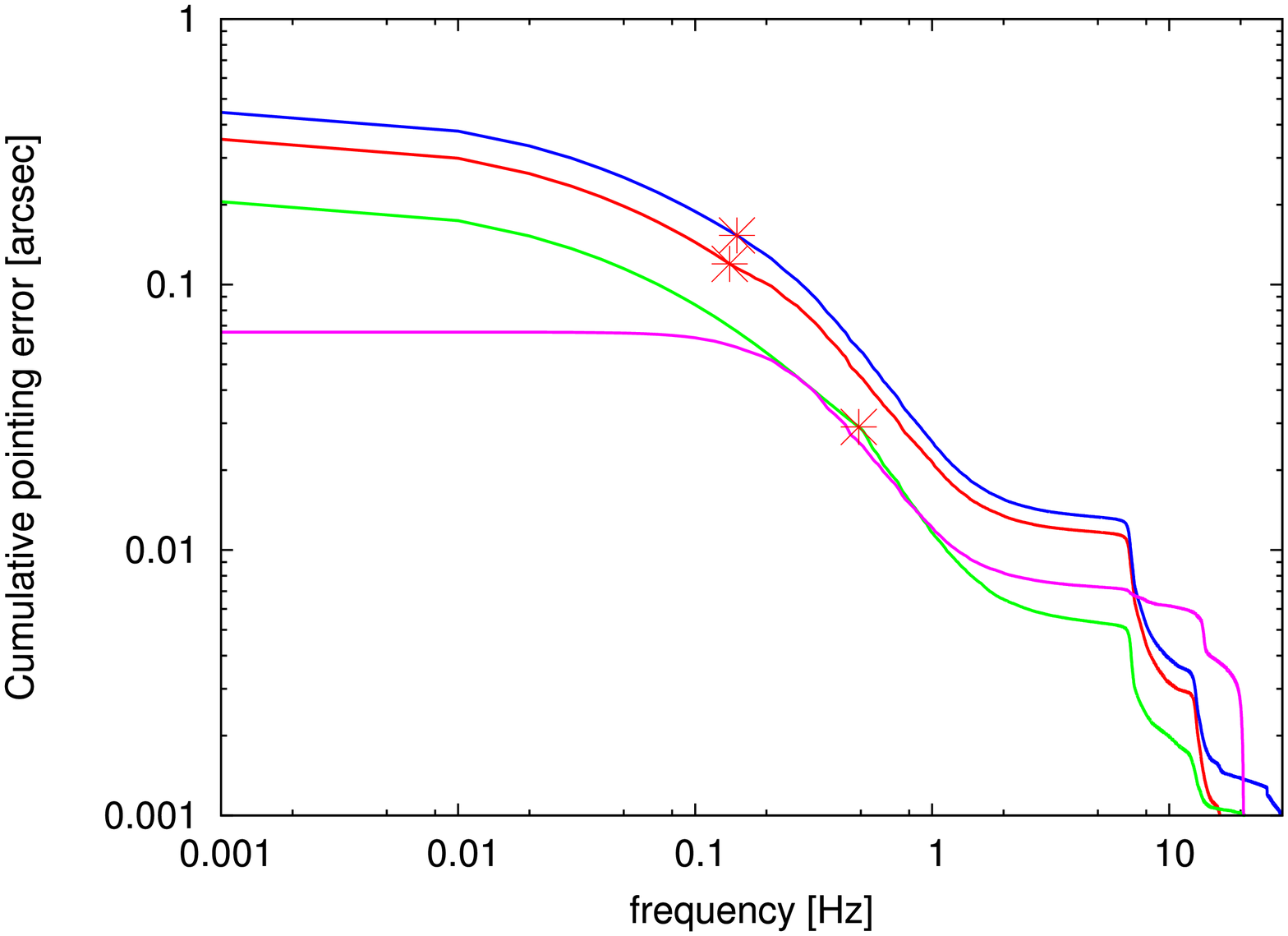}\\[20pt]
\caption{VertexRSI (top) and AEC (bottom) antenna cumulative elevation
  (red), cross-elevation (green), and 
  total (blue) pointing error for extrapolated wind and Chajnantor
  pressure. The pink curve shows the cumulative encoder pointing
  error, which is totally dominated by elevation encoder errors. The
  asterisk shows the point below which the curve has been extrapolated
  to lower frequencies.}
\label{fig:cumpoint}
\end{figure}

\begin{figure}
\centering
\includegraphics[scale=0.4]{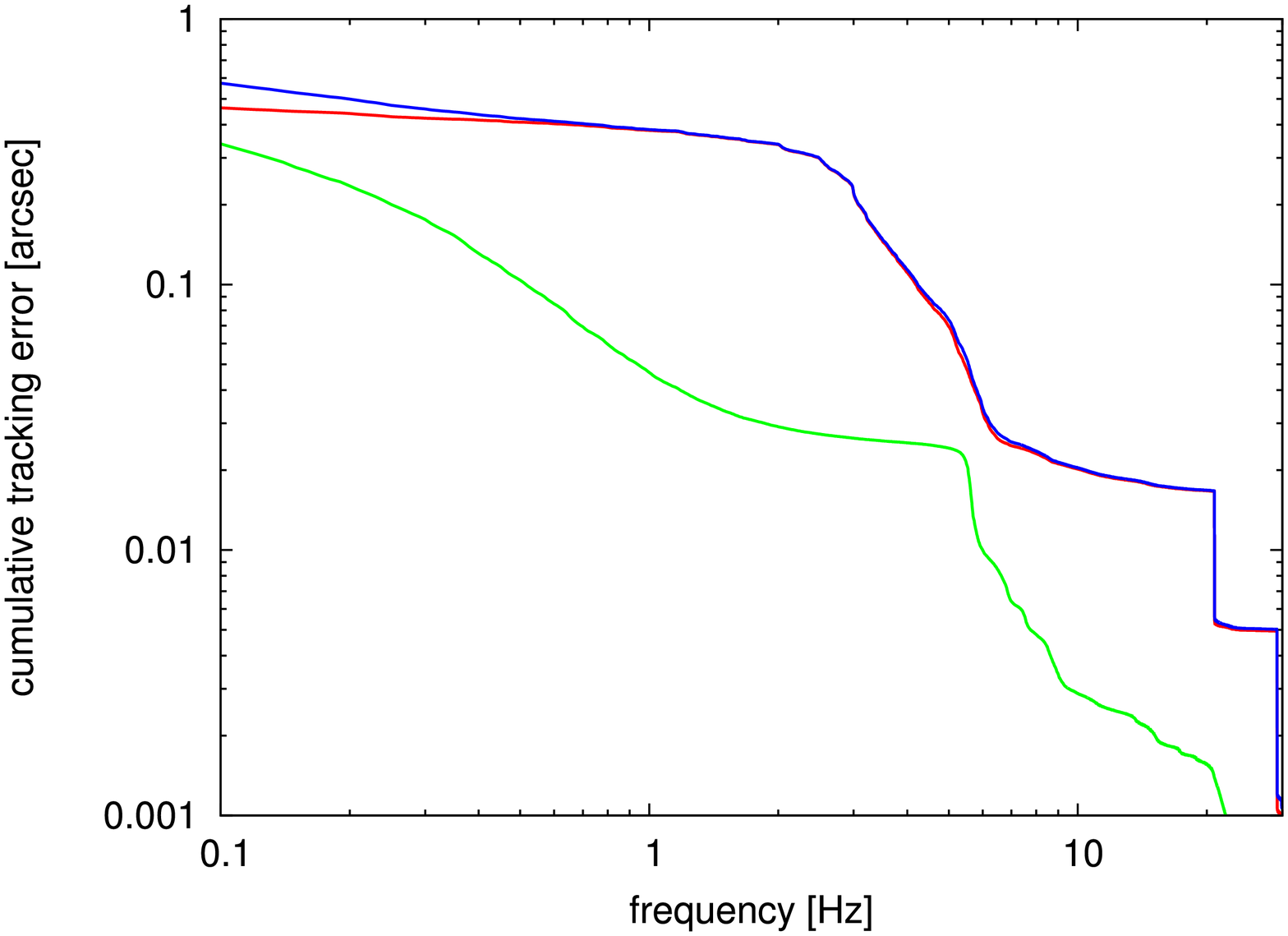}\\[60pt]
\includegraphics[scale=0.4]{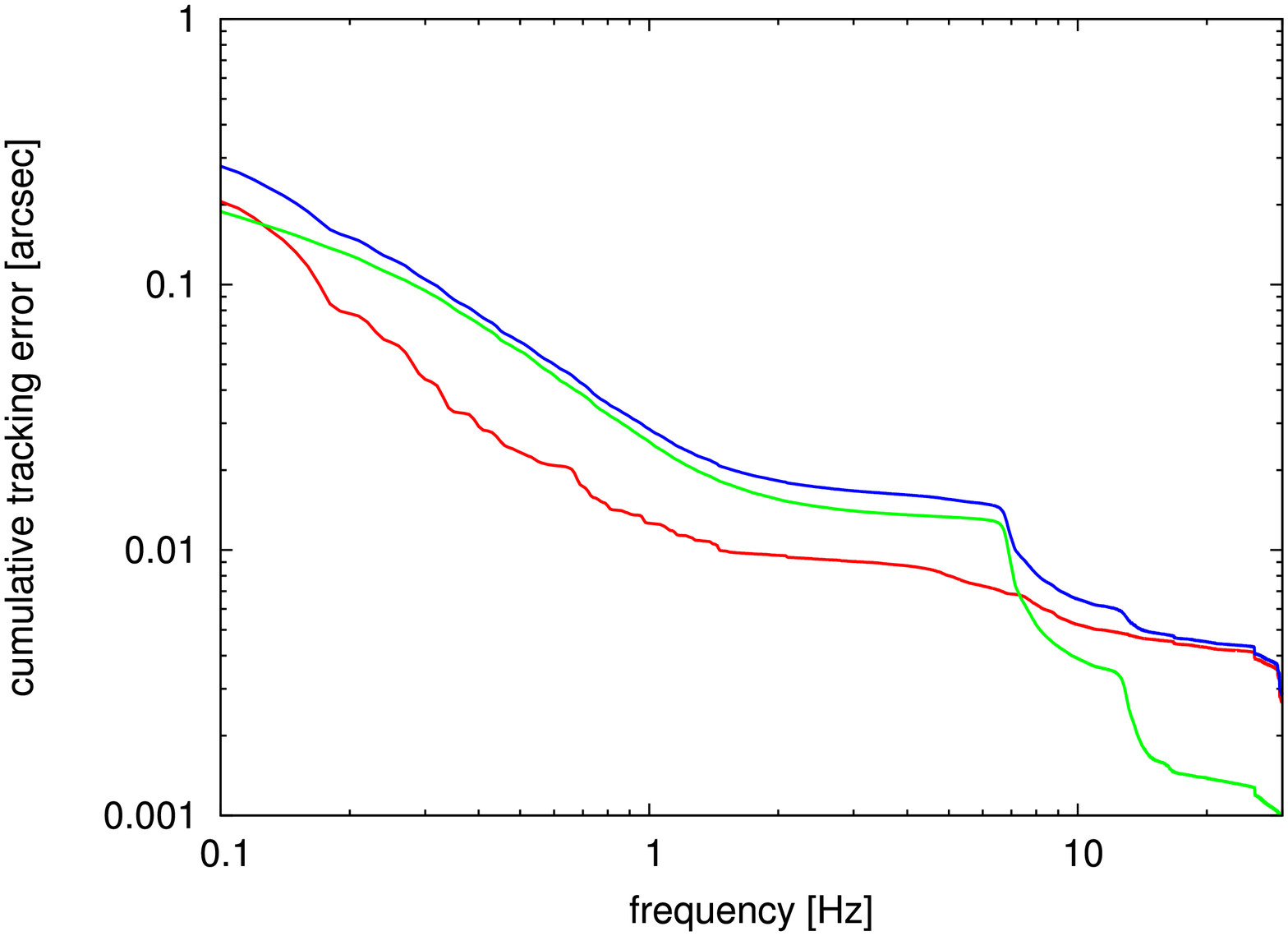}\\[20pt]
\caption{VertexRSI (top) and AEC (bottom) combined cumulative pointing
  error (blue curve) for wind (green) and sidereal tracking (red).}
\label{fig:windtrack}
\end{figure}

The measured pointing errors under the various measurement conditions
are assembled in Table~\ref{tab:accelpoint}.  The cumulative errors as
a function of frequency are shown in Figure~\ref{fig:cumpoint} for the
case of maximum wind (9~m/s) and scaled to the atmospheric situation
at the final ALMA site at 5000~m altitude. As explained above, the
section of the curves below the frequency of the asterisk has been
extrapolated with the estimated constant value of the stiffness
transfer function. Figure~\ref{fig:windtrack} shows a similar plot for
the pointing jitter during sidereal tracking and low wind.

\begin{table}[h!]
\centering
\caption{Accelerometer Measurement of Pointing Errors}
\begin{tabular}{|l|l|l|l|}
\hline
Measurement Condition & $\Delta t$ & \multicolumn{1}{c|}{VertexRSI$^a$} &
\multicolumn{1}{c|}{AEC$^a$} \\
& \multicolumn{1}{c|}{(sec)} & \multicolumn{1}{c|}{(arcsec)} &
\multicolumn{1}{c|}{(arcsec)} \\ 
\hline
Stationary Pointing (9 m/s wind) & 900 & $0.81\pm0.24\pm0.20\pm0.05$ &
$0.45\pm0.10\pm0.11\pm0.02$ \\
Sidereal Tracking (no wind) & 10 & $0.47\pm0.11$ & $0.22\pm0.08$ \\
Tracking With Wind  & 10 & $0.58\pm0.15\pm0.08$ & $0.29\pm0.09\pm0.05$ \\
Tracking With Wind  & 900 & $0.94\pm0.26\pm0.20\pm0.05$ &
$0.50\pm0.13\pm0.11\pm0.02$ \\ 
OTF at 0.5 deg/sec  & 1 & $1.7\pm0.7$ & $0.5\pm0.3$ \\
Mosaicing at 0.05 deg/sec & 1 & $0.8\pm0.5$ & $0.23\pm0.01$ \\		
Encoder Error$^b$ & 900 & 0.14 & 0.07 \\
\hline
\multicolumn{4}{l}{$^a$~Contributions to each quoted value are defined,
  respectively, as pointing variation, spread, } \\
\multicolumn{4}{l}{~~~wind power spectrum
  uncertainty, and wind extrapolation uncertainty.} \\
\multicolumn{4}{l}{$^b$~The encoder error is included in the other
  values, but constitutes a small contribution.} \\
\end{tabular}
\label{tab:accelpoint}
\end{table}

To supplement the information summarized in Table \ref{tab:accelpoint}
we note that:

\begin{itemize}

\item For stationary pointing conditions under high wind loads the BUS
  rim wind shake is dominated by elevation motion for both antennas.

\item For sidereal tracking during low wind conditions the tracking
  jitter of both antennas is largely due to elevation motion,
  with large contribution in the 3-6 Hz range. Largest jitter is
  observed for low elevation, while minimum tracking jitter is seen
  while crossing the meridian.

\item Assuming that the pointing jitter due to wind and that due to
  sidereal tracking are uncorrelated allows for the determination of
  the cumulative pointing error. A check of an independent measurement
  with high wind and sidereal tracking shows that this assumption is
  valid.  Since no tracking jitter at 0.001 Hz could be measured, a
  conservative value equal to the observed tracking jitter at 0.1 Hz
  is used.

\item For the VertexRSI BUS pointing stability during the fast (0.5
  deg/s) OTF scan, the azimuth and elevation at which the scan is
  performed have large impact on the pointing stability during the
  scan, which is reflected in the standard deviation given.

\item For the AEC BUS pointing stability during the fast (0.5 deg/s)
  OTF scan, pointing is affected for some parts of the scan by apex
  rotation feeding back to the BUS motion. This effect died out after
  a minute. The spread of 0.3~arcsec ($1\sigma$) reflects this
  variable pointing stability.

\item The contribution of subreflector motion to the total pointing
  error has been investigated, but not included in the results
  presented here. This anomalous subreflector motion is due to
  excessive and off-axis rotation of the AEC antenna apex structure.
  Our accelerometer system measured only three degrees of apex
  motion, which did not allow for the separation of subreflector
  translation, which directly affects pointing, and subreflector
  on-axis rotation, which has no impact on pointing. The results
  presented here are for BUS pointing only.  
\end{itemize}

We would like to note that measurements of the dynamical behaviour of
the antenna under operational conditions at this level of accuracy are
extremely hard to perform radiometrically. The use of accelerometers
for this purpose provides accurate results quickly and might be
considered for routine checking of pointing stability.

\subsubsection{Pointing Effects Due to Apex Motion}
\label{apexwobble}

The apex structures of both antennas rotate about an axis that is 
assumed to be aligned with the boresight axis.  However, for the AEC 
antenna, this axis is offset by $-1.5$ to $+1.0$~cm
depending on the elevation. Fast switching can excite the rotation
mode, which has translation components on-axis that can amount to
$30 \mu$m peak-to-peak. The rotation and corresponding translation
component is at 5 Hz, and damps out with a 1/e decay time of 5
seconds. With the prime focus plate scale of 34~arcsec/mm, this
translates to radio pointing errors of up to 1~arcsec peak-to-peak in
cross-elevation, provided that the rotation axis is parallel to the
boresight axis, which could not be confirmed with the equipment
available.

For the VertexRSI antenna, minor apex structure rotation was observed,
which did not affect antenna performance since the rotation appears
to be on axis.

\subsection{Pointing Performance Summary}

\begin{enumerate}

\item Due to the remarkably good pointing performance of both
  antennas, neither the optical nor radiometric tests offer sufficient
  resolution to tie down the performance of either antenna
  definitively, especially in the area of offset pointing. 

\item The metrology systems were never properly commissioned and, in
  most cases, did not appear to work properly.

\item For the AEC antenna, the all-sky pointing is just outside
  specification, typically 2.2~arcsec RMS but with occasional 2~arcsec
  RMS results.  The cross-elevation residuals are usually slightly larger
  than the elevation residuals, especially earlier in the test period
  when the elevation performance was particularly good. A sequential
  multiple star measurement was used to estimate the seeing and/or
  short timescale antenna positioning errors.  These measurements
  suggest a seeing/vibration contribution of $\sim$1.5~arcsec, in line
  with the results from radiometric tracking tests (see
  \cite{Holdaway2004}).   Extending the interpretation to account for
  the seeing contribution suggests that the true antenna pointing
  residual is 1.7~arcsec.  This result should be weighed against the
  rather optimistic approach to the analysis presented in this
  document.

\item For the VertexRSI antenna, the all-sky pointing performance was
  typically 1.5~arcsec RMS.

\item The pointing model appears to be quite stable (exceptionally so
  in the case of the azimuth axis tilt terms) on the AEC antenna.
  Pointing model stability was measured to be adequate on the
  VertexRSI antenna.  These measurements suggest that during
  operational use recalibration of each antenna (using perhaps four
  stars around the horizon and one near the zenith) every couple of
  hours would maintain peak pointing performance.

\item The offsetting performance of both antennas sometimes appeared
  to fall just short of the 0.6~arcsecond RMS specification for offsets of
  $2^\circ$ or less, but this is a tentative result and not seen in
  all test runs.  Given that the very stringent antenna specifications
  are a challenge that the measuring techniques struggle to meet, and
  taking into account the fact that the results were achieved without
  the help of the metrology systems, it is perhaps wise to return an
  open verdict on this aspect of the performance of the ALMA prototype
  antennas.

\item Using the accelerometers, the implied requirement of
  0.6~arcsecond \textit{tracking} stability in the offset pointing 
  specification could be checked for sidereal tracking and wind
  shake. The VertexRSI antenna marginally misses this specification
  (0.94~arcsec RMS pointing stability over 15~minutes), while the AEC
  antenna was measured to be more stable at 0.50~arcsec RMS.

\item The wind-induced pointing jitter over timescales up to
  15~minutes at 9~m/s wind was calculated in the antenna error
  budgets to be 0.035 (VertexRSI) and 0.35 (AEC)~arcsec RMS. The
  accelerometers measured these values to be 0.81 (VertexRSI) and 0.45
  (AEC)~arcsec RMS, excluding the effects of subreflector motion. 

\item OTF scanning at 0.5 deg/s lead to an average pointing stability over
  timescales of 1~second amounting to 1.7~arcsec RMS for the VertexRSI
  antenna, and 0.5~arcsec RMS for the AEC antenna. The specification is
  that the position is known at a given time, but the antenna does not
  need to point at the commanded position at that time. Measurements
  confirm that the pointing jitter as seen by the encoders matches
  that as derived from the accelerometers.

\item During interferometric mosaicing with rates of 0.05 deg/s, the
  commanded path is followed within 0.8~arcsec RMS for the VertexRSI
  antenna, and within 0.23~arcsec RMS for the AEC antenna. These are well
  within the specified value of 1~arcsec RMS.

\item The radiometric tests are too few for any firm conclusions to be
  drawn. At the frequency used, the beam width is about 1~arcminute
  FWHM, and the RMS result less than 10\% of this, so there is no
  reason to suppose the underlying pointing and offsetting accuracy is
  poorer than the OPT measurements suggest.

\end{enumerate}

\section{Fast Switching}

\subsection{Optical Pointing Telescope Measurements of Fast Switching
  Performance}
\label{optfs}

OPT offset and encoder data were used to
compare the fast switching capabilities of the VertexRSI and AEC
antennas.  These fast switching measurements are performed by position
switching between two stars that are bright enough to get high SNR
position fits at the highest possible position switching rate (about
0.1~Hz).  The optical pointing offsets as a function of time for 
each slew tell us how quickly the antenna can get on source and how
quickly the pointing settles down after it gets on source.  The
optical pointing offsets include both mechanical antenna pointing and
atmospheric ``seeing'' caused by dry air fluctuations (unlike in the
radio, where the pointing jitter is caused mainly by water vapor
fluctuations).  Reasonable evidence that the encoders accurately
reflect the antenna's mechanical motion is shown in
\cite{Holdaway2004}.  Hence, we expect that the encoder pointing
offsets should also be very useful.

\subsubsection{OPT Fast Switching Measurements}

All OPT fast switching measurements were made on April 20, 2004.  Four
different star pairs were observed with each antenna.  Two star pairs
did not yield valid data for the AEC observations, while all four
pairs of stars yielded good data on the VertexRSI antenna.  Details of
these observations are summarized in Table~\ref{tab:info}.

\begin{table}[h!]
\centering
\caption{OPT Fast Switching Measurements}
\begin{tabular}{|llcccccccc|}
\hline
Run & Antenna & t  & $\Delta t^a_{start}$ & Az  & El  &  $\Delta\theta$  &
$\phi^b$ & SNR$^c$ & $\sigma^d_{atmos}$ \\
& & (UT) & (s) & (deg) & (deg) &  (deg) & (deg) & & (arcsec) \\ \hline
A & VertexRSI  & 6:07:46 & 1294 & -258  &  63  & 0.4 & -25 & 12 &
0.94/1.30/1.38 \\
B & VertexRSI  & 6:30:32 &  998 & -172  &  47  & 1.8 & 86  & 15 &
0.84/1.22/1.45  \\ 
C & VertexRSI  & 6:52:42 & 1459 &   93  &  74  & 1.5 & 88  &  8 &
1.01/1.80/1.78  \\
D & VertexRSI  & 7:18:52 &  680 &  110  &  22  & 1.4 &  5  &  8 &
1.40/2.02/1.44  \\
E & AEC        & 7:43:08 &  552 & -218  &  79  & 0.4 & -2  & 18 &
0.34/0.33/0.97  \\
F & AEC        & 7:52:38 &  553 & -147  &  41  & 1.8 & -71 & 16 &
0.64/0.57/0.89  \\  \hline
\multicolumn{10}{l}{$^a$~Duration of the measurement.} \\
\multicolumn{10}{l}{$^b$~Orientation angle of the vector joining the
  stars (0~degrees being horizontal).} \\
\multicolumn{10}{l}{$^c$~Signal-to-noise ratio cutoff used to
  determine star detection limit.} \\
\multicolumn{10}{l}{$^d$~Atmospheric pointing values are listed as
  Az component, El component, El-Az ratio.} \\
\end{tabular}
\label{tab:info}
\end{table}

\begin{figure}
\centering
\includegraphics[scale=0.30]{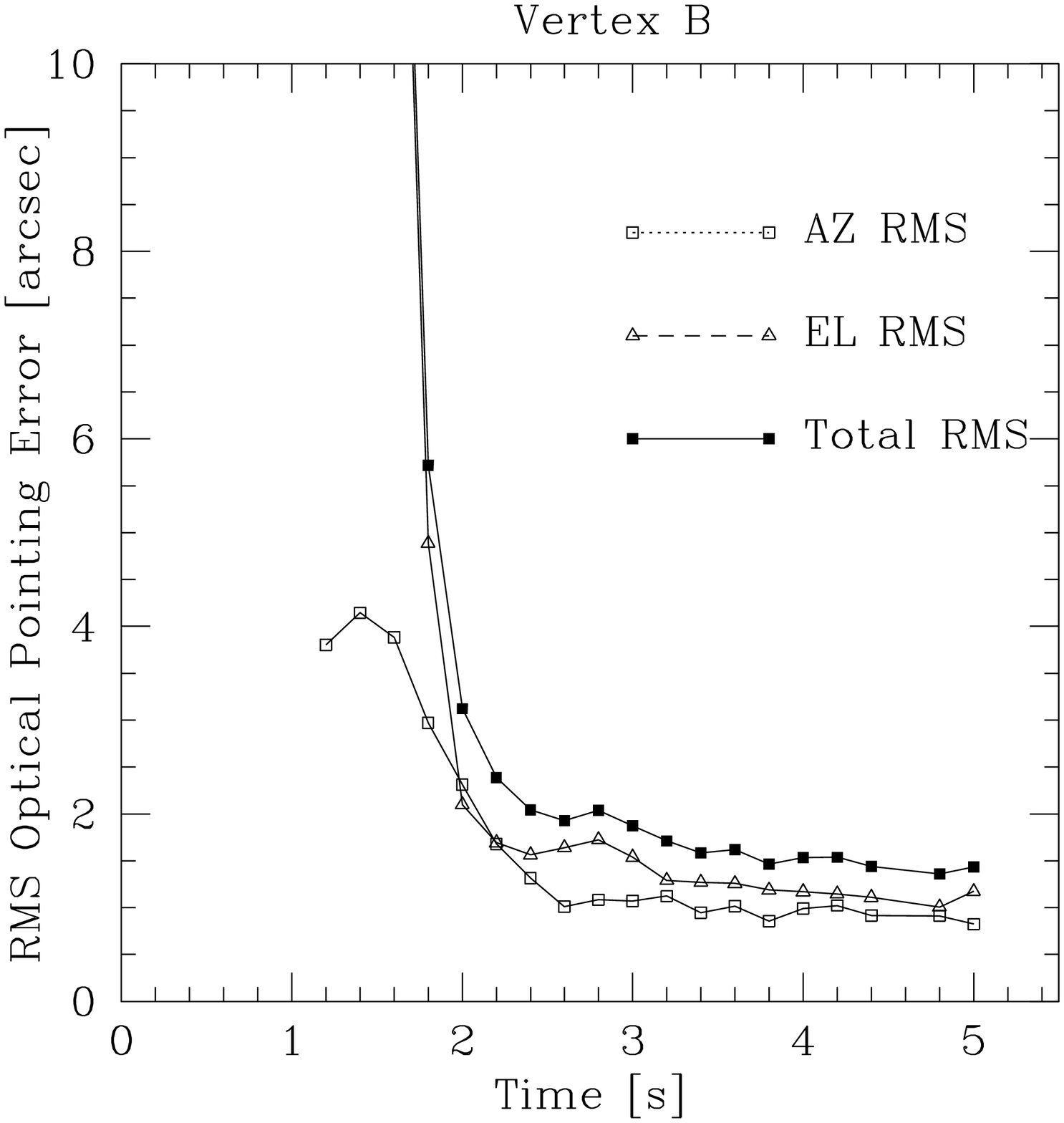} 
\includegraphics[scale=0.30]{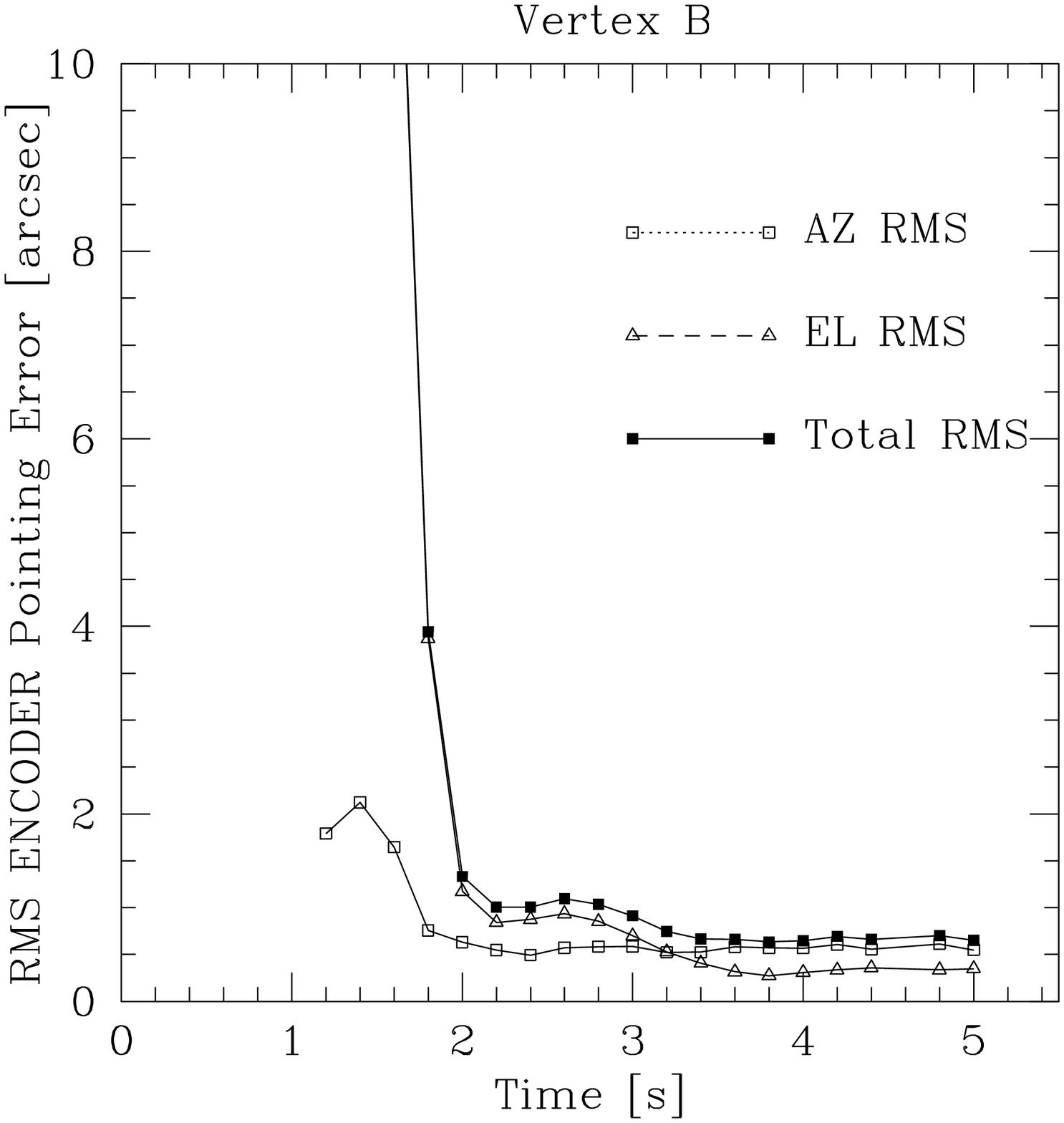} \\
\includegraphics[scale=0.30]{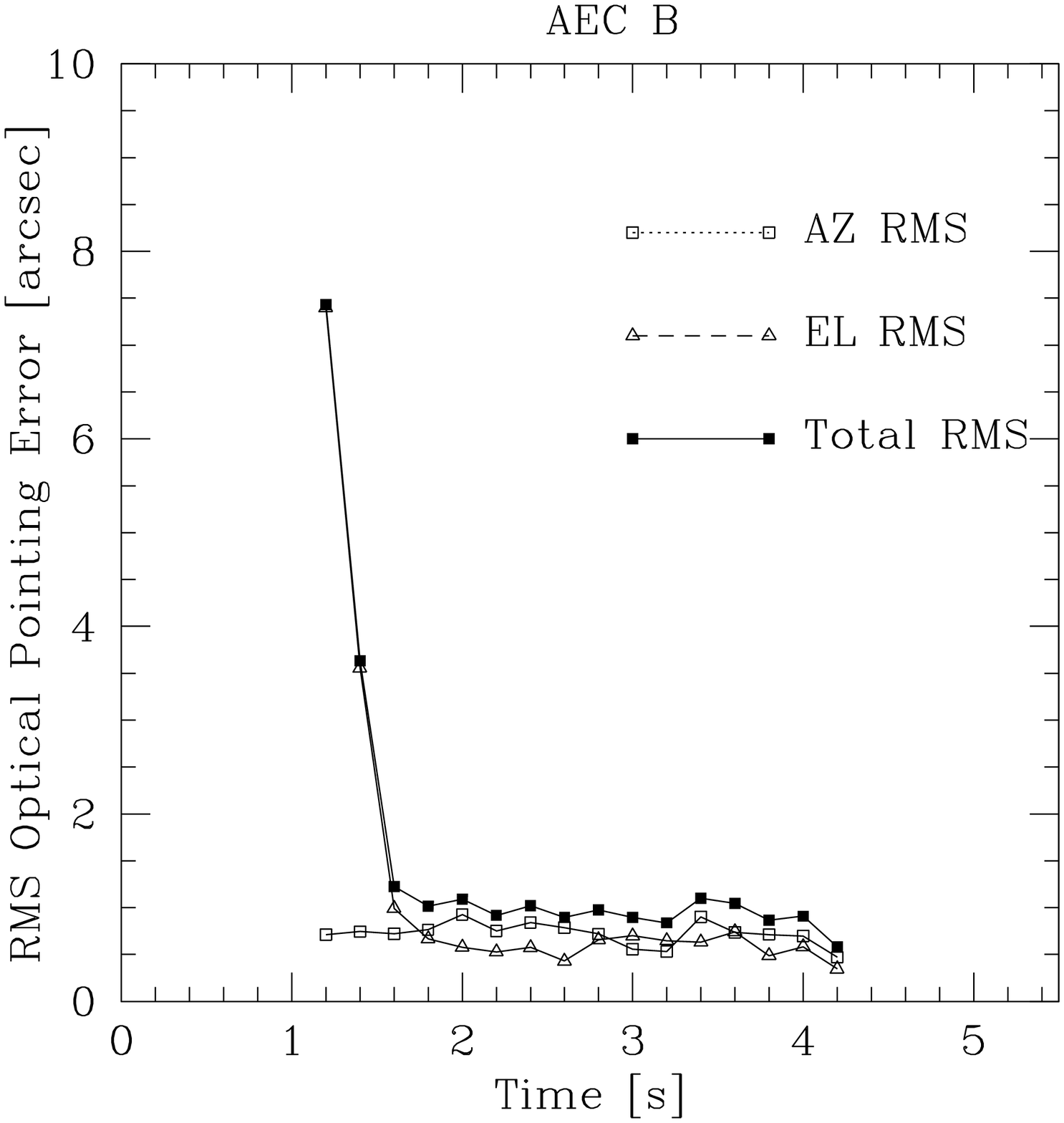}
\includegraphics[scale=0.30]{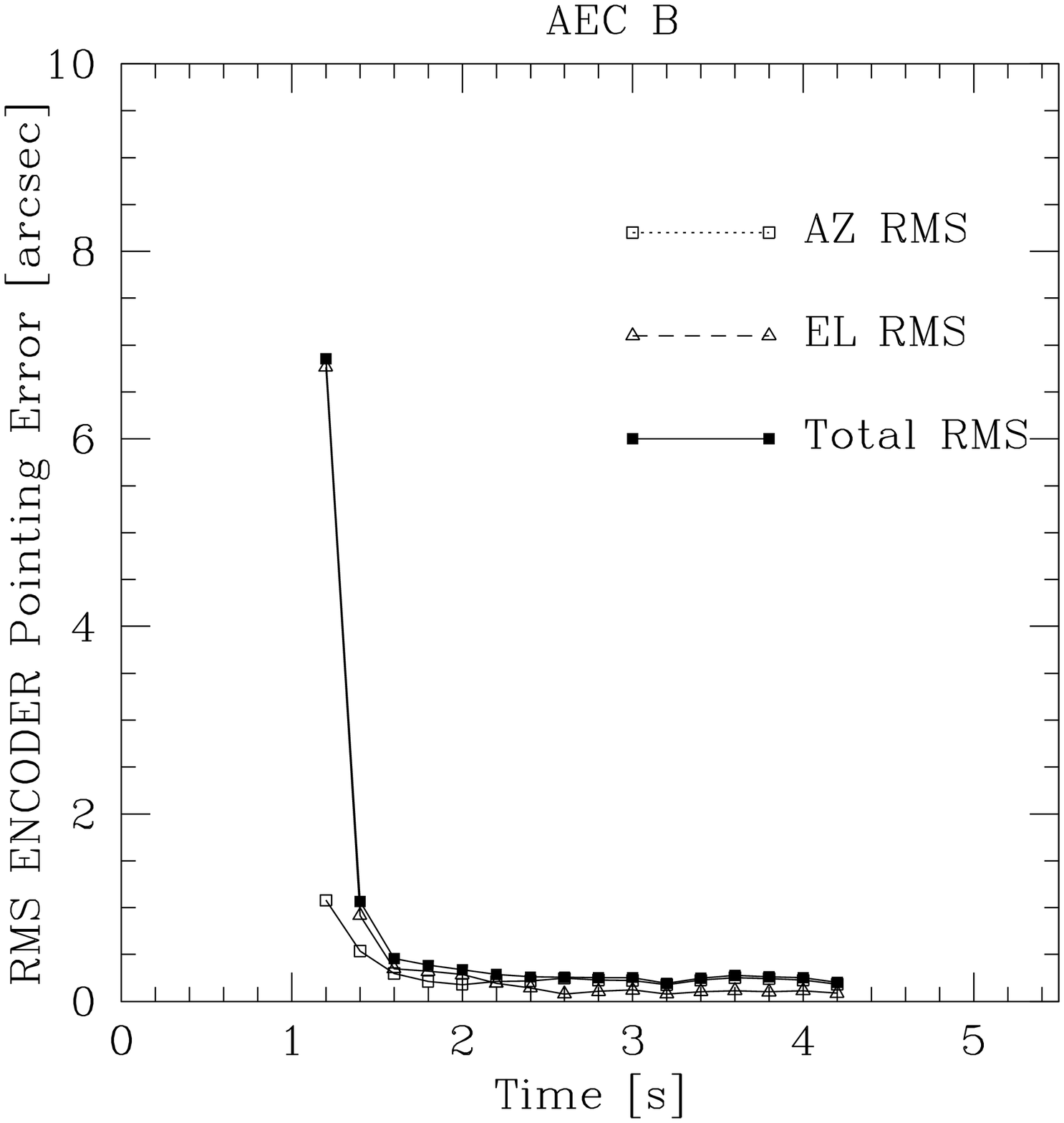}
\caption{RMS optical pointing (left) and encoder (right) offsets as a
  function of time for a typical VertexRSI (top; Run B from
  Table~\ref{tab:info}) and AEC (bottom; Run F from
  Table~\ref{tab:info}) optical fast switching measurement.}
\label{fig:optical}
\end{figure}

For each measurement run, the following analysis was performed:

\begin{enumerate}
\item The 20~Hz sampled encoder Az/El position information was
  linearly interpolated to match the corresponding 10~Hz sampled OPT
  Az/El star position information.
\item Positions from the OPT Az/El data stream were identified as
  ``target'' (on-source) and ``calibrator'' (off-source) positions by
  assigning a star SNR cutoff to the OPT measurements. 
\item Each target-to-target observation consumed about 5 and 4~seconds
  for the VertexRSI and AEC antenna measurements, respectively.  Roughly
  speaking, the first $\sim$1.5~seconds are spent slewing to the
  source, the next second is spent refining the pointing, and the
  remaining time has a nearly stable pointing level.
\item The constant Az/El offset, representing the antenna pointing
  model, is removed from the OPT star position fits to derive the
  Az/El offset for each measurement.
\item For each run a single fourth order polynomial fit to the Az and
  the El encoder readings for the last half of all scans was
  subtracted from the Az and El encoder readings for that run to
  derive the RMS pointing errors.
\item For each slew, we are interested in the RMS pointing error as a
  function of the time since the last measurement from the previous
  slew.  Time bins of width 0.2~seconds (twice the basic data sampling
  rate) representing the time from the end of the previous scan, and
  from the RMS of the Az and El offsets which fall into that time bin
  for all data in a 10-20~minute observation, were constructed.
\item Typically about 2\% of the data points in each run were
  determined to be outliers and removed.
\item Optical RMS Az offsets, RMS El offsets, and the total RMS
  optical and encoder pointing errors as a function of time for each
  run and antenna were then derived.  Figure~\ref{fig:optical} shows
  these results.
\end{enumerate}

Comparing each run Az/El profile (see Figure~\ref{fig:optical}), one
sees general agreement, indicating that these 
two data streams are reliably aligned in time.  However, the
optical pointing RMS generally exceeds the encoder RMS pointing, as
expected, as the optical RMS includes the effects of optical seeing.  
The optical seeing ($\sigma_{atmos}$) will be
uncorrelated with the mechanical pointing errors, so it can be
extracted from the optical pointing contribution by a quadratic
subtraction:

\begin{equation}
  \sigma_{atmos} \simeq \sqrt{ \sigma_{opt}^2 - \sigma_{enc}^2 }.
\end{equation}

\noindent{Table~\ref{tab:info}} shows the inferred atmospheric
pointing contribution for each observation.  In deriving these data, we 
considered optical and encoder data past the 2.5~second point in the
profiles shown in Figure~\ref{fig:optical},
calculated the inferred atmospheric optical pointing for each time
bin, and took the average for all time bins past 2.5~second.  Note that for
the VertexRSI antenna, the ratio of El atmospheric pointing to Az
atmospheric pointing is greater than 1.0, but for the AEC antenna it
is about 1.0.  We suspect that this is a further indication of the
poorer elevation (VertexRSI) and azimuth (AEC) tracking and pointing
performance noted in \S\ref{pointing}.

In conclusion, we find from our OPT measurements of fast switching
performance: 

\begin{itemize}
\item  The VertexRSI antenna typically reaches its target source in
1.5-2.0~s.
\item The VertexRSI antenna can have a significant pointing bounce
in El between 2 and 3~seconds after the slew begins.  Presumably,
the servo is not well-tuned, and this is a correction for overshooting
or something similar.
\item The AEC antenna arrives at the target source after about 
1.5~s, and has a much smaller pointing error after acquiring the
source.  
\item The VertexRSI antenna has larger pointing errors after the
source has been reached than the AEC antenna does, though the encoder
data indicates that the specification of 0.6~arcsec is approximately
met.
\end{itemize}

\subsection{Radiometric Fast Switching Measurements}
\label{radfs}

We simulated radiometric fast switching interferometric observations by
repetition of a blank sky integration, slew, then target source
integration sequence of measurements.  The blank sky integration
simulates a fast switching calibrator measurement.  The detailed
sequence of measurements was as follows:

\begin{enumerate}
\item In order to see pointing fluctuations in both Az and El, we have
  performed observations with Jupiter at both its Az and El half power
  points (Table~\ref{tab:radfsobs}), and with a sequence of 21
  antenna slews to the nearby ``calibrator'' (\ie\ blank sky) of
  1.5~deg in both Az and in El. 
\item These total power measurements utilized beam switching to
  eliminate variable atmospheric emission with a subreflector throw of
  about 80~arcsec nutating at 10~Hz.  The resulting differential total
  power measurements were found to be limited by gain fluctuations in our
  radiometric receiver system rather than incorrect cancellation of
  the atmospheric emission or thermal noise.
\end{enumerate}

\begin{table}[h!]
\centering
\caption{Radiometric Fast Switching Measurements}
\begin{tabular}{|lccccc|} \hline
Scan & Antenna & HPBW Offset  &  Switch Dir & El &
TP$_{on}$ \\ 
&&& (deg) & (deg) & (ampl) \\ \hline %& (ampl) \\ \hline
5639 & VertexRSI & El & $+$1.5 El &  64.6 & 4.89 \\ %& 0.034 \\
5640 & VertexRSI & Az & $+$1.5 Az &  64.5 & 3.08 \\ %& 0.033  \\
5641 & VertexRSI & Az & $-$1.5 Az &  63.5 & 3.05 \\ %& 0.035 \\
5642 & VertexRSI & El & $-$1.5 Az &  62.2 & 4.70 \\ %& 0.036 \\
5643 & VertexRSI & Az & $+$1.5 El &  61.0 & 3.09 \\ %& 0.036 \\
6759 & AEC & Az & $+$1.5 Az &  68.8 & 3.08 \\ %& ??  \\
6762 & AEC & El & $-$1.5 El &  69.4 & 3.13 \\ %& ?? \\
6767 & AEC & Az & $-$1.5 El &  70.0 & 3.08 \\ %& ?? \\
6768 & AEC & El & $+$1.5 Az &  70.2 & 3.03 \\ \hline %& ?? \\  \hline
\end{tabular}
\label{tab:radfsobs}
\end{table}

In summary, the analysis of these measurements proceeded as follows:

\begin{enumerate}
\item To calibrate the conversion from total power level to source
  offset we collected a series of three beam-switched total power
  measurements at offsets of $+35$, $+40$, and $+45$~arcsec from the
  half-power point of Jupiter.  These data indicate the total power
  should change by about 0.160 for every arcsecond of pointing offset
  in the direction of the offset. In the following we adopt this value
  for the change in total power versus position.  The true scaling
  factor will be very close to this for the El scans, but the Az scans
  could have a scaling factor significantly different from this,
  \ie\ as low as 0.11, which will increase the radiometrically
  inferred Az errors by as much as 50\% due to uncertainties in the
  interpretation of the unmeasured peak flux for Jupiter.  Note that
  the RMS of the 0.1~second data samples when OFF of Jupiter were typically
  0.033.  At 0.160 per arcsec, this indicates that the radiometer
  noise is equivalent to a jitter of 0.21~arcsec. Averaging for longer
  times can give us estimates of pointing errors with smaller
  uncertainties.

\item For each 0.1~second time stamp since the start of a slew (there
  are 20 slewing profiles going toward Jupiter per scan) we calculate
  the RMS of the measured total power minus the mean total power
  calculated from the last half of all 20 blocks in the scan, and
  divide by our calibration factor of 0.16 amplitude/arcsec.  All 20
  slew profiles within a given scan are then aligned and averaged to
  derive the radiometrically-inferred pointing error as a function of
  time, all shifted so they align with the slew that got on source
  fastest.  A representative sample of these results is shown in
  Figure~\ref{fig:radfs}.

\begin{figure}[h!]
\centering
\includegraphics[scale=0.30]{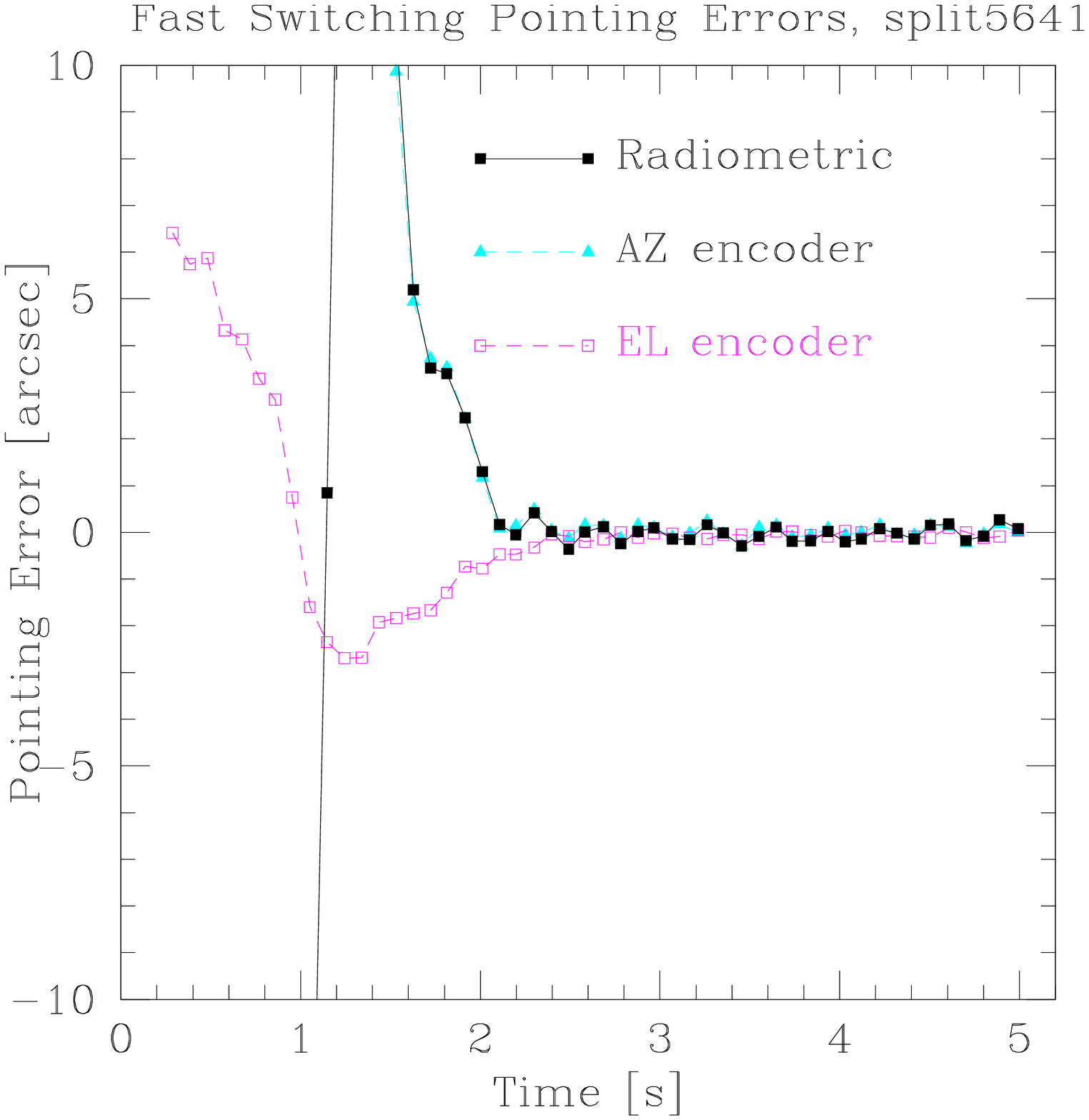}
\includegraphics[scale=0.30]{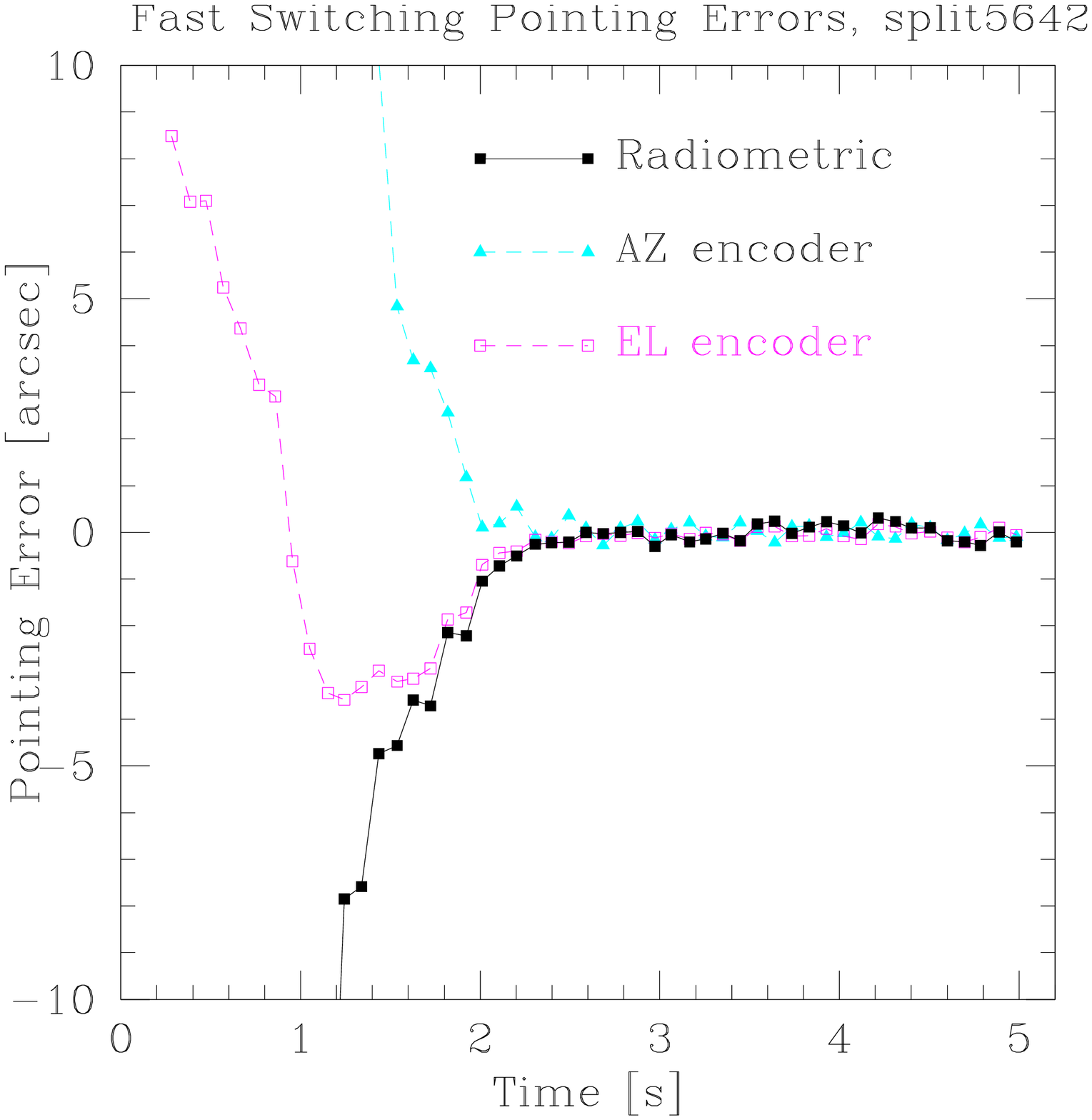} \\
\includegraphics[scale=0.30]{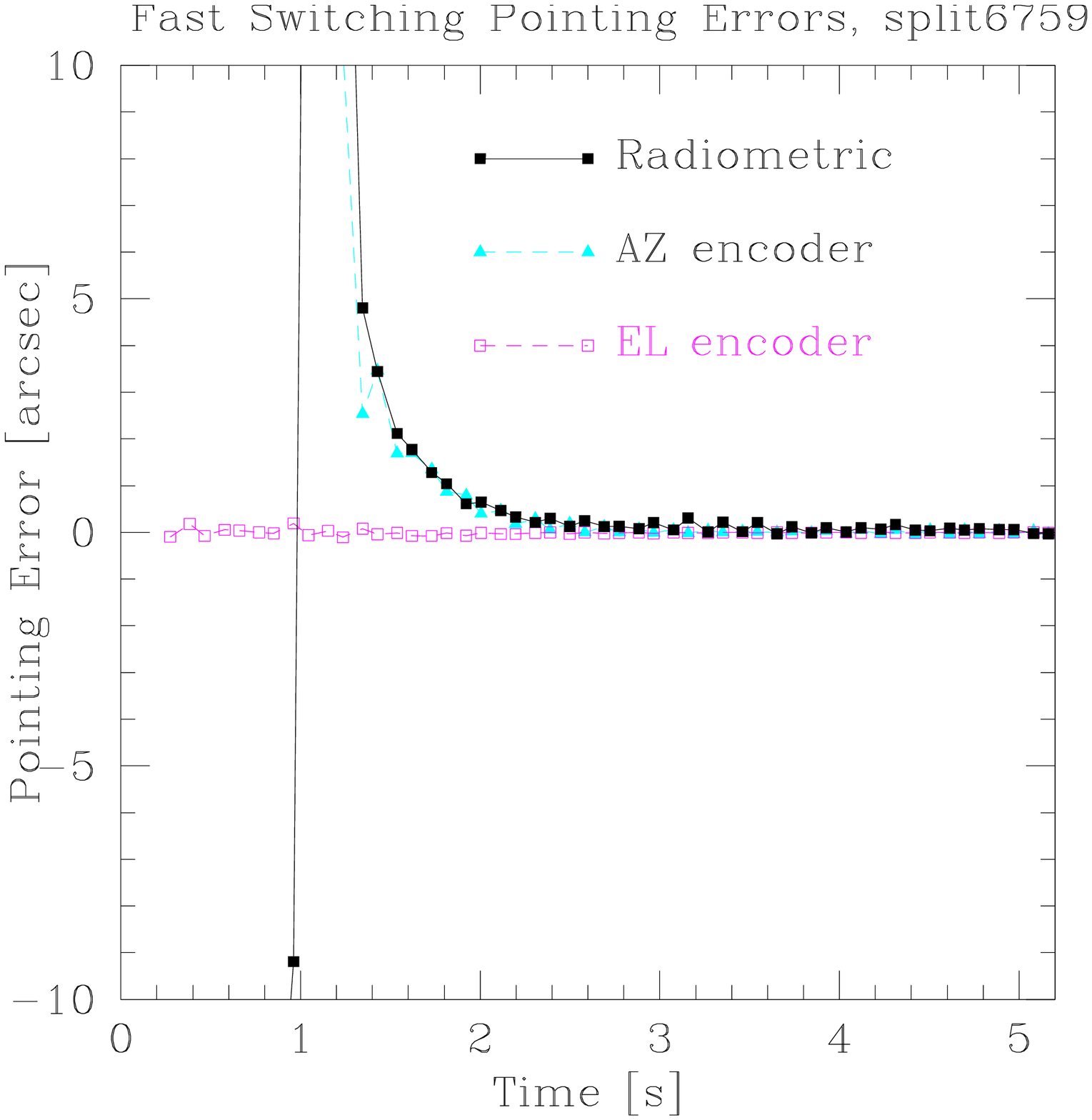}
\includegraphics[scale=0.30]{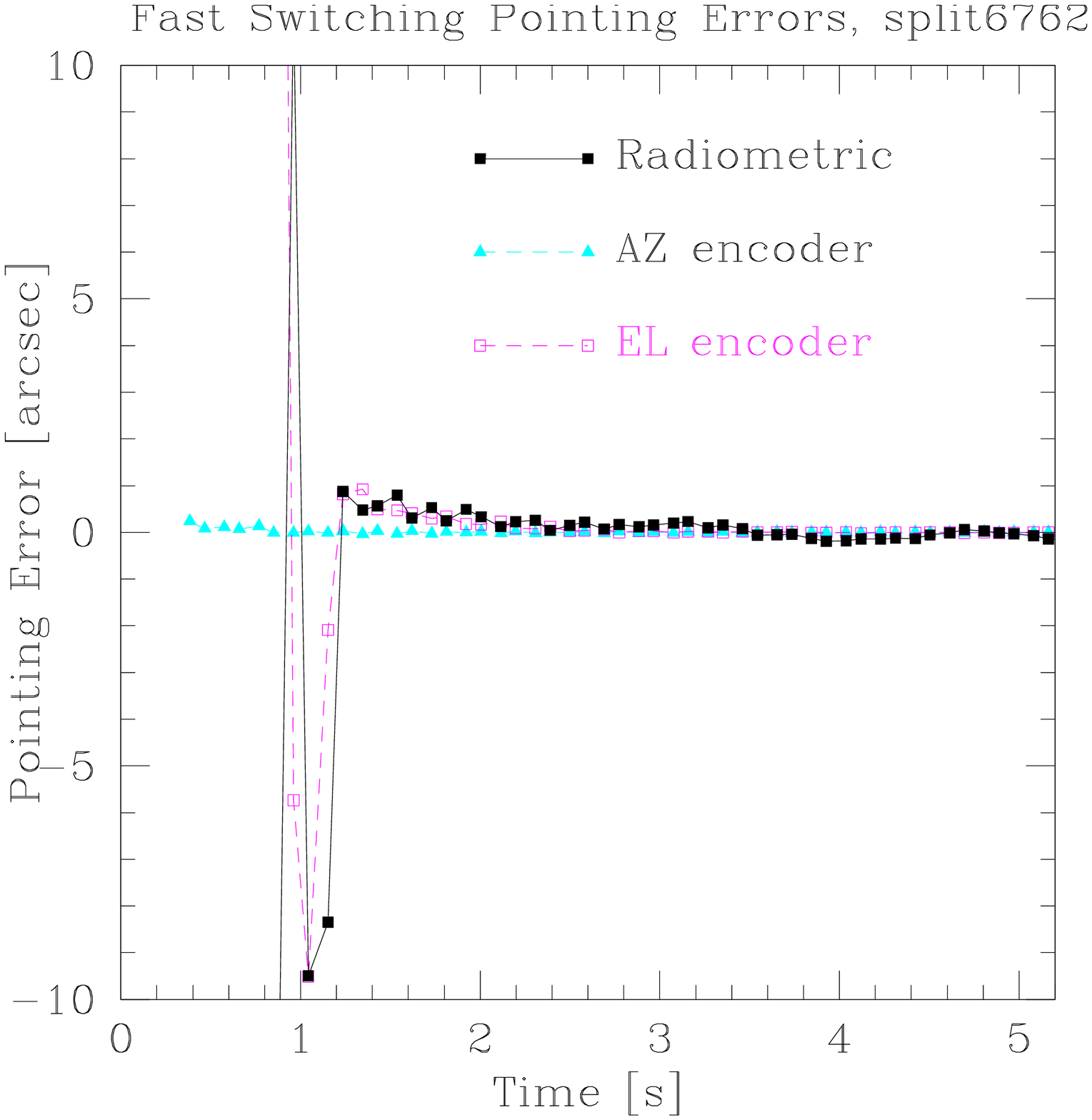}
\caption{Radiometrically-inferred VertexRSI (top) and AEC (bottom) AZ
  (left) and EL (right) pointing errors for an AZ- and El-slew,
  respectively, along with encoder errors. The mean 
  radiometrically-inferred (solid black) and mean azimuth (cyan
  dashed) and elevation (purple dashed) encoder pointing errors as a
  function of time since the start of the 1.5~deg fast switching slew
  are shown.  Note that the noise level in these measurements is the
  time-weighted radiometric system noise of $\sim$
  0.21~arcsec/$\sqrt{20}$ = 0.05~arcsec.}
\label{fig:radfs}
\end{figure}

\item By averaging the approximately 500 samples made during the
  last-half of each scan a measure of the radiometric tracking
  accuracy, which includes contributions from the antenna and
  atmospheric fluctuations, was derived for the VertexRSI antenna.
  Typical values of 0.6~arcsec per axis, or about 0.8~arcsec total,
  were measured.

\item In the AEC antenna measurements we note that there is a 5~Hz
  oscillation in the positioning.  This is a real mechanical effect
  because:
  \begin{itemize}
  \item It has too much power at too small a time scale to fit with
    the atmospheric pointing error structure function,
  \item It is seen in the encoder data (offset by 0.1~s), 
  \item It is not seen in the tracking data,
  \item It seems to get excited as the telescope stops, and dies down
    with time (\ie\ the telescope is settling down).
  \end{itemize}
  The peak to peak of this positioning oscillation is about
  0.4~arcsec. This oscillation has been identified as due to a
  mechanical oscillation of the apex structure (\citet{Snel2006}).
\end{enumerate}

\subsubsection{Radiometric Fast Switching Performance Conclusions}

Using radiometric and encoder data, we have investigated the
mechanical aspects of the fast switching performance of the ALMA
prototype antennas.

\begin{itemize}

\item The VertexRSI antenna basically meets the fast switching
  specification of ``1.5~degrees in 1.5~seconds to an accuracy of
  3~arcseconds''.  The antenna achieves pointing errors of about
  0.8~arcsec RMS if a bit more settle time is permitted.  Minor
  improvements in the software and servo systems could improve the
  fast switching capability.  We find that for night time observing,
  the encoders and the radiometrically-inferred pointing errors agree
  remarkably well. After the slew has been completed and the antenna
  is pointing on source, pointing drifts of under 1~arcsec over
  2~seconds of time, and pointing jitter of amplitude 0.3~arcsec and
  frequency 3~Hz can be seen clearly in both the encoder pointing and
  the radiometrically-inferred pointing.  On average, the encoders and
  radiometrically-determined pointing differs by 0.28~arcsec RMS while
  slewing and 0.15~arcsec after settling.

\item Radiometric and encoder data from the AEC prototype antenna
  indicate that the AEC antenna meets the fast switching motion
  requirements.  We see that as the AEC antenna comes to rest after a
  fast slew, a 5~Hz oscillation is excited by the fast switching
  motion, and that the peak amplitude of this oscillation is about
  0.4~arcsec, which dies out over a few seconds.  Accelerometer
  measurements have identified this oscillation as being due to a
  rotation of the apex structure (\citet{Snel2006}).  For the
  measurements performed in this report, this oscillation does not
  prevent the AEC antenna from meeting the fast switching
  specification.

\end{itemize}

\subsection{Encoder Measurements of Fast Motion and Settling Time}
\label{fsaccel}

From the accelerometer measurements with comparison to the antenna
encoder position information we gain an insight into the settling time
after the fast position switch. We find that the VertexRSI antenna can
perform a fast switch in the prescribed time and positioning limits,
but not for all directions. There are no indications that the AEC
antenna does not meet the fast motion and settling time
specifications.

\section{Path Length Stability}

The path length stability of the antenna structures
was measured using an Automated Precision Incorporated (API) 5-D laser
interferometer (\citet{Greve2006}) and an accelerometer system
(\citet{Snel2006}).  In the following we describe the path length
stability measurement results for both prototype antennas.  Detailed
descriptions of the technical capabilities of each measurement system
are given in the associated references.

\subsection{API5D Laser Interferometer Measurements}
\label{apipath}

An Automated Precision Incorporated 5D (API5D) laser interferometer
was used to measure the individual structural components that compose
the total interferometric path length through the antennas.  Laser
interferometers are used to measure variations in ``straightness''
along a given path.  One component of this ``straightness'' is the
path length variation ($\Delta$\,z).  For the measurements made of the
ALMA prototype antennas the accuracy in ($\Delta$\,z) is better than
1\,$\mu$m for an enclosed path, and 2 to 3\,$\mu$m for an
open--air path.

\begin{figure}
\centering
\includegraphics[scale=0.50]{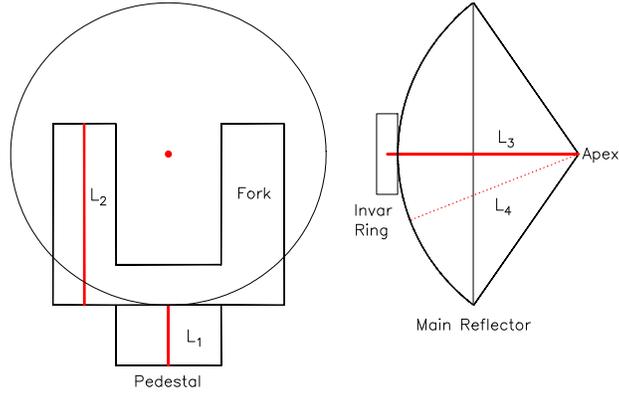}
\caption{Illustration of the path length measurement scheme.  Red
  lines indicate the path length sections measured with the API laser
  interferometer system.}
\label{fig:pathparts}
\end{figure}

Since the total path length variation cannot be made with a single
measurement, the total path length was separated into four parts
(pedestal, fork arm, central receiver flange to subreflector, and
subreflector to surface) and measured during a representative sample
of observing conditions and operation modes. As shown in 
Figure~\ref{fig:pathparts}, these measurements contain

\begin{description}
\item[L1:] path length inside the pedestal =  distance between the foundation 
 (ground) and the lower face of the azimuth\,(AZ)--bearing; it was not possible
 to measure this path length on the AEC antenna;
\item[L2:] path length inside (one) fork arm (left,\,right) = distance between
 the lower part of the traverse and the upper part of the left (right) fork 
 arm, below the elevation (EL)-- bearing; 
\item[L3:] path length along the radio axis = distance between the upper part 
 of the Invar cone (VertexRSI), or the vertex hole (AEC, laser tracker mount) 
 of the main reflector, and the apex of the subreflector; (in the measurements
 the dilatation of the aluminum subreflector support tube (50\,cm length) has 
 been eliminated, as far as possible; 
\item[L4:] path length of the reflector = distance between the Invar 
 cone/vertex hole and a point of the reflector surface, measured via the 
 subreflector. On both antennas this path length measurement was tested but 
 not routinely performed, mainly because of difficult alignment and time 
 limitations. Since, in essence, this path length component originates
 in the CFRP part of the antennas, at least a temperature induced path
 length variation is expected to be small.
\end{description}

\begin{figure}
\centering
\includegraphics[scale=0.5]{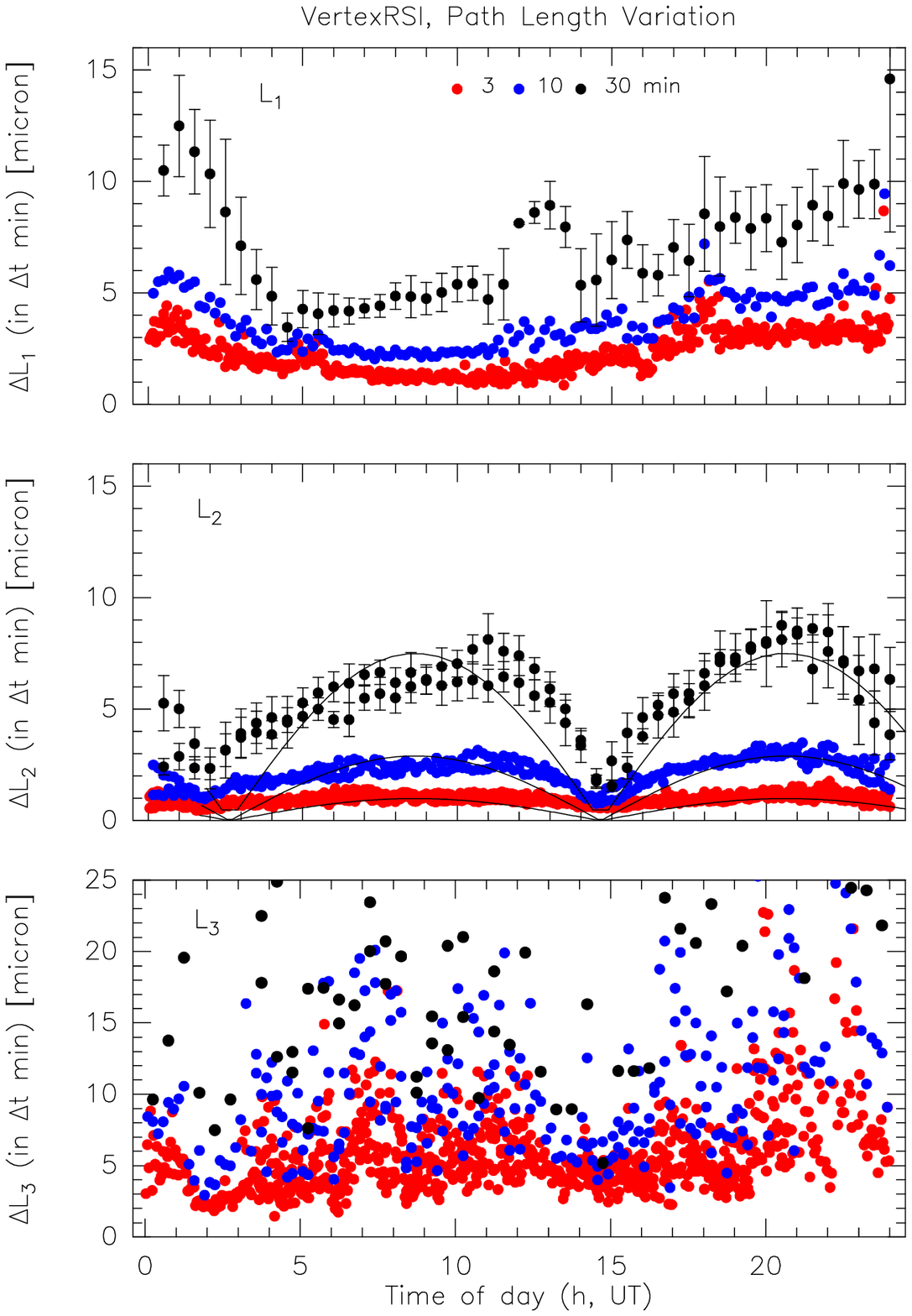}
\includegraphics[scale=0.5]{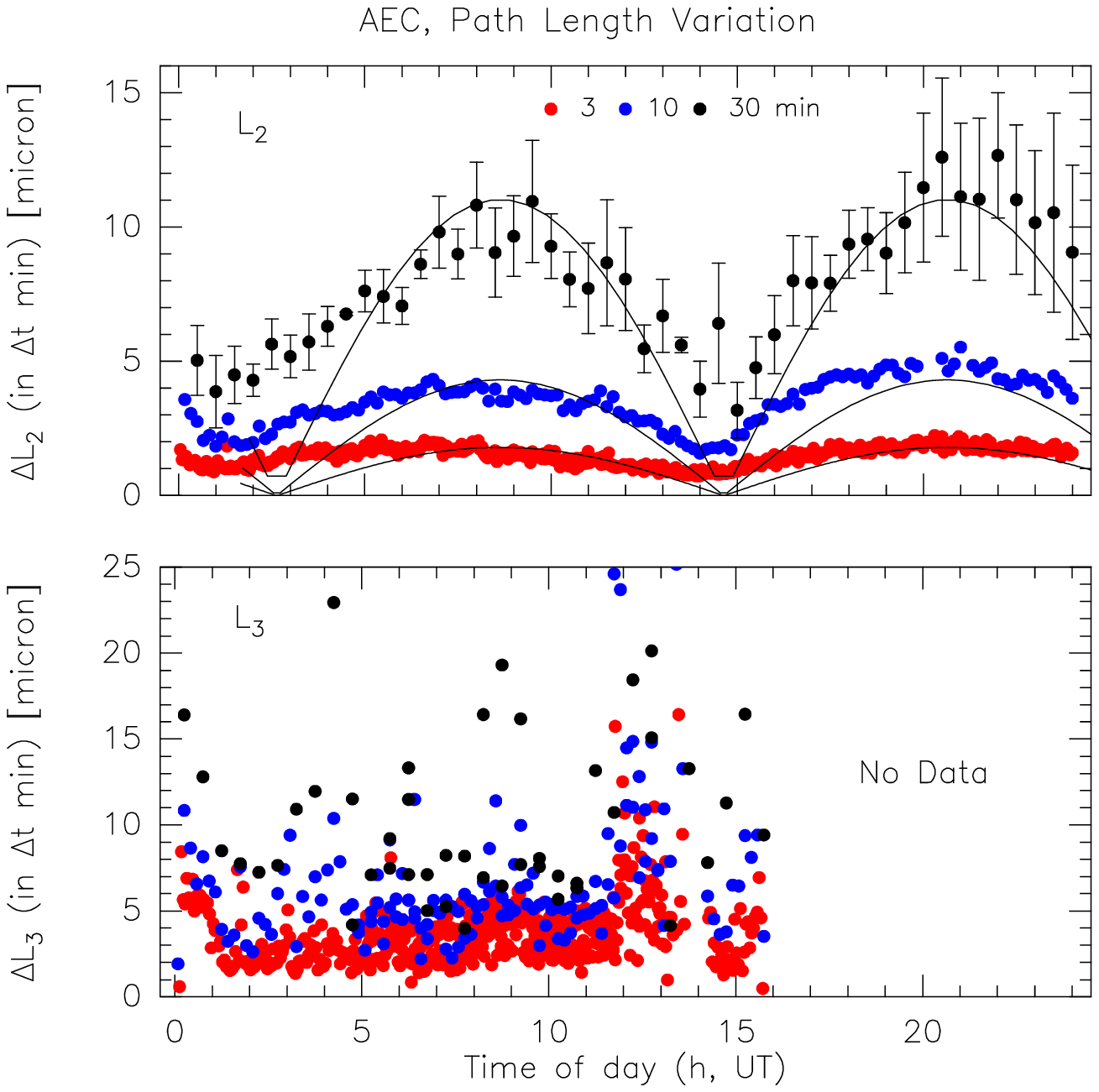}
\caption{VertexRSI (left) and AEC (right) antenna path length
  summaries.  Daily variation of path length $L_1$ (pedestal), of path
  length $L_2$ (fork arm), and of path length $L_3$ (Invar cone to
  subreflector), as function of the time of the day (UT), and 
  within time intervals of 3~minutes, 10~minutes, and 30~minutes
  duration. Note the difference in scale.  The black lines in the
  $\Delta L_2$ panel are sine-function fits to the temporal variation
  in $L_2$, which track the daily ambient temperature variations at
  the VLA site (see \cite{Greve2006} for further information).
}
\label{fig:path}
\end{figure}

\noindent{The} full path length variation is approximately the sum of
the measured components.  Table~\ref{tab:plsummary} lists, while
Figure~\ref{fig:path} shows, the results from these path length
stability measurements for both ALMA prototype antennas.

\begin{table}
\small
\centering
\caption{Path Length Variations Derived From API5D Measurements}
\begin{tabular}{lccccccccc}
\hline\hline
& \multicolumn{4}{c}{VertexRSI} && \multicolumn{4}{c}{AEC} \\
\cline{2-5}
\cline{7-10}
& 3 min & 10 min & 30 min & Tot. Time && 3 min & 10 min & 30 min &
Tot. Time \\  
Path & ($\mu$m) & ($\mu$m) & ($\mu$m) & (hr) && ($\mu$m) & ($\mu$m) &
($\mu$m) & (hr) \\ 
\hline
L$_{1}$ (pedestal) & 3   & 6 & 10 & 250 && $\sim$5 & $\sim$5 & $\sim$5 & Est. \\
L$_{2}$ (fork arm) & 1.5 & 3 & 8  & 360 && 3 & 4 & 10 & 360 \\
L$_{3}$ (quadripod) & 5 & 5 & 5 & 25 && 4 & 5 & $\sim$10 & 25 \\
L$_{4}$ (reflector) & $\sim$5 & $\sim$7 & $\sim$9 & Est. &&
$\sim$5 & $\sim$5 & $\sim$5 & Est. \\
$\Delta$\,z & 15 & 21 & 32 & 635 && 15 & 18 & 30 & 385 \\
\hline
\multicolumn{10}{l}{Note: Each measurement listed as averages over 3,
  10, and 30~minute durations, along with} \\
\multicolumn{10}{l}{~~~the total measurement (or estimate) time, for
  each path.} \\
\end{tabular}
\label{tab:plsummary}
\end{table}

From simple geometrical and material arguments (thermal expansion 
coefficients), and the fact that wind forces act on short time scales, it was 
evident from the beginning that the path length changes are, primarily,
due to thermal dilatation of the antenna components, induced by variation of
the ambient air temperature and solar radiation, buffered by the surface 
finish (paint) and thermal insulation. In the analysis of path length 
variations we have selected time intervals ($\Delta$\,t) of 3~minutes, 
comparable to the time scale of wind, of 10~minutes, comparable to FSW
(fast switch) and OTF modes of observation, and of 30~minutes,
comparable to the time between upgrades of the pointing and
interferometer phases.  The data shown in Figure~\ref{fig:path} were
obtained during long time periods (listed in Table
\ref{tab:plsummary}), and cover a large variety of antenna
motions. For path lengths that could not be measured an estimated
value is entered in Table \ref{tab:plsummary}. 

In the following we investigate the path length dependence of both
prototype antennas under a variety of the individual load conditions
that went into the overall performance quoted above.

\subsubsection{Path Length Variations Influenced by Temperature and Wind}
\label{pathtempwind}

\begin{figure}
\centering
\includegraphics[scale=0.8,angle=-90]{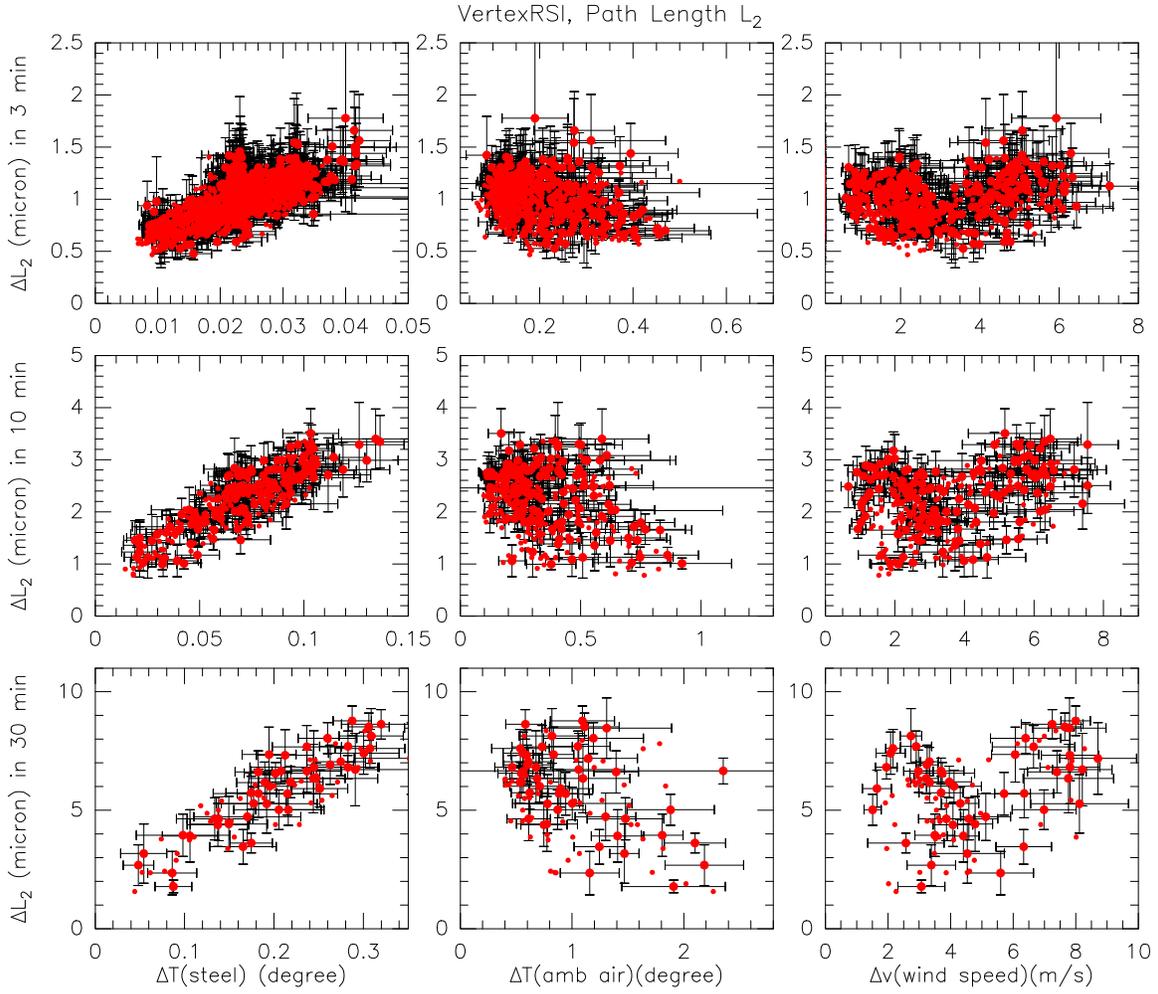}
\caption{VertexRSI antenna dependence of $\Delta$L$_{2}$ on the
  temperature variation of the fork [$\Delta$T\,(steel); left] and
  ambient air [$\Delta$T\,(amb\,air); center] and the variation of the
  wind speed [$\Delta$\,v; right] for time intervals of 3, 10, and 30
  minutes (top to bottom).}  
\label{fig:vertex-corr-t-l2}
\end{figure}

\begin{figure}
\centering
\includegraphics[scale=0.7,angle=-90]{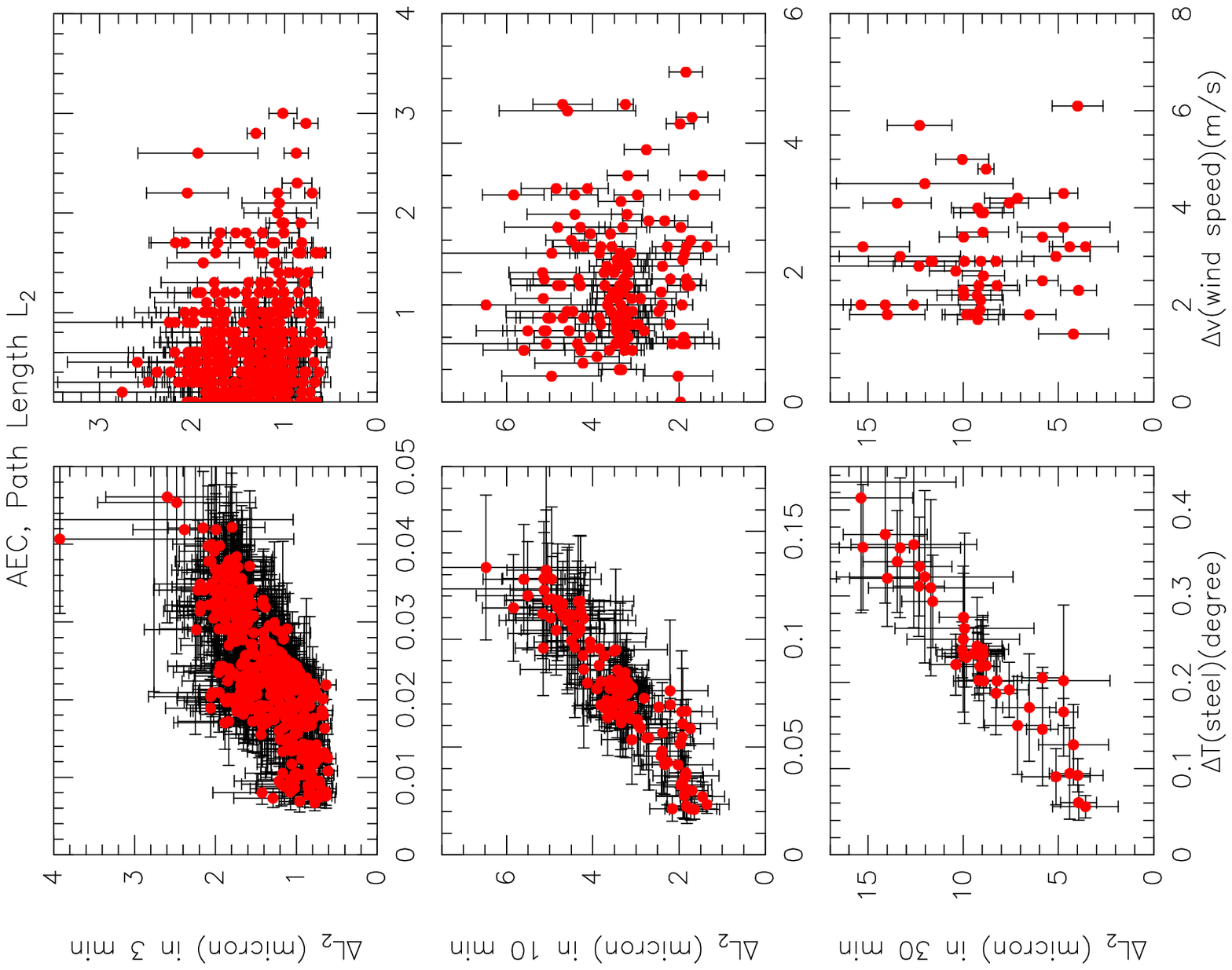}
\caption{Same as Figure~\ref{fig:vertex-corr-t-l2}, but for the AEC
  antenna.  Measurements of the ambient air temperature were not
  available.}
\label{fig:aec-corr-t-l2}
\end{figure}

For purposes of path length prediction, we have searched for
correlation of the path length changes with the steel temperature of
the pedestal and the fork arms, the ambient air temperature, and the
wind speed. The correlation of the fork arm path length change
$\Delta$L$_{2}$ is shown for the VertexRSI antenna in
Figure~\ref{fig:vertex-corr-t-l2}, and for the AEC antenna in
Figure~\ref{fig:aec-corr-t-l2}. As expected, the correlation of the
path length change $\Delta$\,L$_{2}$ with the change of the fork arm
steel temperature is good, and usable for prediction. On the VertexRSI
antenna, a similarly useful correlation was found for the path length
of the pedestal (L$_{1}$; \cite{Greve2006}).  There is
no correlation between path length variation and ambient air
temperature or wind speed variation.

A dedicated investigation of the path length changes with variations
in wind speed was difficult due to logistical reasons. For one
particular day, with changes of the wind speed (v) from $\sim$\,5\,m/s
to $\sim$\,15\,m/s, the path length variation $\Delta$L$_{2}$ (fork arm)
of the VertexRSI antenna was measured, which hinted at a slight
increase in 
$\Delta$L$_{2}$ with increasing wind speed. However, the measured
variation of L$_{2}$ remains within the specification even for the
largest observational wind speed of v = 9\,m/s. There are no data for 
the AEC antenna.

\subsubsection{Path Length Predictions from Steel Temperature Measurements}
\label{pathtemppredict}

\begin{figure}
\centering
\includegraphics[scale=0.8]{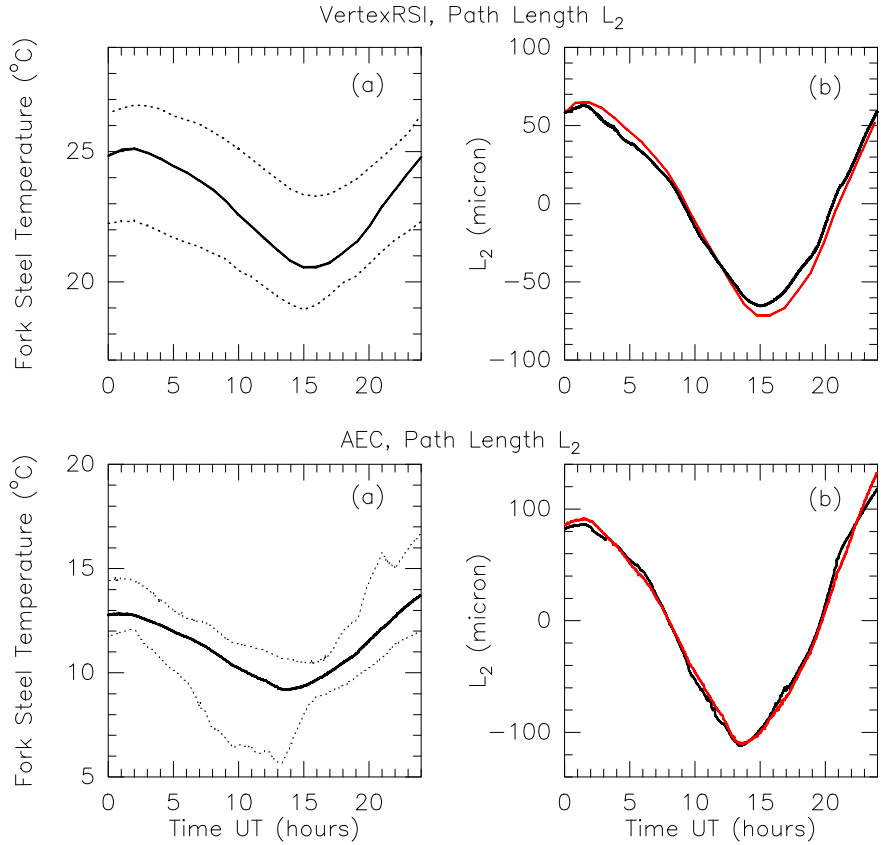}
\caption{Correlation between the path length variation $\Delta$L$_{2}$
  measured with the API laser interferometer and FEM--calculated path
  length derived from the measured fork arm temperatures (14
  sensors) for both prototype antennas. (a) Average (solid line) and
  maximum and minimum temperatures (dashed lines) of the fork arm
  steel derived from the 14 sensors.  (b) Measured path length
  variation (black line), and that calculated from the FEM using the
  average temperature (red line). Similar results were obtained for
  other days.}
\label{fig:path-fem}
\end{figure}

Both prototype antennas have temperature sensors installed on the
steel walls inside the fork arms.  From these recordings we have
derived the average, maximum, and minimum temperature of each fork
arm. The average temperature distribution measured throughout a fork
arm was used in a finite element model (FEM)
analysis to calculate the corresponding thermal
dilatation $\Delta$\,L$_{2}$. Comparing the calculated thermal
dilatation with the path length variation measured directly with the
API5D, we find very good agreement as shown in
Figure~\ref{fig:path-fem}. From this we conclude that the path length 
variations of the steel parts can be predicted, to a high degree of
accuracy, from representative temperature measurements used in the
finite element model, or an empirical relation.

The daily temperature variation of the fork arms and the corresponding path 
length variation $\Delta$L$_{2}$, both shown in Figure~\ref{fig:path-fem}, 
can be approximated with good accuracy by sine--functions. This holds also 
for the ambient air temperature, at least as measured at the VLA site during 
most of the time of the tests. When adopting a sine--function L$_{2}$(t) = 
L$_{\rm o}$ + $\Delta$L$_2\sin(\omega\,t)$, with $\omega$ = 2\,$\pi$/24\,h and
t = time, the path length variation $\Delta$\,L$_{2}$ at a 3,\,10,\,and 30 
minute time interval can be derived by differentiation of the function 
L$_{2}$(t). The corresponding result is shown by solid lines in
Figure~\ref{fig:path-fem}, which gives an explanation of the double peaked
form of the measured variations.  The minima of $\Delta$\,L$_{2}$
occur around 0$^h$ and 14$^h$ UT where the temperature changes of the
ambient air and of the steel of the fork are smallest.

As is evident from Figure~\ref{fig:path-fem}, the total daily path length
variation of the fork arms is of the order of 100\,$\mu$m to
200\,$\mu$m, as fully understandable, and unavoidable, from the height
of the fork arms, the thermal properties of steel, the actual
temperature variation of the steel, and the solar illumination.

\subsubsection{Path Length Variations During Antenna Motion}
\label{pathtmotion}

The ALMA interferometer will use sidereal tracking, OTF mapping, and 
FSW motions between source and calibrator. The OTF, and in particular the FSW 
motions, involve high accelerations of the antenna which may affect the path 
length stability. The path length variations $\Delta$L$_{1}$, $\Delta$L$_{2}$ 
and $\Delta$L$_{3}$ were measured under the following motions of the antennas:
(1) sidereal tracking, as a combined motion in AZ and EL direction; (2) OTF 
motion of 1$^{\rm o}$\,AZ by 1$^{\rm o}$\,EL, at 0.05$^{\rm o}$/s); and (3) 
FSW motion of 1$^{\rm o}$\,AZ by 1$^{\rm o}$\,EL, at 6$^{\rm o}$/s in AZ and 
3$^{\rm o}$/s in EL.

\begin{figure}[h!]
\centering
\includegraphics[scale=1.0]{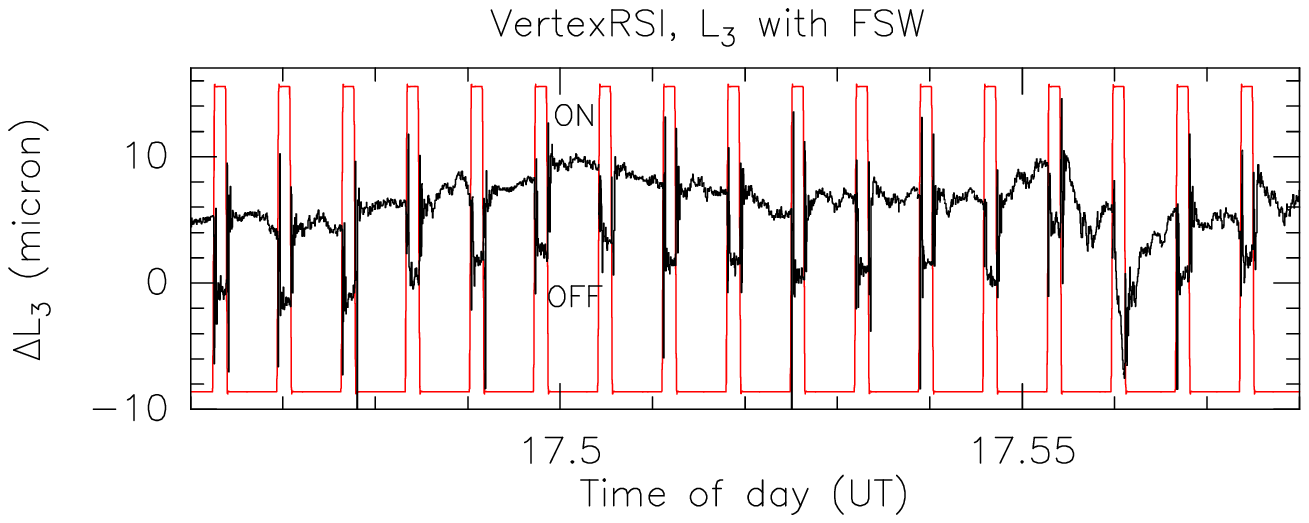}
\includegraphics[scale=1.0]{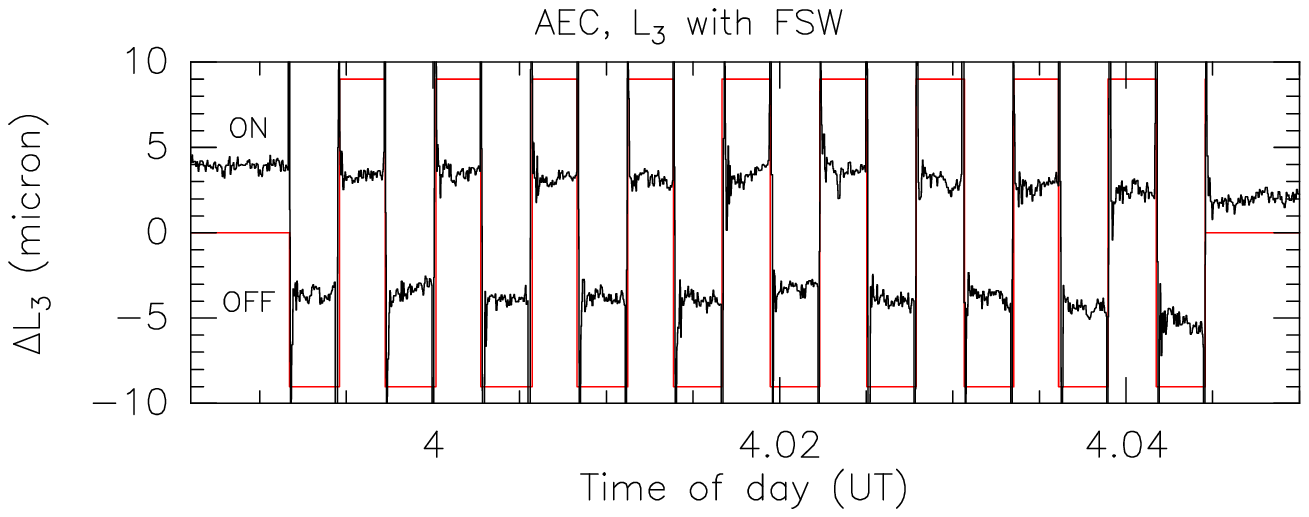}
\caption{VertexRSI (top) and AEC (bottom) antenna path length
  variation $\Delta$L$_{3}$ (Invar cone (VertexRSI) or reflector
  vertex hole (AEC) to subreflector) during fast switching motion.
  The switching cycle is shown by the red line, the path length
  changes by the black line. The ON (10~seconds for both antennas) and
  OFF (2~seconds for VertexRSI, 10~seconds for AEC) positions are 
  indicated.}
\label{fig:path-motion-l3}
\end{figure}

For both antennas, the variations of the path lengths L$_{1}$, L$_{2}$, and
L$_{3}$ are within $\pm$\,2\,$\mu$m for sidereal tracking and OTF motion.
For FSW motion, with the highest acceleration at the subreflector position
of the quadripod, the path length measurements L$_{3}$ are shown in
Figure~\ref{fig:path-motion-l3}. Again, for both antennas the path length 
variation $\Delta$L$_{3}$ at the ON and OFF position is within 
$\pm$\,3\,$\mu$m.

\subsubsection{Influence of Gravity on $L_3$}
\label{l3grav}

\begin{figure}
\centering
\includegraphics[scale=0.9]{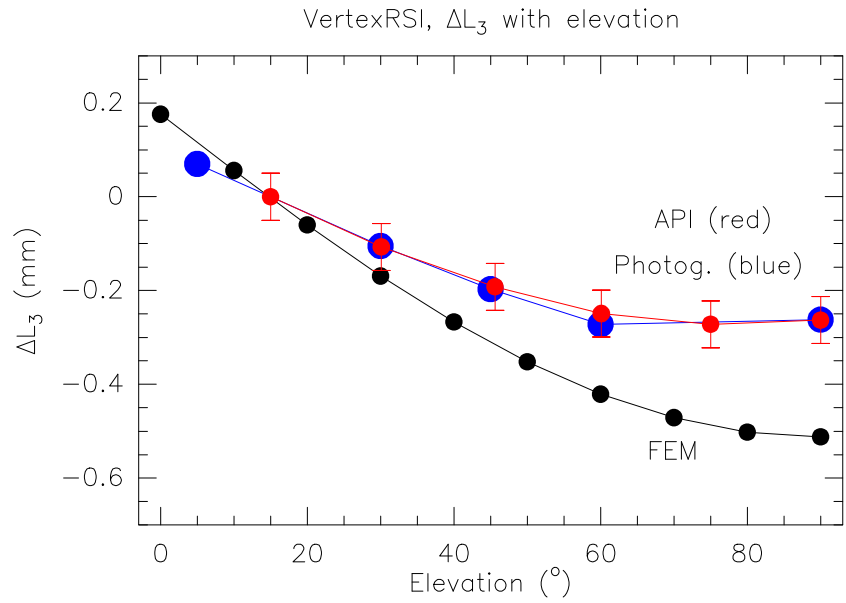}
\caption{VertexRSI antenna. Path length variation $\Delta$L$_{3}$ as
function of elevation of the reflector, measured with the API and by
photogrammetry. The curve indicated FEM is the calculated variation.}
\label{fig:vertex-el-l3}
\end{figure}

\begin{figure}
\centering
\includegraphics[scale=0.9]{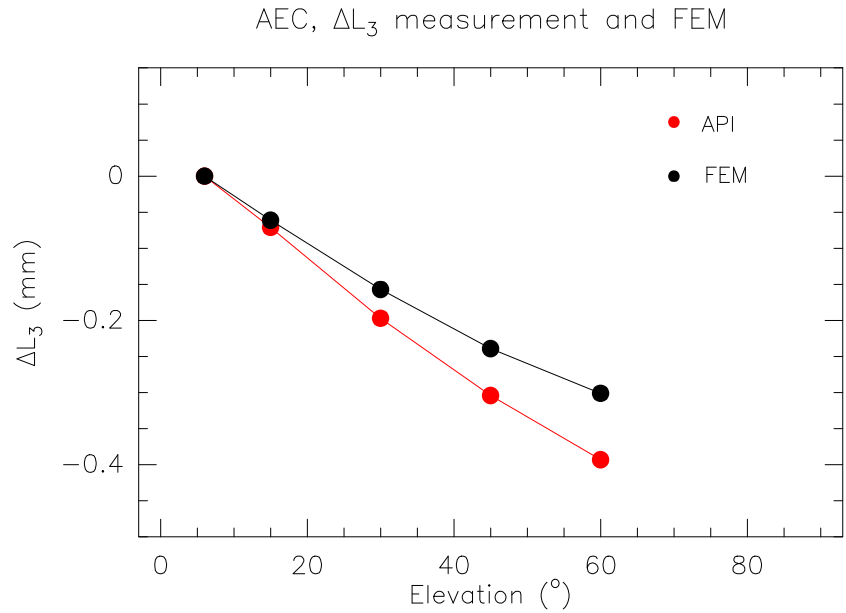}
\caption{AEC antenna. Path length variation $\Delta$L$_{3}$ as
  function of elevation of the reflector. The curve indicated FEM is
  the calculated variation.}
\label{fig:aec-el-l3}
\end{figure}

Because of gravity induced deformations, on both antennas the path length
L$_{3}$ will change with elevation of the reflector. To measure this effect, 
the laser emitter of the API5D was installed on a mount (specially constructed
for the VertexRSI antenna; the laser tracker platform on the AEC
antenna) at the reflector vertex, the retro--reflector was installed
on the subreflector.  The measurements of the L$_{3}$ path length
variation as function of elevation are reliable because the
measurements are insensitive to a small tilt of the laser mount relative
to the antenna structure, and hence of the laser beam reflected in the
retro--reflector. The antennas were tipped in elevation in steps
of 15$^\circ$, between 5$^\circ$ (AEC) or 15$^\circ$ (VertexRSI) and
90$^\circ$ elevation, and the variation of $\Delta$L$_{3}$(E) was
recorded.  These path length 
measurements, along with the path length variation predicted from a
FEM calculation of the antenna structure, are shown in
Figures~\ref{fig:vertex-el-l3} and \ref{fig:aec-el-l3}. The
repeatability of the API5D measurements is approximately $\pm 5\mu$m.
Note that in Figure~\ref{fig:vertex-el-l3} the change in subreflector
position as measured 
by the ALMA Antenna IPT using photogrammetry agrees well with the API
measurement.  Even though there may exist a difference between the
measured and calculated variation of L$_3$ with elevation, the results
indicate that the variation of $\Delta$L$_3$ must be taken into
account in the operation of the interferometer.  They are easily put
into analytic form.

\subsection{Accelerometer Measurements of Path Length Stability}
\label{pathaccel}

The accelerometer system described in \S\ref{accelsurf} and
\cite{Snel2006} was used to derive the \textit{total} path length
stability of the antenna. These measurements were derived from the four 
accelerometers on the rim of the BUS, one accelerometer on the
receiver flange, and an accelerometer on the apex structure.  The
boresight motion of the BUS is subtracted from twice the boresight
motion of the apex structure, which results in the total path length
variation with respect to ground.  As was done for the pointing
results derived from accelerometer measurements (see
\S\ref{accelpoint}) in the following we characterize antenna path
length stability performance for the following situations:

\begin{description}
\item[Stationary Pointing in High Wind:] Wind induced total path length
  variations for both antennas amount to 6~$\mu$m RMS over time scales
  of 15~minutes. These numbers are consistent with the path length
  stability as determined with the API5D instrument (\S\ref{apipath} and
  \cite{Greve2006}). For antenna wake conditions path length stability
  is 11~$\mu$m RMS over 15~minutes. 
\item[Sidereal Tracking:] For sidereal tracking, total path length
  variations over time scales of 1~second remain below 2~$\mu$m for
  the VertexRSI antenna, and below 0.5~$\mu$m for the AEC antenna. 
\item[Fast Motion:] Total path length stability during fast OTF
  scans, with 0.5 deg/s scan rate, averages for the VertexRSI antenna
  to 12~$\mu$m RMS over time scales of 1~second. The AEC antenna
  path length stability averages to 3.3~$\mu$m RMS. When the scan rate is
  reduced to 0.05 deg/s for interferometric mosaicing, total path 
  length stability of the VertexRSI antenna averages to 3.1~$\mu$m RMS
  over time scales of 1~second, and 0.7~$\mu$m RMS for the AEC
  antenna.
\end{description}

\section{Limitations of the Evaluation}

In fulfilling its task the AEG was limited by a number of
circumstances. Most importantly the atmospheric conditions at the VLA
site did not allow the antennas to be tested at their highest operating
frequencies. It is at the edges of the operational range where the
antenna parameters can be most effectively determined by radiometric
measurements. This formed the most serious
limitation to our evaluation program. Instead, we have employed a
number of optical-mechanical measurement devices in an effort to
obtain sufficient material for a valid evaluation of the antennas.

Large delays in the delivery of the antennas shortened the
evaluation process and limited the extent to which performance
specifications could be tested. The VertexRSI antenna was marginally
put at our disposal in March 2003, the AEC antenna in January
2004. Neither antenna was fully functional at those
times. Summarizing, our ability to measure the full performance of the
prototype antennas was limited in the following areas:

\begin{enumerate}
\item The holographic surface measurement was limited to one elevation
  angle. Some data on the reflector surface deformation as a function
  of elevation was obtained with optical methods.
\item The atmospheric conditions at the VLA site are quite different
  from those at the ALMA site on Chajnantor, Chile. Comparing the
  temperature and wind variations encountered at the VLA site with
  those at Chajnantor, it has been possible to draw meaningful
  conclusions regarding the behaviour of the antennas under the
  conditions specified for Chajnantor. However, at the VLA the
  atmospheric transmission only allows observations at the longer
  mm-wavelengths and over a limited (winter) period of the year. This
  has significantly impeded the quantity and quality of the
  radiometric tests. Moreover, we found that the VLA site has poor
  seeing - seldom better than 4~arcsec. This made the optical
  measurements of the important tracking and offsetting performance
  below 1~arcsec difficult to determine.
\item Radiometric pointing measurements are the final criterion for
  the determination of the pointing behaviour. The system temperatures
  of the evaluation receivers are excellent; however the limits set by
  gain fluctuations were factors of order 20 (at 3mm) and 10 (at 1mm)
  above the thermal fluctuations. Thus our limiting sensitivity was
  entirely set by gain stability at the 10 Hz switching frequency,
  restricting the number of available sources to $\sim 20$ at 3mm
  wavelength. The nutator could only operate reliably under quiet or
  moderately windy conditions (a few m/s). Only a limited set of data
  could be collected at the VertexRSI antenna, and even less on the
  AEC antenna. 
\item Some specifications fell outside our specific charter, in
        particular:
  \begin{enumerate}
  \item Time needed to remove and attach the antenna to its
        foundation.
  \item Alignment errors after a drop of the antenna during
        transportation.
  \item Mechanical and electrical requirements.
  \end{enumerate}
\end{enumerate}

Considering these circumstances, it is gratifying to report that we
were able to collect an amount of data that allows us to draw
meaningful conclusions as to the performance of the ALMA prototype
antennas. The test program on the VertexRSI antenna began in March
2003, but really useful data were collected only from October
2003 onwards. The AEC antenna program was started in January 2004. The
entire program was concluded by the end of May 2004. Clearly, this
time span did not allow us to make reliable statements on the
long--term behaviour of the antennas.

\section{Conclusions}

The overall design and performance of the ALMA prototype antennas
makes either of them attractive options for production antennas that
will satisfy the stringent ALMA requirements.

\textit{Acknowledgements: The contributions of the following individuals were
  necessary for the success of the ALMA prototype antenna evaluation
  process: Marc Rafal (Commonwealth Technical Associates LLC); Fritz
  Stauffer, Nicholas Emerson, Jinquan Cheng, Jack Meadows (NRAO);
  Angel Ot\'arola (ESO); Jos\'e Lopez-Perez (OAN); 
  David Smith (MERLAB); Michael Bremer (IRAM); and Henry Matthews
  (HIA).}

\end{document}